\documentclass[aps,prmat,reprint,floatfix]{revtex4-1}
\usepackage[T1]{fontenc}       

\usepackage{hyperref}
\usepackage{graphicx}
\usepackage{multirow}
\usepackage{color}
\usepackage{amsmath,amsfonts,amssymb}
\let\oldAA\AA
\renewcommand{\AA}{\text{\normalfont\oldAA}}

\newcommand{\tas}{TaS$_3$}
\newcommand{\nbse}{NbSe$_3$}

\begin{document}

\title{Electron-hole response function of transition metal trichalcogenides \texorpdfstring{NbSe$_3$}{NbSe3} and monoclinic-\texorpdfstring{TaS$_3$}{TaS3}}

\author{Bogdan Guster}
\altaffiliation[Current address: ]{Institute of Condensed Matter and Nanosciences, Universit\'{e} Catholique de Louvain, Chemin des \'{e}toiles 8, bte L07.03.01, B-1348 Louvain-la-Neuve, Belgium}
\affiliation{Catalan Institute of Nanoscience and Nanotechnology (ICN2), CSIC and The Barcelona Institute of Science and Technology, Campus Bellaterra, 08193 Barcelona, Spain}

\author{Miguel Pruneda}
\affiliation{Catalan Institute of Nanoscience and Nanotechnology (ICN2), CSIC and The Barcelona Institute of Science and Technology, Campus Bellaterra, 08193 Barcelona, Spain}

\author{Pablo Ordej\'on}
\affiliation{Catalan Institute of Nanoscience and Nanotechnology (ICN2), CSIC and The Barcelona Institute of Science and Technology, Campus Bellaterra, 08193 Barcelona, Spain}

\author{Enric Canadell}
\affiliation{Institut de Ci\`encia de Materials de Barcelona (ICMAB-CSIC), Campus Bellaterra, 08193 Barcelona, Spain}

\author{Jean-Paul Pouget}
\affiliation{Laboratoire de Physique des Solides, CNRS UMR 8502, Universit\'e de Paris-Sud, Universit\'e Paris-Saclay, 91405 Orsay, France}

\begin{abstract}
    NbSe$_3$ and monoclinic-TaS$_3$ ($m$-TaS$_3$) are quasi-1D metals containing three different types of chains and undergoing two different charge density wave (CDW) Peierls transitions at T$_{P_1}$ and T$_{P_2}$. The nature of these transitions is discussed on the basis of first-principles DFT calculation of their electron-hole Lindhard response function. As a result of stronger inter-chain interactions the Fermi surface (FS) and Lindhard function of NbSe$_3$ are considerably more complex than those for $m$-TaS$_3$; however a common scenario can be put forward to rationalize the results. The intra-chain inter-band nesting processes dominate the strongest response for both type I and type III chains of the two compounds.  Two well-defined maxima of the Lindhard response for NbSe$_3$ are found with the (0$a$*, 0$c$*) and (1/2$a$*, 1/2$c$*) transverse components at T$_{P_1}$ and T$_{P_2}$, respectively, whereas the second maximum is not observed for $m$-TaS$_3$ at T$_{P2}$. Analysis of the different inter-chain coupling mechanisms leads to the conclusion that FS nesting effects are only relevant to set the transverse $a$* components in NbSe$_3$. For the transverse coupling along $c$* in NbSe$_3$ and along both $a$* and $c$* for $m$-TaS$_3$, one must take into account the strongest inter-chain Coulomb coupling mechanism. Phonon spectrum calculations show the formation of a giant 2$k_F$ Kohn anomaly in $m$-TaS$_3$. All these results support the weak coupling scenario for the Peierls transition of transition metal trichalcogenides.      
\end{abstract}

\keywords{Transition metal trichalcogenides, charge density waves, density functional theory, Lindhard response function}
\pacs{}

\maketitle

\section{Introduction}\label{sec:intro}

Since the discovery of the charge density wave (CDW) instability in several families of one- (1D) and two-dimensional (2D) conductors such as the Krogmann salts~\cite{Comes1973}, organic charge transfer salts~\cite{Jerome1982} and transition metal dichalcogenides~\cite{Wilson1975}, the unconventional physics associated with these instabilities~\cite{Gorkov1989} as well as the search for new families of CDW materials (for recent reviews see~\cite{Monceau2012,Pouget2016}) has been the focus of continued attention. The basic mechanism of the CDW instability is well understood for 1D metals~\cite{Peierls1955}. Due to their simple band structure the Lindhard response function, which depends of the electronic dispersion in the vicinity of the Fermi level~\cite{Chan1973}, exhibits a sharp maximum at the 2$k_F$ wave vector ($k_F$ is the Fermi wave vector of the 1D electron gas) which induces a CDW (electron-hole) modulation with precisely this 2$k_F$ wave vector. Almost simultaneously the CDW triggers a periodic lattice distortion (PLD) of the lattice through the electron-phonon coupling. Since this coupling generally occurs with acoustic-like phonon branches, the PLD observed in 1D conductors usually consists of a modulation wave of bond distances known as bond order wave (BOW) in the literature~\cite{Pouget2016}. In 1D metals the coupled 2$k_F$ CDW/BOW instability drives a metal-insulator transition predicted by Peierls~\cite{Peierls1955} long time before its discovery in the Krogmann salts~\cite{Comes1973}. Since the charge density is modulated with the 2$k_F$ wave vector which depends on the band filling, the CDW is often incommensurate with respect to the lattice periodicity. Such incommensurate CDWs thus can collectively slide under the action of an external electric field, as predicted by Fr\"ohlich~\cite{Frohlich1954} and observed for the first time in NbSe$_3$~\cite{Monceau1976} and later in other quasi-1D metals like the blue bronze~\cite{Dumas1983}.

\begin{figure}[!htbp]
    \centering
    \includegraphics[width=0.45\textwidth]{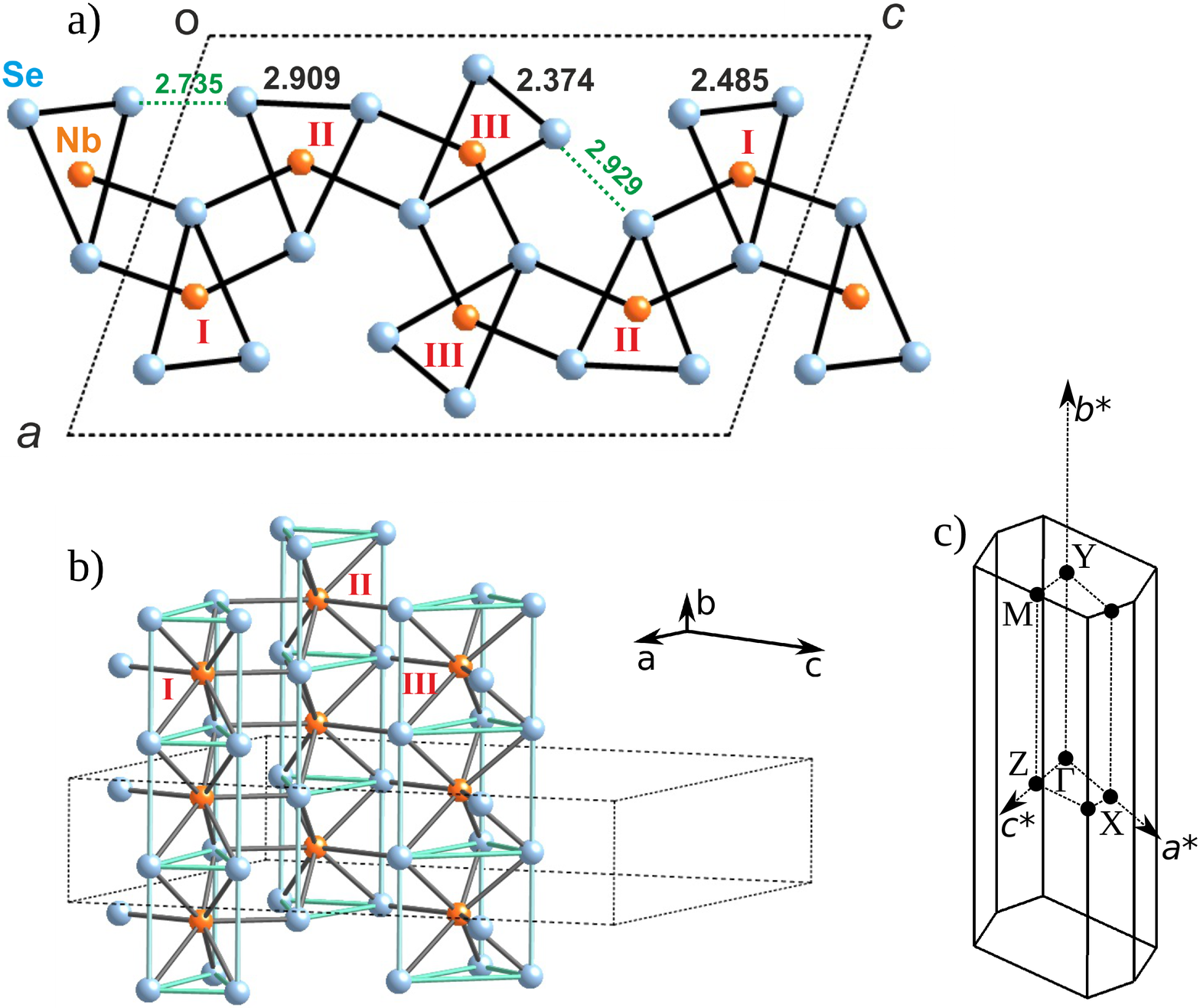}
    \caption{(a) and (b) Crystal stucture of \nbse~at room temperature. The labels I, II and III refer to the three different types of chains discussed in the text. The $m$-\tas~structure is completely equivalent. All distances are expressed in \AA. (c) Brillouin zone for NbSe$_3$ and $m$-\tas.}
    \label{fig:struct_2}
\end{figure}

NbSe$_3$ and monoclinic-TaS$_3$ rank among the most studied CDW materials. NbSe$_3$, whose structure is shown in Fig.~\ref{fig:struct_2}~\cite{Hodeau1978}, is a paradigmatic example of the unique physics of pseudo-1D metals. Monoclinic-TaS$_3$~\cite{Meerschaut1981} (from now on simply $m$-TaS$_3$) exhibits the same structure where three different types of MX$_3$ trigonal prismatic chains lead to MX$_3$ layers in the ($b$,$c$) plane through the formation of interchain M-X bonds. Both low-dimensional solids are room temperature metals and undergo two CDW instabilities when lowering the temperature (for a recent review see~\cite{Monceau2012}). For $m$-TaS$_3$ the first modulation, with wave vector $q_1$= (0, 0.254(3) $b$*, 0), occurs at T$_{P1}$= 240 K whereas the second, with wave vector $q_2$= ($a$*/2, 0.245(3) $b$*, $c$*/2), occurs at T$_{P2}$= 160 K~\cite{Roucau1980}. NbSe$_3$ experiences two successive Peierls transitions at T$_{P1}$ = 144 K and T$_{P2}$ = 59 K associated with structural modulations with wave vectors $q_1$ = (0, 0.243(3) $b$*, 0) and $q_2$ = ($a$*/2, 0.259(3) $b$*, $c$*/2), respectively~\cite{Hodeau1978,Fleming1978}. Local NMR~\cite{Devreux1982,Ross1986} and STM studies~\cite{Brun2009} as well as the structural refinement of the modulated structures~\cite{Smaalen1992,Smaalen1993} establish that the first transition affects mainly type III chains (see Fig.~\ref{fig:struct_2} for the labeling), while the second transition affects mostly type I chains. An important difference between the two systems is that after the two CDW transitions $m$-TaS$_3$ is semiconducting whereas NbSe$_3$ keeps its metallic character.  

For a longtime most of the theoretical studies of CDW materials were based on model hamiltonians~\cite{Gorkov1989}. Only recently first-principles calculations of the band structure, phonon spectra and electron-hole Lindhard response function based on the real crystal structure of the materials have been used to quantitatively understand the CDW instability.~\cite{JMH06,Guster2019} Very recently, we have performed such calculations for the blue bronze, K$_{0.3}$MoO$_3$~\cite{Guster2019} and based on the results we have been able to show that its metal-insulator transition can be well accounted for within the framework of the weak electron-phonon coupling theory of the Peierls transition. However, this is not necessarily the case for other CDW materials. In fact, the nature of many CDW instabilities, as those of transition metal di- and tri-chalcogenides, is still debated after almost forty years of intense research. For instance, the first-principles Lindhard response calculated for both bulk~\cite{JMH06} and single-layer 2$H$-NbSe$_2$~\cite{Guster2019NbSe2} clearly show that there is no clear maximum that can account for the nearly 3$\times$3 modulation of this material so that a weak coupling mechanism does not seem to be appropriate. 

Here we report and analyse the first-principles Lindhard function calculation for NbSe$_3$ and $m$-TaS$_3$ for which the mechanism of the Peierls transition is still far from being understood. Although the electronic structure of these solids has been the subject of several studies~\cite{Bullett1979,Hoffmann1980,Shima1982,Shima1983,Canadell1990,Schafer2001,Nicholson2017,Valbuena2019}, the Lindhard response function has never been reported hampering a full discussion of the microscopic origin of the $q_1$ and $q_2$ modulations. This question has been raised again by several recent experimental investigations of the electronic structure in particular via ARPES measurements. While the first ARPES measurements pointed out the importance of Fermi surface (FS) nesting processes ~\cite{Schafer2001,Schafer2003}, more recent investigations emphasized the role of intra-~\cite{Nicholson2017} and inter-chain~\cite{Valbuena2019} Coulomb interactions. The present study usefully complements our recent work on the blue bronze~\cite{Guster2019} since both materials are quasi-1D metals and exhibit non-linear conductivity yet many experimental results suggest that the mechanism of the CDW instability in the two materials must differ significantly~\cite{Monceau2012}.

In this work, as well as in recent studies~\cite{JMH06,Guster2019} considering the charge response of a low dimension electron gas to an external potential caused by the coupling to the phonon field, the Lindhard response is taken as a scalar quantity (note that a tensorial form of the Lindhard response should be used to describe the inter-atomic response when phonon dynamics is considered). $\chi(q,\omega)$ is generally defined as a complex quantity whose real part for $\omega$= 0, probes the tendency of the system to exhibit a CDW instability and whose imaginary part corresponds to the density of states of ($q,\omega$) electron-hole excitations. In the limit $\omega \rightarrow$ 0, the imaginary part exhibits maxima for nesting conditions of the FS: $\epsilon_i({k})$= $\epsilon_j({k}+{q})$= E$_F$~\cite{JM08}. For 2D metals such as tellurides and dichalcogenides, the maxima of the real and imaginary parts of the Lindhard function are found to be different~\cite{JM08}. Thus, for these materials the simple consideration of the best $q$ nesting conditions of the FS does not imply that the system should undergo a CDW instability at this particular $q$ wave vector. In fact the CDW instability occurs for the $q$ wave vector at which the $\omega$= 0 real part of the Lindhard function (simply called Lindhard function below and given by Eq.~\ref{eq:chi}) exhibits a low temperature divergence; such $q$ divergence is built from multiple connections between $\mid i,k$> and $\mid j,k+q$> electronic states over a large $k$ range connecting $\epsilon_i({k})$ and $\epsilon_j({k}+{q})$ energies from each side of the Fermi level (and not only at the Fermi level). This is the reason why, in spite of previous considerations of nesting properties of the strongly hybridized multisheet FS of NbSe$_3$ probed by ARPES~\cite{Schafer2001,Schafer2003}, we have undertaken the direct calculation of the Lindhard function for transition metal trichalcogenides. 

\section{Computational details}\label{sec:computational_details}

DFT calculations~\cite{HohKoh1964,KohSha1965} were carried out using a numerical atomic orbitals approach, which was developed for efficient calculations in large systems and implemented in the \textsc{Siesta} code~\cite{SolArt2002,ArtAng2008}. We have used the generalized gradient approximation (GGA) to DFT and, in particular, the functional of Perdew, Burke and Ernzerhof~\cite{PBE96}. Only the valence electrons are considered in the calculation, with the core being replaced by norm-conserving scalar relativistic pseudopotentials~\cite{tro91} factorized in the Kleinman-Bylander form~\cite{klby82}. The non-linear core-valence exchange-correlation scheme~\cite{LFC82} was used for all elements. We have used a split-valence double-$\zeta $ basis set including polarization functions~\cite{arsan99}. The energy cutoff of the real space integration mesh was 550 Ry. To build the charge density, the Brillouin zone (BZ) was sampled with the Monkhorst-Pack scheme~\cite{MonPac76} using grids of (21$\times$89$\times$21) {\it k}-points. The phonon band structure for $m$-TaS$_3$ was calculated using the finite differences method within a 1$\times$11$\times$1 supercell considering a \textit{k}-point grid of 5$\times$3$\times$3, an energy cutoff of the real space integration of 2000 Ry and a 50 K Fermi-Dirac smearing. The unit cell was previously relaxed until the forces on the atoms were below 3$\times$10$^{-4}$ meV/$\AA$.

The Lindhard response function,

\begin{equation}\label{eq:chi}
\chi(q)=-\sum_{i,j}\sum_{k}\frac{f_F(\epsilon_i({k}))-f_F(\epsilon_j({k}+{q}))}{\epsilon_i({k})-\epsilon_j({k}+{q})},
\end{equation}

\begin{figure*}[!hptb]
    \centering
    \includegraphics[width=0.875\textwidth]{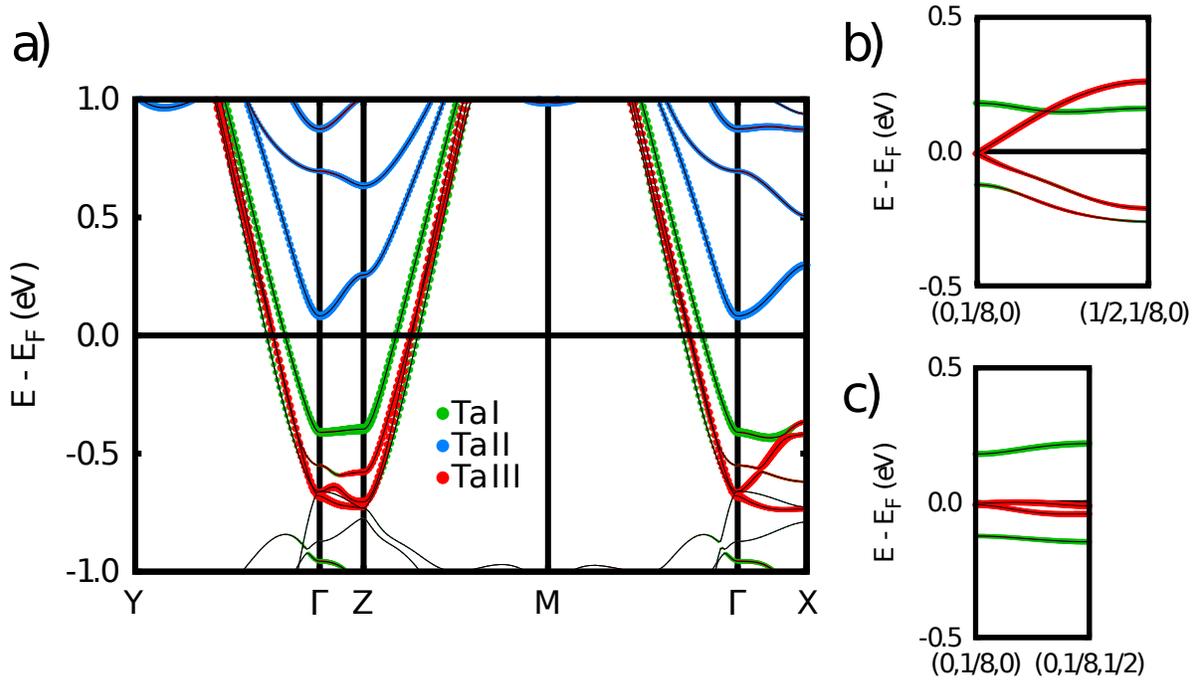}
    \caption{ DFT band structure of $m$-\tas. $\Gamma$= (0, 0, 0), X= (1/2, 0, 0), Y= (0, 1/2, 0), M= (0, 1/2, 1/2) and Z=(0, 0, 1/2) in units of the monoclinic reciprocal lattice vectors are defined in Fig.~\ref{fig:struct_2}c (a). Dispersion relations calculated along the (0, 1/8, 0) to (1/2, 1/8, 0) (b) and (0, 1/8, 0) to (0, 1/8, 1/2) (c) lines of the Brillouin zone. The size of the green, blue and red dots are proportional to the Ta$_I$, Ta$_{II}$ and $Ta_{III}$ character, respectively.}
    \label{fig:tas3_bs}
\end{figure*}

\noindent
was obtained from the computed DFT  band eigenvalues $\epsilon_i({k})$. The integral over {\it k}-points of the BZ was approximated by a direct summation over a dense, regular grid of points. As the Lindhard function is more sensitive to the accuracy of the BZ integration than the total energy, especially in very anisotropic systems, and/or in the presence of hot spots in the band structure (e.g. saddle points with the corresponding van Hove singularity in the DOS), the {\it k}-points grid used for its calculation must be more dense than in the standard self-consistent determination of the charge density and Kohn-Sham energy. The calculations are done, nevertheless, using the eigenvalues obtained in the DFT calculation for the coarser grid, and interpolating their values in the denser grid, using a post-processing utility available within the \textsc{Siesta} package. In this work, for the calculation of the Lindhard response function, the BZ was sampled using a grid of (64$\times$256$\times$64) {\it k}-points. The four partially filled bands for \tas, respectively five for \nbse~were those taken into account in the calculations. Note that Eq.~\ref{eq:chi} is strictly valid for plane waves. In the case of Bloch wave functions each numerator of this equation should incorporate the squared matrix element $\mid <i,k \mid$ exp$(iqr)\mid j,k+q>\mid ^2$~\cite{Ziman1972}. In Section ~\ref{sec:Lindhard} we use the plane wave approximation as is currently used in the literature and we discuss the validity of this approximation in Sect.~\ref{sec:matrix_elements}.

\section{Electronic vs. Crystal structure}\label{sec:electronic_structure}

Although the electronic structures of $m$-TaS$_3$ and NbSe$_3$ have already been reported in the literature~\cite{Hoffmann1980, Canadell1990,Bullett1979,Shima1982,Shima1983,Schafer2001,Nicholson2017,Valbuena2019} it is essential to understand how the details of the crystal structure are related to the band structure and FS in order to fully grasp the information contained in their Lindhard response functions. As shown in Fig.~\ref{fig:struct_2}a, the unit cell of \nbse~ contains six chains of Nb atoms trigonally coordinated with Se atoms running along $b$. As mentioned above, there are three different types of \nbse~ chains; in those of type I and type III one of the Se-Se triangular sides is very short and compatible with a Se-Se bond. However, in chains of type II such distance is too long to be associated with a Se-Se bond. It is important to note (see Fig.~\ref{fig:struct_2}b) that since two adjacent chains are displaced by half the repeat vector along the chain direction ($b$), every transition metal atom is coordinated to six Se atoms of its own chain $and$ two additional Se atoms of the neighboring chains. i.e. they are really eight-coordinated, thus leading to ($b,c$) NbSe$_3$ layers. There are several Se...Se contacts, both intra-layer and inter-layer ones, shorter than twice the van der Waals radii of Se (i.e. 3.8 \AA) conferring some 3D character to this structure.

For electron counting purposes the isolated Se atoms must be considered as Se$^{2-}$ but those involved in Se-Se bonds as (Se$_2$)$^{2-}$. Consequently, the system can be formulated as 2 $\times$ [Nb$_I$(Se$^{2-}$)(Se$_2^{2-}$) + Nb$_{II}$(Se$^{2-}$)$_3$ + Nb$_{III}$(Se$^{2-}$)(Se$_2^{2-}$)]. In other words, there are two electrons to fill the low-lying bands of six \nbse~ chains. Since the Nb atoms of chains II are formally $d^0$ the two electrons will fill the low-lying bands of chains I $and$ III. 
For a transition metal atom in a trigonal prismatic coordination there are three low-lying $d$ orbitals. With a local coordinate axis with the $z$ direction along the chain direction (i.e., $b$) and the bisector of the $x$ and $y$ axes lying on the approximate bisector plane of the chain, these orbitals are $d_{z^2}$, $d_{xy}$ and $d_{x^2-y^2}$. Because the Nb atom of one chain lies on the same plane as the Se atoms of the two neighboring chains (Fig.~\ref{fig:struct_2}b), the $d_{xy}$ and $d_{x^2-y^2}$ orbitals strongly interact with the $p_x$ and $p_y$ of the two Se capping atoms and the two $d$ orbitals are pushed to high energies. Under such circumstances, the only low-lying Nb $d$ orbitals remaining are the $d_{z^2}$ of Nb$_I$ and Nb$_{III}$. Consequently, there are just two electrons to fill the four low-lying bands of NbSe$_3$ which are based on the $d_{z^2}$-type orbitals of the Nb atoms in chains I $and$ III, i.e. a set of four quarter-filled $d_{z^2}$-type bands.    

All these structural and electronic features of NbSe$_3$ are shared by $m$-TaS$_3$. However, since the  sulphur orbitals are less spread than those of selenium, the inter-chain and inter-layer interactions in $m$-TaS$_3$ are weaker thus leading to simpler, less warped FSs. Thus, we will start our analysis considering the electronic structure of $m$-TaS$_3$. The calculated band structure around the Fermi level is shown in Fig.~\ref{fig:tas3_bs}a. In this figure we also present a fatband analysis of the band composition: the size of the green, blue and red circles is proportional to the Ta$_{I}$, Ta$_{II}$ and Ta$_{III}$ character, respectively. It is clear from this figure that the bands based on the Ta$_{II}$S$_3$ chains lie higher than the Fermi level and thus should not be primarily affected by the CDW modulations, and that the Ta$_{III}$S$_3$ and Ta$_{I}$S$_3$ chains lead to the two inner (red) and outer (green) partially filled bands, respectively. 

\begin{figure}[!hptb]
    \centering
    \includegraphics[width=0.30\textwidth]{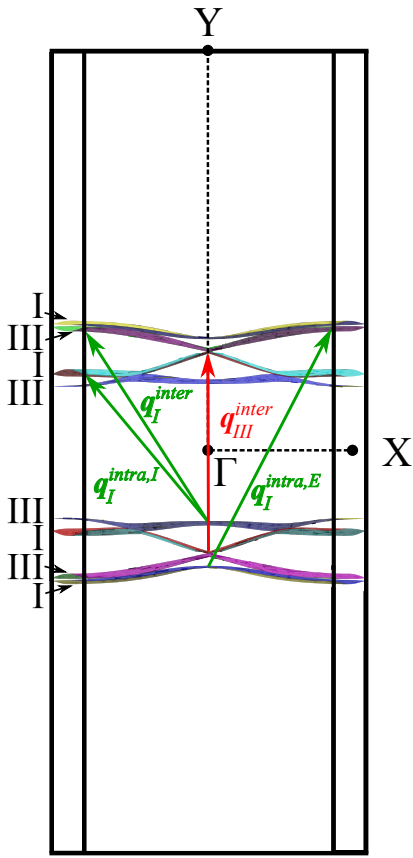}
    \caption{Fermi surface of \tas~(the Brillouin zone is shown in Fig.~\ref{fig:struct_2}c).  The different nesting wave vectors discussed in the text are noted. The labels I/III at the left of the different portions of the FS indicate that these portions originate from chains I/III of the structure.}
    \label{fig:tas3_fs}
\end{figure}

The calculated FS is shown in Fig.~\ref{fig:tas3_fs}. As expected, it contains four pairs of sheets. The two inner ones, originating from the Ta$_{III}$S$_3$ chains, are considerably warped whereas the two outer ones, originating from the Ta$_{I}$S$_3$ chains, are very flat. It is somewhat unexpected that the really warped sheets of the Fermi surface are those associated with the tilted Ta$_{III}$S$_3$ chains whose $q_1$-CDW exhibits nil components along the inter-chain ($c$) and inter-layer ($a$) directions  whereas the very flat sheets are those associated with the Ta$_{I}$S$_3$ chains which exhibit a $q_2$-CDW with 1/2 component along both directions. Looking at the band structure of Fig.~\ref{fig:tas3_bs}a along the $\Gamma \rightarrow$ X direction, it is clear that one of the red bands (Ta$_{III}$S$_{3}$ chains) exhibits a quite sizable dispersion whereas the green ones are considerably flatter (one must be careful when looking at these bands because along $\Gamma$ to X and $\Gamma$ to Z there are some avoided crossings and the top part of two additional mostly sulphur based valence bands also occur around $\Gamma$). This suggests that the chains of type III undergo non negligible inter-chain interactions along $a$ whereas the chains of type I are subject to weaker inter-chain interactions along this direction. In Figs.~\ref{fig:tas3_bs}b and c we show the dispersion of the red and green bands along the $a$* and $c$* directions for a $b$* component of 1/8 ( i.e. practically at the Fermi level). It is clear that the warping of the inner sheets, associated with the red bands, is largely dominated by the interaction along the $a$* (inter-layer) direction. In contrast, for the outer sheets the small warping seems to be due to smaller interactions in both $a$* and $c$* directions.

\begin{figure}[!hptb]
    \centering
    \includegraphics[scale=0.40]{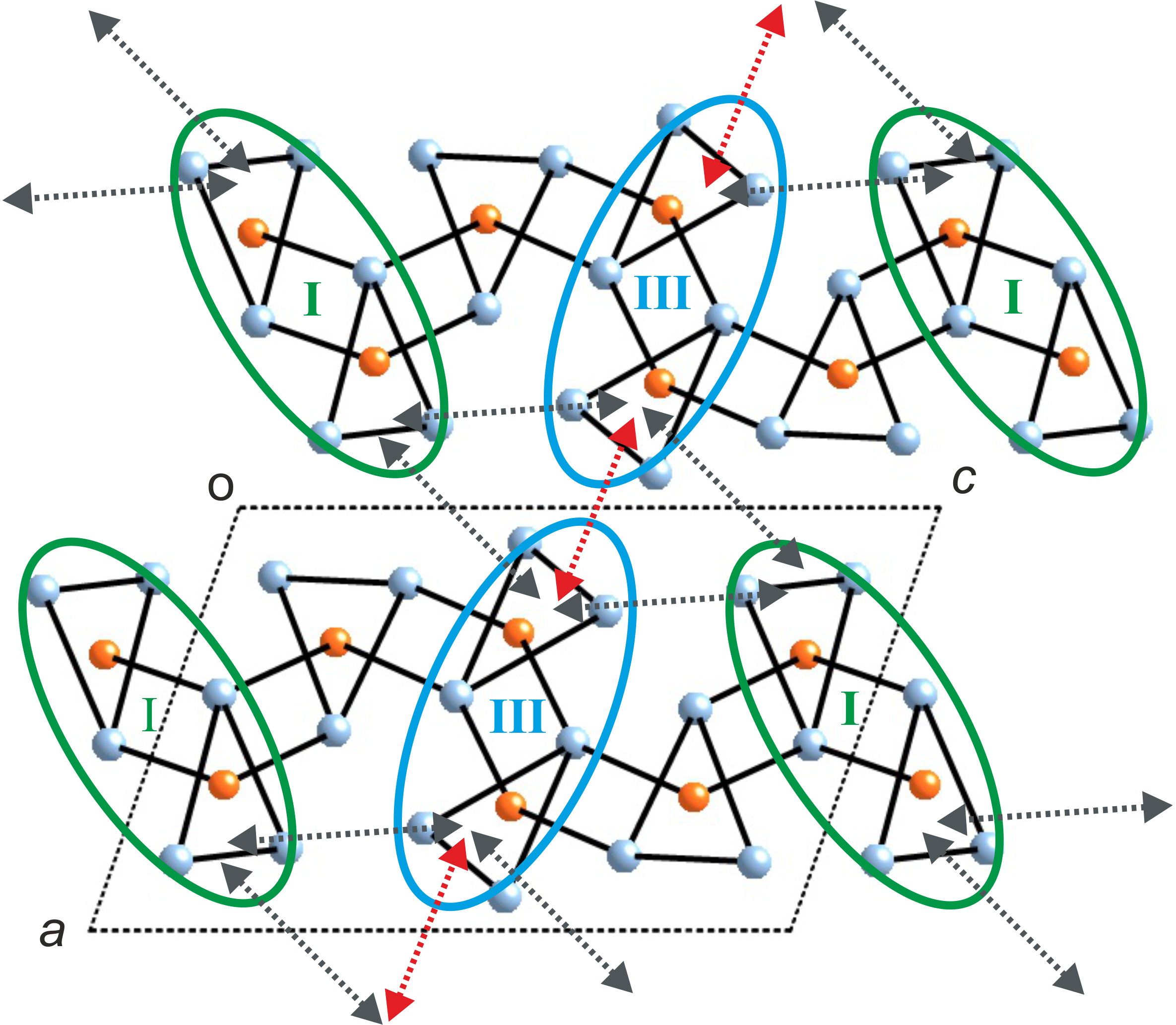}
    \caption{Inter- and intra-layer interactions associated with X...X (X: S or Se) contacts shorter than the sum of the van der Waals radii between the pairs of chains of type I and/or III in the crystal structure of $m$-TaS$_3$ and NbSe$_3$.}
    \label{fig:coupling}
\end{figure}

\begin{figure*}[!hptb]
    \centering
    \includegraphics[width=0.875\textwidth]{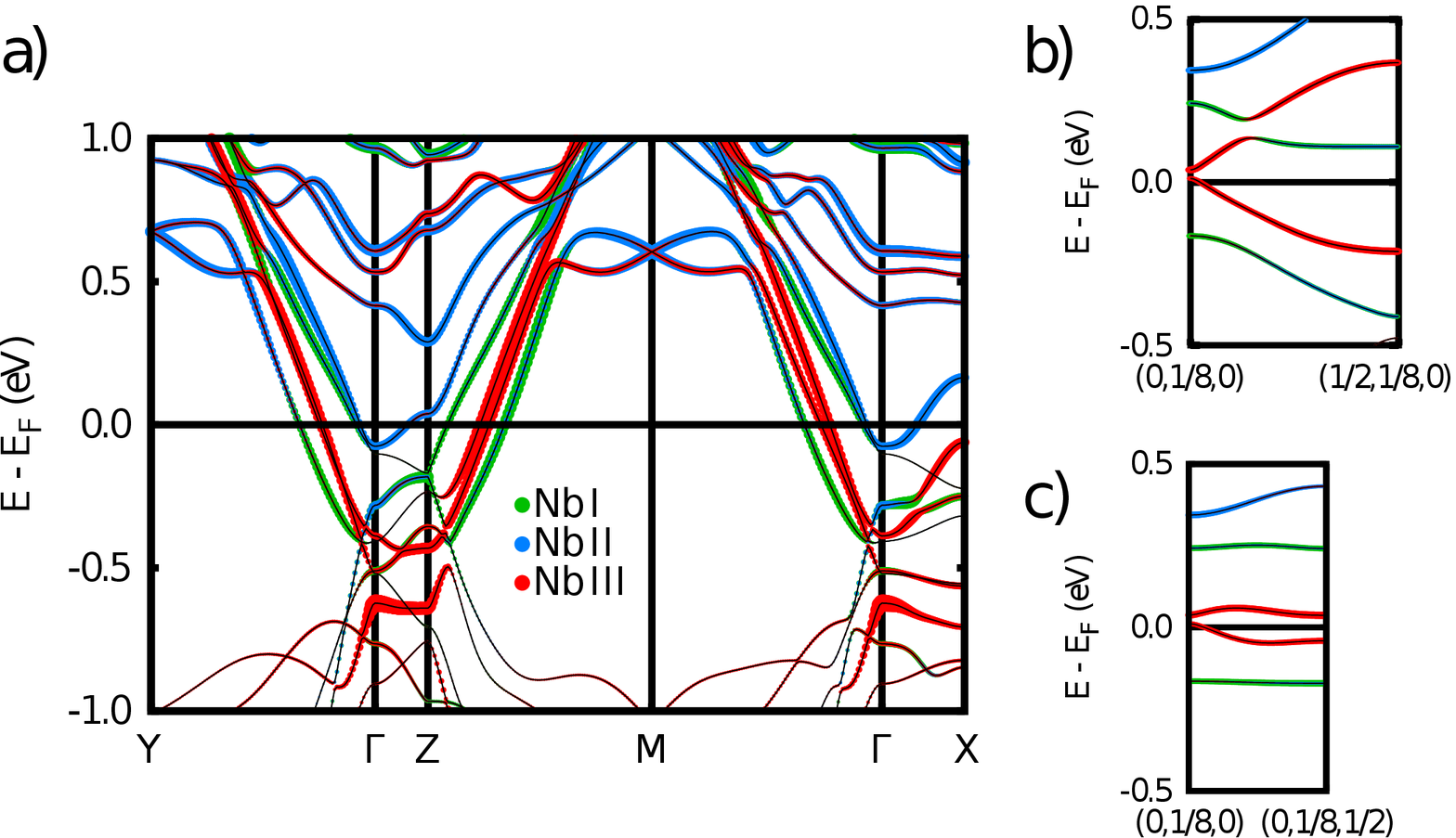}
    \caption{DFT band structure of \nbse. $\Gamma$= (0, 0, 0), X= (1/2, 0, 0), Y= (0, 1/2, 0), M= (0, 1/2, 1/2) and Z=(0, 0, 1/2) in units of the monoclinic reciprocal lattice vectors are defined in Fig.~\ref{fig:struct_2}c (a). Dispersion relations calculated along the (0, 1/8, 0) to (1/2, 1/8, 0) (b) and (0, 1/8, 0) to (0, 1/8, 1/2) (c) lines of the Brillouin zone. The size of the green, blue and red dots is proportional to the Nb$_I$, Nb$_{II}$ and Nb$_{III}$ character, respectively.}
    \label{fig:nbse3_bs}
\end{figure*}

\begin{figure}[!hptb]
    \centering
    \includegraphics[width=0.35\textwidth]{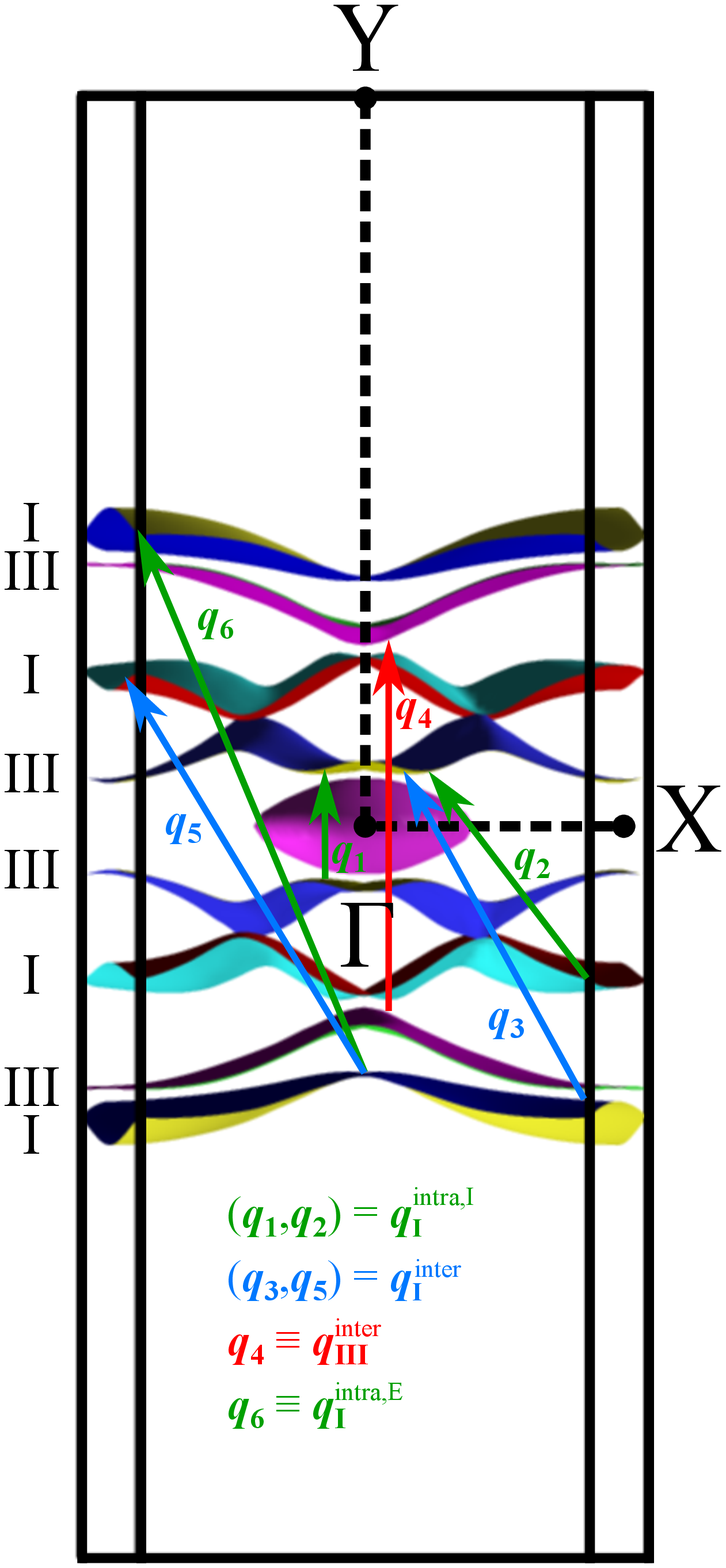}
    \caption{Fermi surface of \nbse~(the Brillouin zone is shown in Fig.~\ref{fig:struct_2}c) The different nesting wave vectors discussed in the text are noted. The labels I/III at the left of the different portions of the FS indicate that these portions originate from chains I/III of the structure.}
    \label{fig:nbse3_fs}
\end{figure}

Analysis of the S...S inter- and intra-layer interactions in $m$-TaS$_3$ (see Fig.~\ref{fig:coupling}) provides useful hints to understand the warping of the Fermi surface sheets. The high temperature transition, occurring on the Ta$_{III}$S$_3$ chains, is due to the coupling between the inner sheets of the FS. Although every one of these sheets is clearly warped, the fact that they have opposite warping makes the two pairs of sheets well nested by a vector with nil $a$* and $c$* components (i.e. the red nesting vector $q_{III}^{inter}$ in Fig.~\ref{fig:tas3_fs}). The reason for this opposite curvature is that the orbitals of the two Ta$_{III}$S$_3$ chains of one layer  lead to in-phase and out-of-phase combinations which thus, when interact directly along the inter-layer $a$* direction (red dotted arrows in Fig.~\ref{fig:coupling}) through several S...S contacts shorter than the sum of the van der Waals radii, they must acquire opposite curvature. In contrast, they are practically non-dispersive along the inter-chain direction $c$* because they are separated by the quartets of Ta$_{II}$S$_3$ and Ta$_{I}$S$_3$ chains. Thus, even if there are quite noticeable inter-chain interactions along the inter-layer direction, the nesting vector has only a $b$* component.

The pairs of Ta$_{I}$S$_3$ chains interact through several S...S short contacts only indirectly through the pairs of chains Ta$_{III}$S$_3$ along $c$ and $\sim$($a$/2)+$c$ (black dotted arrows in Fig.~\ref{fig:coupling})  so that the interaction is weaker. However, the interaction within the pair of chains I is now stronger, as shown by the fact that the two green bands in Fig.~\ref{fig:tas3_bs}a are separated while the red ones (Ta$_{III}$S$_3$) are practically degenerate. This is essentially due to the shorter S...S contacts between the inner S$^{2-}$ atoms in the Ta$_{I}$S$_3$ chains. Note that the larger warping of the inner sheets leads to an unexpected complication: there are very weakly avoided crossings between the inner and outer sheets. Consequently, although the Fermi surface is made of two pairs of slightly warped sheets and every pair can be clearly associated with either chains III or chains I, the existence of these real or avoided crossings as well as regions where the contributions of the two chains practically overlap, blur somewhat the attribution of the nesting wave vectors to specific chains. This will be especially so for NbSe$_3$ because of the stronger Se...Se interactions.

The calculated band structure and Fermi surface for NbSe$_3$ are shown in Figs.~\ref{fig:nbse3_bs} and \ref{fig:nbse3_fs}, respectively. As anticipated, the inter-chain interactions are stronger leading to considerably more warped FSs and a notably larger separation of the two green bands  (Nb$_I$Se$_3$ chains). Yet the main picture correlating the structural and electronic features is still at work. The main difference with the case of TaS$_3$ is that in the present case a fifth band associated with the Nb$_{II}$Se$_3$ chains slightly crosses the Fermi level leading to the appearance of an additional closed pocket around $\Gamma$ in the Fermi surface (see Fig.~\ref{fig:nbse3_fs}). If this additional pocket should occur or not in a perfectly stoichiometric material is still unclear from the experimental viewpoint (see for instance the different experimental results recently reported in refs. \cite{Nicholson2017} and \cite{Valbuena2019}). As a matter of fact, the analysis of the Lindhard response does not lead to any significant variation when this pocket is included or not in the calculation~\cite{note1}. Otherwise the present results concerning the band structure are in very good agreement with previous ARPES studies~\cite{Schafer2001,Nicholson2017,Valbuena2019,Schafer2003} as well as tight-binding [28] and DFT results~\cite{Shima1982,Shima1983,Schafer2001,Nicholson2017,Valbuena2019}.

\begin{table*}[!hptb]
\caption{Wave-vector component and HWHM ($1/\xi_{eh}^b$) along $b$* determined via the three Lorentzians sum fitting of the Lindhard responses of $m$-\tas~along the (0, $q$, 0) and (1/2, $q$, 1/2) directions at 10 K, 200 K and 400 K. The error is indicated in parenthesis.}
\begin{tabular}{|c|c|c|c|c|c|c|}
\hline
             & \multicolumn{2}{c|}{Lorentzian 1}                                                                                        & \multicolumn{2}{c|}{Lorentzian 2}                                                                                      & \multicolumn{2}{c|}{Lorentzian 3}                                                                                        \\ \hline
(0,$b^*$,0)     & $q_I^{intra,I}$ ($b^*$ units) & $1/\xi_{eh}^b (\AA^{-1})$ & $q_I^{inter}$ ($b^*$ units) & $1/\xi_{eh}^b (\AA^{-1})$ & $q_I^{intra,E}$ ($b^*$ units) & $1/\xi_{eh}^b (\AA^{-1})$ \\ \hline
10 K         & 0.192(0)                   & 0.059(1)                                                                                    & 0.247(0)                 & 0.049(1)                                                                                    & 0.303(0)                   & 0.071(1)                                                                                    \\ \hline
200 K        & 0.191(0)                   & 0.061(1)                                                                                    & 0.247(0)                 & 0.063(1)                                                                                    & 0.303(1)                   & 0.076(1)                                                                                    \\ \hline
400 K        & 0.189(1)                   & 0.067(2)                                                                                    & 0.245(0)                 & 0.081(3)                                                                                    & 0.299(2)                   & 0.092(2)                                                                                    \\ \hline
(1/2,$b^*$,1/2)  & $q_I^{intra,I}$ ($b^*$ units) & $1/\xi_{eh}^b (\AA^{-1})$ & $q_I^{inter}$ ($b^*$ units) & $1/\xi_{eh}^b (\AA^{-1})$ & $q_I^{intra,E}$ ($b^*$ units) & $1/\xi_{eh}^b (\AA^{-1})$ \\ \hline
10 K         & 0.195(0)                   & 0.061(1)                                                                                    & 0.243(0)                 & 0.043(1)                                                                                    & 0.295(0)                   & 0.064(1)                                                                                    \\ \hline
200 K        & 0.192(1)                   & 0.061(1)                                                                                    & 0.242(0)                 & 0.061(2)                                                                                    & 0.295(1)                   & 0.070(1)                                                                                    \\ \hline
400 K        & 0.190(1)                   & 0.068(3)                                                                                    & 0.239(0)                 & 0.077(4)                                                                                    & 0.291(1)                   & 0.087(2)                                                                                    \\ \hline
\end{tabular}
\label{table:I}
\end{table*}

\section{Analysis of the Lindhard function}\label{sec:Lindhard}

In this section we first describe the Lindhard function of $m$-\tas~ which exhibits more regular and less hybridized warped open FSs associated with the two pairs of chains of type III and I (compare Figs.~\ref{fig:tas3_fs} and \ref{fig:nbse3_fs}). Then we will analyze the more complex case of \nbse~ where warping and hybridization effects between the various sheets of the FS are stronger and where the presence of a closed FS component, associated with a 5th band in the vicinity of the $\Gamma$ point, perturbs the dispersion (see Fig.~\ref{fig:nbse3_bs}). How these results are related to the available experimental information is discussed in detail in Sect.~\ref{sec:discussion}. 

As noted above, different sheets of the FS can be associated with different types of chains of the structure therefore we will refer to the FS portions originating from chain j as type j FS. Since these sheets occur in pairs (there are two chains of each type in the unit cell), it is essential to clearly state the meaning of the different nesting vector labels that will be used along the discussion (see Fig.~\ref{fig:tas3_fs}; note that the vectors shown in the figure are only meant to indicate the FS sheets related by the vector). First, we will use a subscript to indicate the chain to which they are associated. Second, the terms $inter$ or $intra$ will refer to inter-band or intra-band nesting  $within$ a pair of bands associated with the same type of chain. Third, for the $intra$ case we will use an additional label to differentiate the intra-band nesting associated with the two internal (I) or external (E) sheets of a given pair.

\subsection{\texorpdfstring{$m$-TaS$_3$}{TaS3}}

\begin{figure}[!hptb]
    \centering
    \includegraphics[width=0.4710\textwidth]{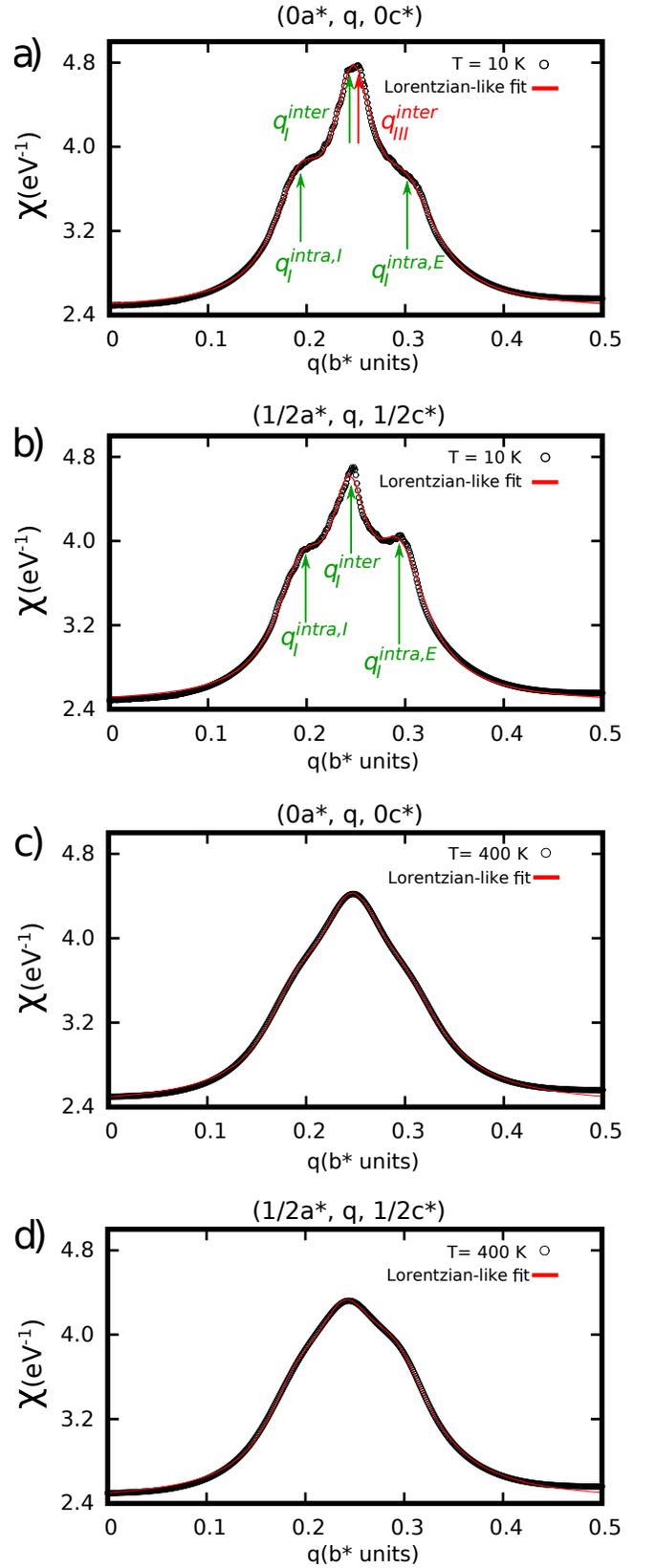}
    \caption{Longitudinal scans of the (0, \textit{q}, 0) and (1/2, \textit{q}, 1/2) Lindhard responses of $m$-\tas~at 10 K ((a) and (b)), and at 400 K ((c) and (d)) together with their fit by the sum of 3 Lorentzians.}
    \label{fig:tas3_lrf_scans}
\end{figure}

Figs.~\ref{fig:tas3_lrf_scans}a and b show (0, $q$, 0) and (1/2, $q$, 1/2) scans of the Lindhard response at 10 K, respectively.  Each scan exhibits three superposed but clearly separated peaks revealing the presence of three well-defined responses. This is also visible for (1/2, $q$,  0) and (0, $q$, 1/2) scans not shown here. As a consequence, the Lindhard response in the $b$* direction can be nicely fitted by the sum of three Lorentzians (the $q$ dependence of an individual electron-hole response has a Lorentzian shape for independent particles~\cite{Jerome1982}). For each scan the strongest response, observed at around the "2$k_F$" $\approx$ 0.25$b$* in-chain component, corresponds to the interband nesting processes $q_i^{inter}$  (i= I or III). Note that:

-	for the (0, $q$, 0) scan, there is a near superposition of the inter-band nesting processes between type III FS and type I FS leading to a plateau of maxima (Fig.~\ref{fig:tas3_lrf_scans}a), 

-	for the (1/2, $q$, 1/2) scan, the dominant inter-band nesting processes between type I FS give rise to a sharper maxima (Fig.~\ref{fig:tas3_lrf_scans}b).

Two weakest responses appear as shoulders at each side of the 2$k_F \approx$ 0.25$b$* maxima. They correspond to two possible intra-band nesting processes for the double sheets associated to type I chains:

-   at 2$k_F \approx$ 0.19$b$* for nesting of the internal FS ($q_I^{intra,I}$).

-	at 2$k_F \approx$ 0.30$b$* for nesting of the external FS ($q_I^{intra,E}$).

The different $q_{III}^{inter}$, $q_I^{inter}$, $q_I^{intra,I}$ and $q_I^{intra,E}$ nesting wave vectors are marked in Fig.~\ref{fig:tas3_fs} and Figs.~\ref{fig:tas3_lrf_scans}a and b. Note that the finding of nearly identical $q_I^{inter}$, $q_I^{intra,I}$ and $q_I^{intra,E}$ peak positions for the Lindhard function in both (0, $q$, 0) and (1/2, $q$, 1/2) scans means that the (weak) transverse dispersion of the FS along $a$* and $c$* does not appreciably change the longitudinal components of the FS nesting instabilities for type I chains. This is not the case for the type III chains where $q_{III}^{inter}$ is detected only for the longitudinal (0, $q$, 0) scan direction. Good Lorentzian fits have been obtained from all Lindhard functions calculated between 10 K and 400 K. For example, the (0, $q$, 0) and (1/2, $q$, 1/2) Lindhard functions calculated at 400 K are shown in Figs.~\ref{fig:tas3_lrf_scans}c and d, respectively. Note that at this temperature one still distinguishes bumps at the position of the two intra-band nesting processes. 

An interesting quantity which can be extracted from these longitudinal fits is the half-width at half-maximum (HWHM) of the Lorentzian of each individual response. As we will see in the discussion (i.e. Sect.~\ref{sec:longitudinal_fluctuations}) the HWHM of the Lorentzian response centered at $q_i$ gives the inverse electron-hole coherence length in the chain direction, 1/$\xi_{eh}^{b}$, associated with the $q_i$ nesting process. The $b$* peak position and its HWHM of the individual Lorentzians fitting the total response are reported in Table~\ref{table:I} for selected temperatures. Note that the HWHM of the three Lorentzians remains well defined at 400 K. Table~\ref{table:I} also shows that fits of the (0, $q$, 0) and (1/2, $q$, 1/2) responses lead to consistent results. We defer to Sect.~\ref{sec:longitudinal_fluctuations} the discussion of the thermal dependence of 1/$\xi_{eh}^{b}$ for $q_{III}^{inter}$ (given in Fig.~\ref{fig:tas3_thermal_dependence_inv_eh}).

\begin{figure}[!hptb]
    \centering
    \includegraphics[width=0.45\textwidth]{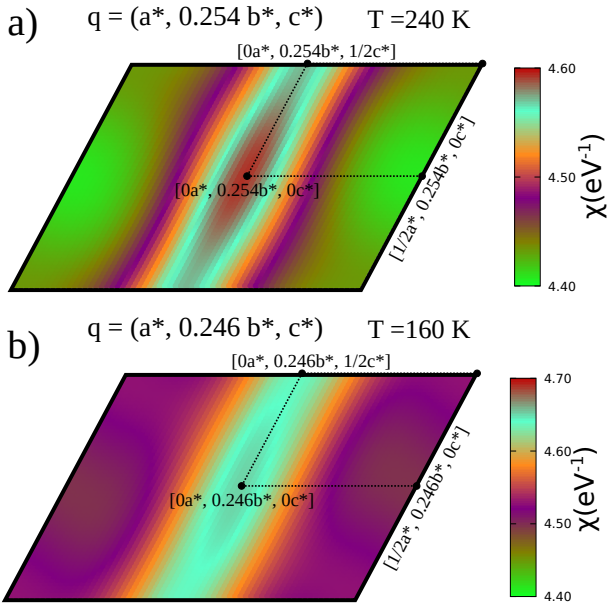}
    \caption{2D transverse plot of the Lindhard function at the 2$k_F$ critical wave vector of: (a) the upper Peierls transition of $m$-\tas~at 240 K, and (b) the lower Peierls transition of $m$-\tas~at 160 K (b).}
    \label{fig:tas3_lrf_maps}
\end{figure}

\begin{figure}[!hptb]
    \centering
    \includegraphics[width=0.45\textwidth]{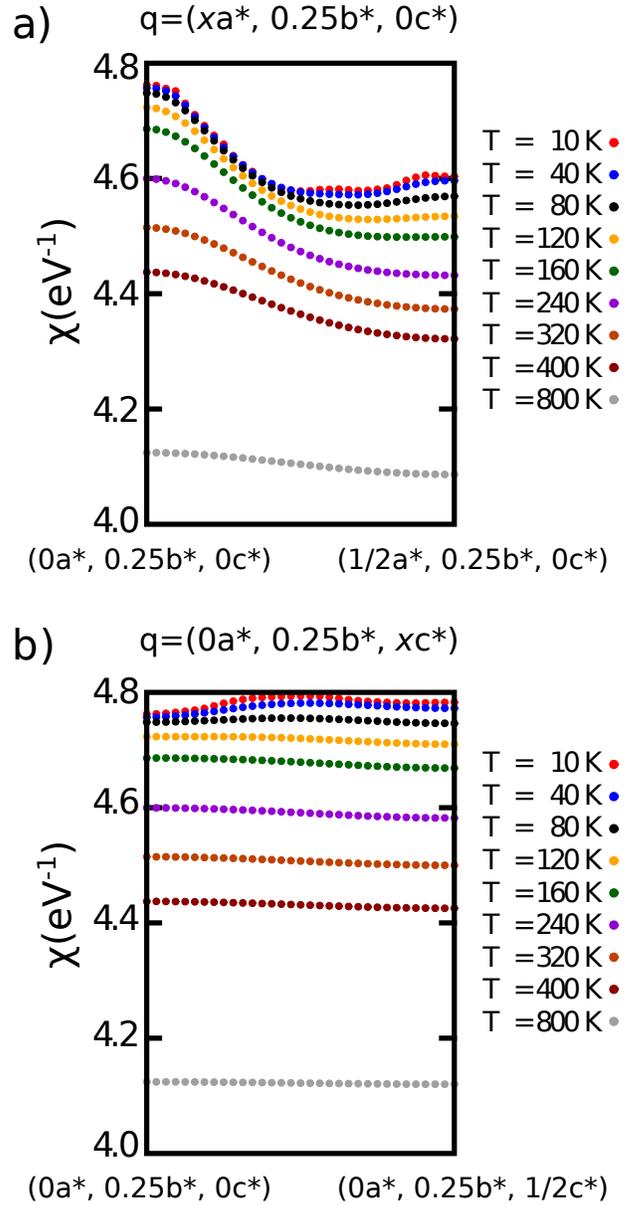}
    \caption{Transverse $a^{*}$ (a) and $c^{*}$ (b) scans across the 0$a$* maximum of the Lindhard function of $m$-\tas~as a function of temperature.}
    \label{fig:tas3_lrf_scan_aa_cc}
\end{figure}

Fig.~\ref{fig:tas3_lrf_maps}a and b give 2D ($a$*, $c$*) transverse plots of the Lindhard response for the critical 2$k_F$= 0.254$b$* wave vector of the T$_{P1}$= 240 K upper Peierls transition and for the critical 2$k_F$= 0.246$b$* wave vector of the T$_{P2}$= 160 K lower Peierls transition of $m$-TaS$_3$, respectively. Note that: 

- At T$_{P1}$= 240 K there is a broad line of maximum intensity centered at about 0$a$* (Fig.~\ref{fig:tas3_lrf_maps}a). This maximum is clearly revealed by $a$* transverse scans (see Fig.~\ref{fig:tas3_lrf_scan_aa_cc}a). This should be contrasted with the result for the $c$* transverse scans (Fig.~\ref{fig:tas3_lrf_scan_aa_cc}b) which do not reveal any appreciable maximum at 0$c$* above T$_{P1}$. 

- At  T$_{P2}$=160 K, when the T$_{P2}$ Peierls transition occurs, the maximum of the Lindhard response expected with 1/2$a$* and 1/2$c$* components  is not observed (Fig.~\ref{fig:tas3_lrf_maps}b). In order to sustain this finding we have performed diagonal ($a$*$\pm c$*) scans (see Fig.~S1 in Supplementary Information (SI)) which show that a secondary maximum located in ($a$*$\pm c$*)/2 appears upon cooling, but only below T$_{P2}$= 160 K.

The thermal dependence of the HWHM of the $a$* response displayed in Fig.~\ref{fig:tas3_lrf_scan_aa_cc}a which amounts to the inverse electron-hole coherence along $a$* (1/$\xi_{eh}^{a^*}$, given in Fig.~\ref{fig:inv_eh_tas3}) will be discussed in Sect.~\ref{sec:transversal_fluctuations}. 

\subsection{\texorpdfstring{NbSe$_3$}{NbSe3}}

\begin{figure}[!hpbt]
    \centering
    \includegraphics[width=0.4825\textwidth]{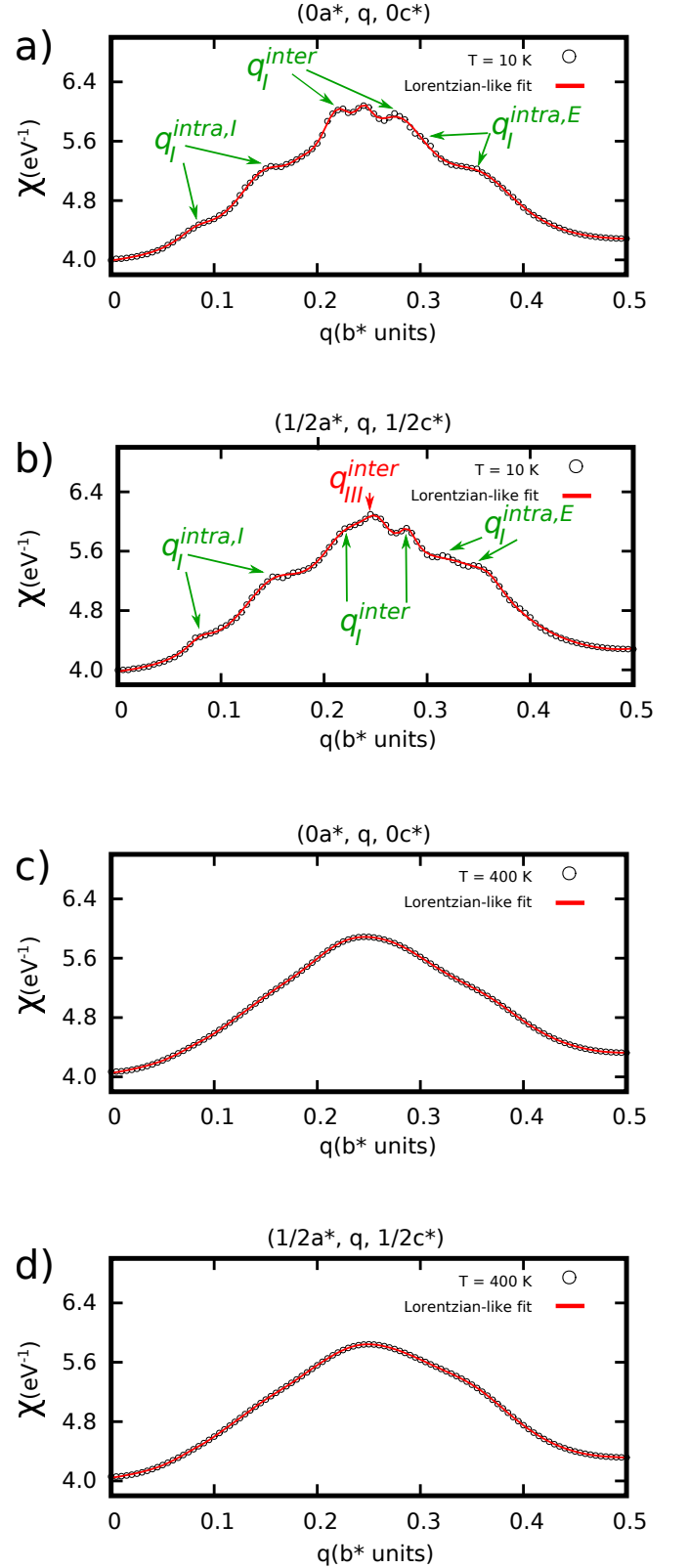}
    \caption{Longitudinal scans of the (0, \textit{q}, 0) and (1/2, \textit{q}, 1/2) Lindhard responses of \nbse~at 10 K ((a) and (b)), and at 400 K ((c) and (d)) together with their fit by the sum of 6/7 Lorentzians at 10 K and 3 Lorentzians at 400 K.}
    \label{fig:nbse3_lrf_scans}
\end{figure}

The Lindhard function of NbSe$_3$, although more complex, keeps the basic features of that for $m$-TaS$_3$. This can be seen by looking at the (0, $q$, 0) and (1/2, $q$, 1/2) scans at 400 K (Figs.~\ref{fig:nbse3_lrf_scans}c and d) which strongly resemble those of $m$-TaS$_3$ (Figs.~\ref{fig:tas3_lrf_scans}c and d). However the longitudinal scans for NbSe$_3$ are much broader than those for the $m$-TaS$_3$. The difference can be clearly realized from the low temperature data where the NbSe$_3$ (0, $q$, 0) and (1/2, $q$, 1/2) scans (see Figs.~\ref{fig:nbse3_lrf_scans}a and b for 10 K) exhibit more maxima than those for $m$-TaS$_3$. Such difference originates from a more complex FS with a larger transverse dispersion along $a$* and numerous band hybridizations as discussed in Sect.~\ref{sec:electronic_structure} (see Fig.~\ref{fig:nbse3_fs}).  

The NbSe$_3$ (0, $q$, 0) and (1/2, $q$, 1/2) scans shown in Figs.~\ref{fig:nbse3_lrf_scans}a and b, respectively, exhibit several overlapping but still distinguishable peaks revealing the presence of 6 or 7 distinct responses. Thus the 10 K (0, $q$, 0) and (1/2, $q$, 1/2) longitudinal Lindhard responses can be fitted by the sum of 6 and 7 Lorentzians, respectively, whose individual $b$* peak positions and HWHM are reported for selected temperatures in Tables~\ref{table:II} and ~\ref{table:III}, respectively.  

\begin{table*}[!hptb]
\caption{Wave-vector component and HWHM ($1/\xi_{eh}^b$) along $b$* determined via Lorentzians sum fitting of the Lindhard response of \nbse~along the (0, $q$, 0) direction at 10 K, 160 K and 400 K. The error is indicated in parenthesis.}
\begin{tabular}{|c|c|c|c|c|c|c|}
\hline
         & \multicolumn{2}{c|}{Lorentzian 1} & \multicolumn{2}{c|}{Lorentzian 2} & \multicolumn{2}{c|}{Lorentzian 3} \\ \hline
(0,$b^*$,0) & $q_1$ ($b^*$ units) & $1/\xi_{eh}^b (\AA^{-1})$ & $q_2$ ($b^*$ units) & $1/\xi_{eh}^b (\AA^{-1})$ & $q_3$ ($b^*$ units) & $1/\xi_{eh}^b (\AA^{-1})$ \\ \hline
10 K     & 0.080(1)        & 0.034(7)        & 0.153(1)        & 0.089(5)        & 0.220(0)        & 0.051(1)        \\ \hline
160 K    & 0.081(1)        & 0.044(3)        & 0.149(1)        & 0.090(3)        & 0.223(1)        & 0.081(4)        \\ \hline
400 K    & -               & -               & 0.136(2)        & 0.095(11)       & -               & -               \\ \hline
         & \multicolumn{2}{c|}{Lorentzian 4} & \multicolumn{2}{c|}{Lorentzian 5} & \multicolumn{2}{c|}{Lorentzian 6} \\ \hline
(0,$b^*$,0) & $q_4$ ($b^*$ units) & $1/\xi_{eh}^b (\AA^{-1})$ & $q_5$ ($b^*$ units) & $1/\xi_{eh}^b (\AA^{-1})$ & $q_6$ ($b^*$ units) & $1/\xi_{eh}^b (\AA^{-1})$ \\ \hline
10 K     & 0.247(1)        & 0.017(3)        & 0.280(0)        & 0.079(5)        & 0.359(1)        & 0.097(9)        \\ \hline
160 K    & 0.248(1)        & 0.023(8)        & 0.282(1)        & 0.091(3)        & 0.361(1)        & 0.099(3)        \\ \hline
400 K    & 0.246(0)        & 0.209(6)        & -               & -               & 0.365(1)        & 0.123(8)        \\ \hline
\end{tabular}
\label{table:II}
\end{table*}

\begin{table*}[!hptb]
\caption{Wave-vector component and HWHM ($1/\xi_{eh}^b$) along $b^*$ determined via Lorentzians sum fitting of the Lindhard response of \nbse~along the (1/2, $q$, 1/2) direction at 10 K, 160 K and 400 K. The error is indicated in parenthesis.}
\begin{tabular}{|c|c|c|c|c|c|c|}
\hline
         & \multicolumn{2}{c|}{Lorentzian 1} & \multicolumn{2}{c|}{Lorentzian 2} & \multicolumn{2}{c|}{Lorentzian 3} \\ \hline
(1/2,$b^*$,1/2) & $q_1$ ($b^*$ units) & $1/\xi_{eh}^b (\AA^{-1})$ & $q_2$ ($b^*$ units) & $1/\xi_{eh}^b (\AA^{-1})$ & $q_3$ ($b^*$ units) & $1/\xi_{eh}^b (\AA^{-1})$ \\ \hline
10 K     & 0.080(1)        & 0.030(4)        & 0.149(1)        & 0.082(4)        & 0.221(0)        & 0.070(8)        \\ \hline
160 K    & 0.081(1)        & 0.044(3)        & 0.147(1)        & 0.090(4)        & 0.222(6)        & 0.097(11)        \\ \hline
400 K    & -               & -               & 0.139(2)        & 0.113(9)        & -               & -               \\ \hline
         & \multicolumn{2}{c|}{Lorentzian 4} & \multicolumn{2}{c|}{Lorentzian 5} & \multicolumn{2}{c|}{Lorentzian 6} \\ \hline
(1/2,$b^*$,1/2) & $q_4$ ($b^*$ units) & $1/\xi_{eh}^b (\AA^{-1})$ & $q_5$ ($b^*$ units) & $1/\xi_{eh}^b (\AA^{-1})$ & $q_6$ ($b^*$ units) & $1/\xi_{eh}^b (\AA^{-1})$ \\ \hline
10 K     & 0.251(1)        & 0.036(6)        & 0.281(1)        & 0.021(3)        & 0.323(1)/0.359(1)        & 0.145(11)/0.033(8)        \\ \hline
160 K    & 0.251(2)        & 0.055(19)       & 0.283(2)        & 0.060(11)       & 0.361(1)/0.350(1)        & 0.032(10)/0.103(3)        \\ \hline
400 K    & 0.245(1)        & 0.197(14)       & -               & -               & \phantom{aaaaaa.}-/0.351(1)               & \phantom{aaaaaa..}-/0.128(6)                \\ \hline
\end{tabular}
\label{table:III}
\end{table*}

The maxima of the Lindhard response scans can be correlated with different nesting processes of the FS (Fig.~\ref{fig:nbse3_fs}) in the following way (the different $q_i$'s are highlighted in both Fig.~\ref{fig:nbse3_fs} and Figs.~\ref{fig:nbse3_lrf_scans}a and b). It appears that :

-	$q_4 \approx$ 0.248$b$* corresponds to the inter-band III FS nesting ($q_{III}^{inter}$)

-	$q_3 \approx$ 0.221$b$*and $q_5 \approx$ 0.281$b$*, whose average is 0.251$b$*, correspond to partial inter-band I FS nesting ($q_I^{inter}$)

-	$q_1 \approx$ 0.08$b$* and $q_2 \approx$ 0.15$b$*, correspond to partial internal intra-band I FS nesting ($q_I^{intra,I}$)

-	$q'_6 \approx$ 0.32$b$* and $q''_6 \approx$ 0.36$b$* correspond to the external intra-band I FS nesting ($q_I^{intra,E}$)

Thus, one recovers the same FS nesting processes discussed for $m$-TaS$_3$ with the addition of a splitting of some nesting wave vectors essentially caused by the strongly perturbed FS sheets associated with the type I bands because of the stronger Se...Se inter-chain interactions. As for $m$-TaS$_3$, one observes nearly identical split sets of $q_I^{inter}$, $q_I^{intra,I}$, and $q_I^{intra,E}$ peak positions for the longitudinal (0, $q$ ,0) and (1/2, $q$, 1/2) scans of the Lindhard function for NbSe$_3$ (Fig.~\ref{fig:nbse3_lrf_scans}). This means that, due to the strongly hybridized nature of the transverse band dispersion along $a$* and $c$*, the transverse components of the split intra- and inter-band nesting processes for the chain I FS sheets are poorly defined. Thus, one should consider that the indication in Fig.~\ref{fig:nbse3_fs} of intra- and inter-band nesting wave vectors between chain I sheets is only indicative.

The (0, $q$, 0) and (1/2, $q$, 1/2) longitudinal responses have been followed upon heating. When T increases the individual responses broaden, so that their separation becomes more difficult to estimate.  However the fit with 6/7 Lorentzians is still reasonable until about 200 K. Fitting of the (0, $q$, 0) and (1/2, $q$, 1/2) scans gives the same $q_i$ peak position, although with a significant dispersion of their HWHMs, especially for the $q_4$ peak. The error on the width of the individual Lorentzians is enhanced when reaching 200 K. Consequently, the fit with 3 Lorentzians, leading to separate maxima at the $q_2$, $q_4$ and $q_6$ positions, is more reliable for T > 200 K. Figs.~\ref{fig:nbse3_lrf_scans}c and d show the three Lorentzians fit of the (0, $q$, 0) and (1/2, $q$, 1/2) longitudinal scans obtained at 400 K. The result of these fits is reported for selected temperatures in Tables~\ref{table:II} and ~\ref{table:III}. Note that the fit with three Lorentzians leads to a considerable jump of the HWHM for the $q_4$ peak (which is not the case for the $q_2$ and $q_6$ peaks) probably because the central Lorentzian now includes the $q_3$, $q_4$ and $q_5$ peaks. 
\begin{figure}[!hptb]
    \centering
    \includegraphics[width=0.405\textwidth]{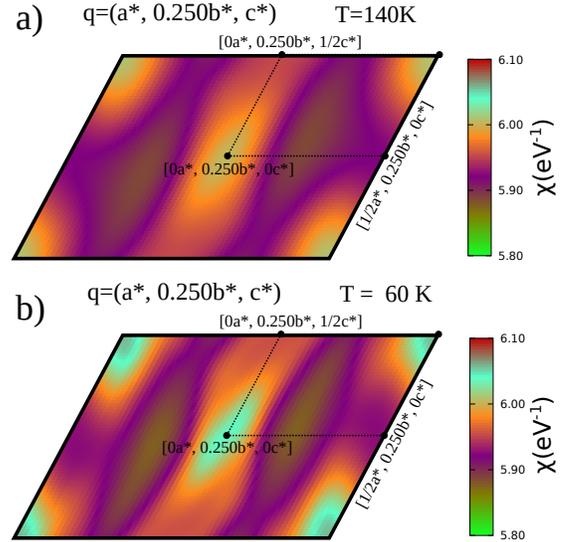}
    \caption{2D transverse plots of the Lindhard function of NbSe$_3$ for <<2$k_F$>> = 0.25 $b^*$ at 140 K(a) and at 60 K(b).}
    \label{fig:nbse3_2d_trans_map}
\end{figure}
\begin{figure}[!hptb]
    \centering
    \includegraphics[width=0.45\textwidth]{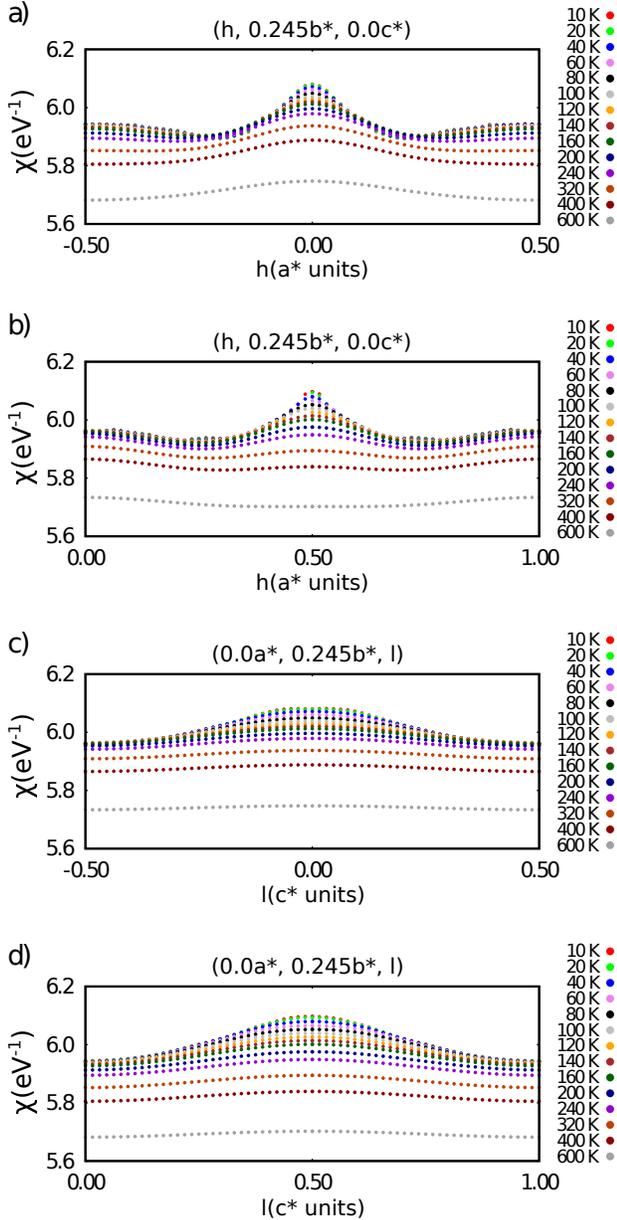}
    \caption{Transverse $a^*$ scans for different $b$* components across the (0$a$*, 0$c$*) and (1/2$a$*, 1/2$c$*) maxima of the Lindhard function of \nbse~as a function of temperature (a),(b), respectively. Transverse $c^*$ scans across the (0$a$*, 0$c$*) and (1/2$a$*, 1/2$c$*) maxima of the Lindhard function of \nbse~as a function of temperature (c),(d), respectively.}
    \label{fig:nbse3_astar_cstar_trans}
\end{figure}

Following the observation of an average of maxima of longitudinal response between  $q_{III}^{inter}$ and $q_I^{inter}$ at "2$k_F$"$\approx$ 0.25$b$*, we report in Fig.~\ref{fig:nbse3_2d_trans_map} the ($a$*, $c$*) 2D plot of the Lindhard response for this wave vector at 140 K and 60 K, close to the upper (T$_{P1}$= 144 K) and lower (T$_{P2}$= 59 K) Peierls transition temperatures. One can clearly observe two well-defined maxima at (0$a$*, 0$c$*) and (1/2$a$*, 1/2$c$*). These transverse components are those giving the best nesting conditions for the different FS shown in Fig.~\ref{fig:nbse3_fs}. The (0$a$*, 2$k_F^{III}$, 0$c$*)  "longitudinal" maximum accounts for the experimental $q_1$-BOW/CDW modulation stabilized at T$_{P1}$ while the (1/2$a$*, 2$k_F^I$, 1/2$c$*) "staggered" maximum accounts for the experimental $q_1$-BOW/CDW modulation stabilized at T$_{P2}$. The (0$a$*, 2$k_F^{III}$, 0$c$*) maximum is more localized in reciprocal space in NbSe$_3$ than in $m$-TaS$_3$. Note that the (1/2$a$*, 2$k_F^I$, 1/2$c$*) maximum is not observed for $m$-TaS$_3$ at T$_{P2}$ (compare Figs.~\ref{fig:nbse3_2d_trans_map} and \ref{fig:tas3_lrf_maps}).

$a$* transverse scans for different $b$* components show that the (0$a$*, 2$k_F^{III}$, 0$c$*) maxima is the strongest for 2$k_F^{III}$= 0.245$b$*. As shown in Fig.~\ref{fig:nbse3_astar_cstar_trans}a, $a$* scans starting from (0$a$*, 2$k_F^{III}$, 0$c$*) exhibit a quite well-defined maximum for 0$a$*. 
The thermal dependence of the HWHM of the $a$* response, corresponding to the inverse electron-hole coherence length along $a$*, $1/\xi_{eh}^{a^*}$, plotted in Fig.~\ref{fig:inv_eh_tas3}, will be discussed in Sect.~\ref{sec:transversal_fluctuations}. The $c$* transverse scans starting from (0$a$*, 2$k_F^{III}$, 0$c$*) exhibit a quite flat maximum around $c$*= 0 (see Fig.~\ref{fig:nbse3_astar_cstar_trans}c). From the HWHM of the $c$* response one gets the inverse electron-hole coherence length along $c$*, $1/\xi_{eh}^{c^*}$, for type III chains of NbSe$_3$. The coherence length thus obtained, $ \xi_{eh}^{c^*} \approx$ 11 \AA~at 140 K varies weakly with temperature. 

Fig.~\ref{fig:nbse3_2d_trans_map} shows that, in contrast with $m$-TaS$_3$, another strong (1/2$a$*, 2$k_F$, 1/2$c$*) zone boundary maximum of the Lindhard response is already clearly visible at T$_{P1}$ and, with an enhanced intensity, at T$_{P2}$. At the latter temperature it is more intense than the (0$a$*, 2$k_F$, 0$c$*) maximum. The relative intensity variation of these two peaks can be more precisely considered by looking at the diagonal transverse scans along the ($a$*$\pm c$*) directions (Fig. S2 in SI). The (1/2$a$*, 2$k_F$, 1/2$c$*) zone boundary maximum, already detected at 400 K, strongly increases upon cooling and becomes stronger than the (0$a$*, 2$k_F$, 0$c$*) maximum below about 100 K. Finally, $a$* transverse scans starting from (1/2$a$*, 2$k_F$, 1/2$c$*) (Fig. S2 in SI) exhibit a well defined maximum for 1/2$a$*. From the HWHM of the $a$* response one can obtain the inverse electron-hole coherence length along $a$* for type I chains. Its thermal dependence, plotted in Fig.~\ref{fig:inv_eh_tas3}, follows $1/\xi_{eh}^{a^*}$ for the type III chains. The $c$* transverse scans starting from (1/2$a$*, 2$k_F$, 1/2$c$*) (Fig. S2 in SI) exhibit a flat maximum around 1/2$c$* and from the HWHM of the $c$* response one gets the inverse electron-hole coherence length along $c$* ($1/\xi_{eh}^{c^*}$) for type I chains of NbSe$_3$. The coherence length thus obtained, $\xi_{eh}^{c^*}\approx $ 11 \AA~ at 60 K, varies weakly with temperature and amounts to the measured value of $\xi_{eh}^{c^*}$ for type III chains. These transverse coherence lengths will be discussed in Sect.~\ref{sec:transversal_fluctuations}. 

\section{Discussion}\label{sec:discussion}

We can now use the Lindhard response function results to examine the relevance of the spatial coupling of electron-hole pairs in driving the CDW fluctuations and the bond-order-wave (BOW) fluctuations preceding the two successive Peierls instabilities of the NbSe$_3$ and $m$-TaS$_3$. This will also allow us to quantitatively discuss the nature of the inter-chain coupling in achieving the successive T$_{P1}$ and T$_{P2}$ Peierls transitions. Finally, we will present some general considerations concerning the strength of the electron-phonon coupling and the critical Peierls lattice dynamics 

\subsection{General shape of the electron-hole response.}\label{sec:general_shape}

The shape of the Lindhard response can be simply explained for $m$-TaS$_3$. Let us first consider its variation in the $b$ chain direction (Fig.~\ref{fig:tas3_lrf_scans}): it consists of a central inter-band response surrounded by two intra-band responses, resembling that previously found for the blue bronze~\cite{Guster2019}. The central response is practically the superposition of two inter-band FS nesting processes for type III and type I chains (see Fig.~\ref{fig:tas3_fs}). The 2$k_F$ wave vectors $\sim$ 0.25$b$* agree with those experimentally determined for $m$-TaS$_3$~\cite{Roucau1980}: 2$k_F^1$= 0.254 $b$* and 2$k_F^2$= 0.245 $b$* at the T$_{P1}$ = 240 K and T$_{P2}$ = 160 K transitions  involving type III and type I chains, respectively. The two other responses, at smaller and higher 2$k_F$ values correspond respectively to the intra-band nesting processes between internal and external pairs of the FS associated with type I chains, which are quite well separated in reciprocal space (Fig.~\ref{fig:tas3_fs}). Such intra-band processes do not occur for pairs of FS associated with chains III which, because of their particular transverse dispersion, cross at a common $k_F$ wave vector already involved in the inter-band nesting process.

The shape of the Lindhard response of NbSe$_3$ is more complex. The longitudinal scans shown in  Figs.~\ref{fig:nbse3_lrf_scans}a and b reveal maxima which are spread over a $q$ range which is around twice broader than the longitudinal response of $m$-TaS$_3$. In addition, 6-7 singularities can be distinguished in the low temperature longitudinal response of NbSe$_3$ instead of only 3 in $m$-TaS$_3$. At high temperatures the Lindhard response of NbSe$_3$ and $m$-TaS$_3$ are similar. This means that the response should be basically decomposed into intra- and inter-band processes as for $m$-TaS$_3$. However, the splitting of low temperature maxima is related to the occurrence of a quite complex FS nesting mechanism (Fig.~\ref{fig:nbse3_fs}) due to the stronger inter-chain interactions. 

By analogy with the interpretation of the Lindhard response of $m$-TaS$_3$ we suggest that:

1. The central longitudinal response at $q_4\approx$ 0.248 - 0.245$b$* corresponds to the inter-band III nesting process ($q_{III}^{inter}$) whose value corresponds to the experimental 2$k_F$ CDW modulation on type III chains measured as 0.2445(1)$b$* at T$_{P1}$,~\cite{Moudden1990}.

2. There is apparently no single response corresponding to the inter-band I FS nesting process ($q_I^{inter}$). Instead, one can find at each side of $q_{III}^{inter}$ two responses at $q_3\approx$ 0.221$b$* and $q_4\approx$ 0.281$b$*. These responses should correspond to partial inter-band I nesting processes between different portions of the FS as shown in Fig.~\ref{fig:nbse3_fs}. Note that the average of these two wave vectors, 0.251$b$*, is close to the 2$k_F$ component, 0.259(3)$b$*, of the experimental CDW type I chain modulation occurring at the T$_{P2}$ transition of NbSe$_3$~\cite{Hodeau1978}.

3. $q_1\approx$ 0.08$b$* and $q_2\approx$ 0.15$b$* seem to correspond also to partial intra-band nesting processes ($q_I^{intra,I}$) between the inner FS of type I chains. The large difference between these wave vectors is due to the quite sizable warping of the internal FS. 

4. $q'_6\approx$ 0.32$b$* and $q''_6\approx$ 0.36$b$* most likely  correspond to the intra-band nesting processes ($q_I^{intra,I}$) between the external FS of type I chains. 

Note that the average value between $q_1$ and $q_2$, 0.12$b$*, is smaller than $q_I^{intra,I}$ = 0.19$b$* for $m$-TaS$_3$ and that the average value of $q'_6$ and $q''_6$, 0.34$b$*, is larger than $q_I^{intra,E}$ = 0.30$b$* for $m$-TaS$_3$. This is a consequence of the larger separation between the FSs of type I chains in NbSe$_3$ because of the stronger inter-chain interactions (Sect.~\ref{sec:electronic_structure}).

In section~\ref{sec:Lindhard} we have decomposed the total Lindhard response $\chi(q)$ of the trichalcogenides into several components $\chi_i(q)$ which individually exhibit a maximum at $q_i$: 
\begin{equation}\label{eq:chi2}
\chi(q)= \sum_{i}\chi_i (q).
\end{equation}
\noindent Near each $\chi_i$ maximum the $q$ dependence of $\chi_i (q)$ can be expanded in powers of ($q - q_i$)$_j$ along the three directions j of an orthogonal frame. For monoclinic trichalcogenenides the decomposition along the orthogonal frame ($a$, $b$, $c$*) of proper directions should be used~\cite{Rouziere1996}. In this frame there are no cross terms in the $q$ expansion. Thus in the vicinity of the maximum $q_i$ one gets
\begin{equation}\label{eq:chi3}
\chi_i(q)= \frac{\chi_i(q_i)}{1+\sum_{j}\xi_j^2(q-q_i)_j^2}.
\end{equation}
\noindent $\chi_i(q)$ has a Lorentzian shape where each term in the ($q - q_i$)$_j^2$ development along the proper direction $j$ involves a coefficient which is homogeneous to a square length
\begin{equation}\label{eq:chi4}
\xi_j^2 = -\Bigg[\frac{\delta ln\chi_i(q)}{\delta q_j^2}\Bigg]_{q_i}.
\end{equation}
\noindent For the component $\chi_i(q)$ of the Lindhard function, $\xi_j$ given in Eq.~\ref{eq:chi4} (and noted $\xi_{eh}^j$ below) is the electron-hole coherence length in the $j$ direction. The $\xi_{eh}^j$s used in this paper are obtained from the best fit of the various component of the DFT Lindhard function with a Lorentzian profile. 1/$\xi_{eh}^j$ of $\chi_i(q)$ is thus directly given by the HWHM along $j$ of the Lorentzian fit of this component. The thermal dependence of the electron-hole coherence length in the chain direction $b$ will be analyzed in Sect.~\ref{sec:longitudinal_fluctuations}, while those in the two transverse directions $a$ or $a$* (close to $a$) and $c$* will be analyzed in Sect.~\ref{sec:transversal_fluctuations}. In particular, the comparison of a transverse coherence length along $j$ with inter-stack distances along the same direction allows to determine how electron-hole pairs located on neighboring distant chains are coupled.

\subsection{Influence of the matrix elements $\mid <i,k \mid$ exp$(iqr)\mid j,k+q>\mid ^2$ in the numerator of the Lindhard response.}\label{sec:matrix_elements}

The electron-hole responses analyzed in Sect.~\ref{sec:general_shape} have been calculated assuming that all matrix elements $\mid <i,k \mid$ exp$(iqr)\mid j,k+q>\mid ^2$ in the numerator of the Lindhard response are equal to the unity (i.e. plane wave approximation of Eq.~\ref{eq:chi}). The matrix element $\mid <i,k \mid$ exp$(iqr)\mid j,k+q>\mid ^2$ takes into account the spatial overlap of the $\mid i,k>$ and $\mid j,k+q>$ Bloch functions of bands $i$ and $j$ respectively. In transition metal trichalcogenides the conduction band structure is primarily built from $d_{z^2}$ orbitals located either on type III or type I chains (see Sect.~\ref{sec:electronic_structure}). As indicated in Fig.~\ref{fig:tas3_bs}, the two inner conduction bands are basically located on chains III and the two outer ones on chains I. Due to the fact that chains III and chains I are spatially separated in the structure (see Fig.~\ref{fig:struct_2}a) and as the $d_{z^2}$ orbitals are directed along the chain direction, the matrix elements associated with the overlap of chain III and chain I wave functions should be much smaller than those associated with the overlap of two chain III or two chain I wave functions. Thus the Lindhard function should be primarily the sum of the separate contributions of the individual chains III and I. Note that the analysis of the Lindhard function in Sect.~\ref{sec:Lindhard} is based on such a decoupling.

In order to check the validity of this assumption, we have separately calculated the intra-chain III and intra-chain I Lindhard responses of $m$-TaS$_3$ at 10 K in the (0, $q$, 0) and (1/2, $q$, 1/2) longitudinal directions by separating the contribution of the inner and the outer bands in the dispersion shown in Fig.~\ref{fig:tas3_bs}a. The intra-chain III and intra-chain I contributions still obtained in the plane wave approximation are shown in Figs. S3a and S3b of the supplementary information, respectively. These partial Lindhard responses exhibit maxima of intensity which are more resolved than those of the total Lindhard function shown in Figs.~\ref{fig:tas3_lrf_scans}a and b. However the $\ll 2k_F\gg$ peaks are located at about the same wave vector. More precisely the intra-chain III response is the strongest for $q= 0.254b$* in the (0, $q$, 0) direction. This value nicely corresponds to the experimental $q_1$ value. The intra-chain I response in the (0, $q$, 0) and (1/2, $q$, 1/2) directions exhibits maxima of similar intensity, but located at slightly different $q$ values: $q= 0.246b$* and $q= 0.254b$* for the two scans, respectively. These maxima also correspond to the experimental $q_2$ and $q_1$ values respectively. This shows that the nesting condition for the FS of chain I is not so well defined as for chains III. However chains I could undergo a single $q_2$-CDW instability at T$_{P2}$ after the removal of the $q_1$ instability by the onset of a $q_1$-CDW below the upper T$_{P1}$ Peierls transition. 

The intra-chain I response exhibits two well defined secondary maxima in the (1/2, $q$, 1/2) scan at 0.18$b$* and 0.31$b$*, which correspond to $q_I^{intra,I}$ and $q_I^{intra,E}$ in Fig.~\ref{fig:tas3_fs}. However such secondary maxima are also observed in the (0, $q$, 0) scan, which means that the intra-band I nesting processes are loosely defined. Secondary maxima can be also guessed in the intra-chain III response. A possible explanation is that they are due to some mixing existing between the inner and outer sets of conduction bands primarily built with chains III and I, as already mentioned in section III. This seems to be particularly the case near the Brillouin zone boundary when FSs primarily associated with chains III and chains I become tangent to each other (see Fig.~\ref{fig:tas3_fs}). We have not performed separate intra-band calculations of the Lindhard response of NbSe$_3$ because there is more mixing between type III and type I bands due to the larger hybridization between the different sets of bands (see Figs.~\ref{fig:nbse3_bs} and~\ref{fig:nbse3_fs}) and because in this case the attribution of the maxima of the Lindhard function to a given set of chain would be uncertain.

Although the calculation of separate intra-chain contributions of the Lindhard response basically validates the nesting scenario proposed in Sect.~\ref{sec:Lindhard} for $m$-TaS$_3$, the calculation of the true Lindhard function incorporating the matrix elements would be important. However such a calculation is difficult for many of the low-dimensional systems of interest. Only in some very recent works such matrix elements have been included~\cite{Divilov2020,Heil2014}. The work of Divilov et al. \cite{Divilov2020} shows that the inclusion of the matrix elements does not appreciably change the $2k_F$ instability of 1D metals (as for instance (CH)$_x$) primarily obtained with a Lindhard response calculated with constant matrix elements. By analogy, one expects that the observation of well defined $2k_F$ maxima in the Lindhard response of 1D metals such as the transition metal trichalcogenides (this work) and the blue bronze~\cite{Guster2019} will persist after inclusion of the matrix elements. In contrast, the work of ref~\cite{Divilov2020} shows that the inclusion of the matrix elements completely alters (washes out) the structure of the response function of 2D metals such as VSe$_2$. The same situation certainly occurs in other 2D transition metal dichalcogenides such as $2H$-NbSe$_2$ where the calculation of individual intra-band components of the response function in the plane wave approximation was unable to exhibit clear-cut $\ll2k_F\gg$ maxima~\cite{JMH06}. In fact, the need for the inclusion of the matrix elements in the calculation of the response function of transition metal dichalcogenides as $2H$-NbSe$_2$ is understandable. The important bands in this case result from the hybridization of different types of orbitals (the  $d_{z^2}$ and $d_{x^2-y^2}/d_{xy}$ ~\cite{JMH06}) and consideration of the different matrix
elements clearly influences the calculated response \cite{Divilov2020}. This is also the case of 3D system like Cr \cite{Heil2014}.

\subsection{Longitudinal electron-hole fluctuations.}\label{sec:longitudinal_fluctuations}

\begin{figure}[!hptb]
    \centering
    \includegraphics[width=0.4825\textwidth]{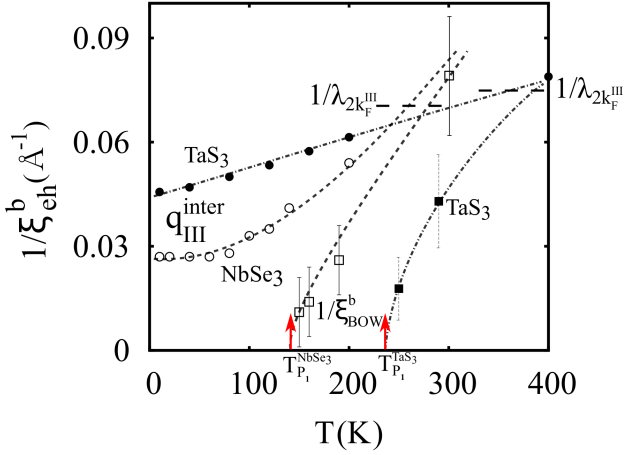}
    \caption{Thermal dependence of the inverse electron-hole coherence length $1/\xi_{eh}^b$ of $m$-\tas~and \nbse~for the responses associated to the $q_{III}^{inter}$ inter-band electron-hole instability (full and empty circles, respectively). The experimental dependence of the inverse BOW correlation length, $1/\xi_{BOW}^{b}$, measured for the upper T$_{P1}$ Peierls transition of $m$-\tas~and \nbse~is also shown (full squares from \cite{Pouget1989} and empty squares from \cite{Moudden1990}). The inverse of the 2$k_{F}^{III}$ wave length for both compounds is indicated. $1/\xi_{BOW}^{b}$ of $m$-TaS$_3$ is extrapolated above 300 K using the square root thermal dependence of Gaussian fluctuations.}
    \label{fig:tas3_thermal_dependence_inv_eh}
\end{figure}

Tables~\ref{table:I},~\ref{table:II} and~\ref{table:III} report the calculated inverse electron-hole coherence length in the chain direction ($1/\xi_{eh}^{b}$) for the different $q_i$ electron-hole singularities. The thermal dependence $1/\xi_{eh}^{b}$ for the inter-band response of $m$-TaS$_3$ is shown in Fig.~\ref{fig:tas3_thermal_dependence_inv_eh} and is compared with the inverse experimental correlation length for the T$_{P1}$ CDW/BOW transition ($1/\xi_{BOW}$) of $m$-TaS$_3$ driven by the $q_{III}^{inter}$ electron-hole instability~\cite{Pouget1989}. The 1D BOW fluctuations have been clearly detected in this material both by electron~\cite{Roucau1980} and X-ray~\cite{Moret} scattering at 300 K. The experimental $1/\xi_{BOW}$ tends asymptotically towards $1/\xi_{eh}^{b}$ around 400 K, which is also the temperature at which $\xi_{eh}^{b}$ amounts to the 2$k_F$ wave length $\lambda_{2k_F}\approx$ 4$b$ =13 \AA. Above this temperature, when $\xi_{eh}^{b}$< $\lambda_{2k_F}$, the CDW fluctuations are not well defined.

The analysis of the longitudinal electron-hole fluctuations due to the $q_{III}^{inter}$ electron-hole instability of NbSe$_3$, gives a quite inaccurate estimation of the inverse electron-hole coherence length $1/\xi_{eh}^{b}$ for the type III chains (see Sect.~\ref{sec:Lindhard}). The average of the two estimations obtained from the fit of the (0, $q$, 0) and (1/2, $q$, 1/2) scans is plotted for temperatures lower than 200 K in Fig.~\ref{fig:tas3_thermal_dependence_inv_eh}. In this figure, the thermal variation of $1/\xi_{eh}^{b}$ is also compared with that of the inverse experimental correlation length ($1/\xi_{BOW}$) of the fluctuations preceding the T$_{P1}$ CDW/BOW transition of NbSe$_3$ involving type III chains~\cite{Pouget1989,Moudden1990}. Note that 1D BOW fluctuations on both type III and type I chains have been clearly detected at 300 K by X-ray scattering methods~\cite{Pouget1983Nb}. $1/\xi_{eh}^{b}$ extrapolates to $1/\xi_{BOW}$ at about 300 K, which is also the temperature at which $\xi_{eh}^{b}$ reaches the 2$k_F$ wave length $\lambda_{2k_F}\approx$ 4$b$= 13 \AA. Above this temperature, when $\xi_{eh}^{b}$< $\lambda_{2k_F}$, the CDW fluctuations are not well defined.

Fig.~\ref{fig:tas3_thermal_dependence_inv_eh} shows that $1/\xi_{BOW}$ on type III chains departs from $1/\xi_{eh}^{b}$ below about 400 K for $m$-TaS$_3$ and 300 K for NbSe$_3$. These temperatures are approximately 150 K above T$_{P1}$ for both compounds. The enhanced growth of $\xi_{BOW}$ with respect to $\xi_{eh}^{b}$ is due to the critical effect of the electron-phonon coupling to achieve the BOW/Peierls instability. A somewhat similar behaviour was previously reported for the blue bronze~\cite{Guster2019}.  Note that if the electron-phonon coupling is not strong enough to drive a Peierls instability, $\xi_{BOW}$ follows the thermal dependence of  $\xi_{eh}^{b}$, as found in the Bechgaard salts~\cite{Guster2020}, and a Peierls instability is not achieved.  

\begin{figure}[!hptb]
    \centering
    \includegraphics[width=0.425\textwidth]{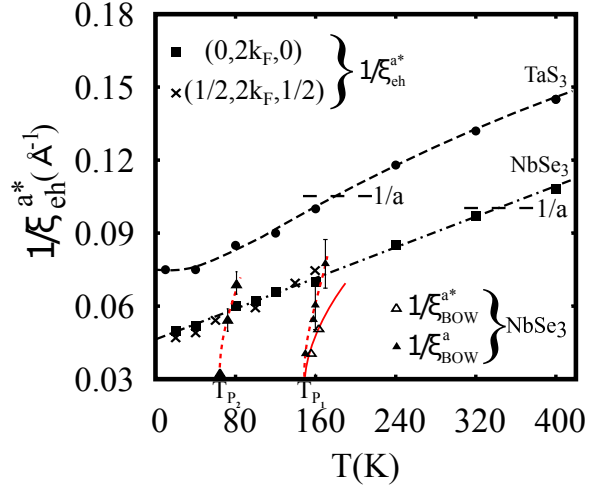}
    \caption{Thermal dependence of the inverse electron-hole coherence length along $a^*$ (1/$\xi_{eh}^{a^*}$) for the type III chains of $m$-(full circles) and for both the type III and type I chains of \nbse~(full squares and crosses, respectively). These values are compared with the inverse CDW/BOW correlation length measured along $a^*$ for \nbse~(1/$\xi_{BOW}^{a^*}$) above T$_{P_1}$ (empty triangles)~\cite{Moudden1990} and the inverse CDW/BOW correlation length measured along $a$ (1/$\xi_{BOW}^{a}$) above T$_{P_1}$ and T$_{P_2}$ \cite{Rouziere1996} for \nbse~(full triangles). The inverse value of the lattice vector $a$ is indicated for both compounds}.
    \label{fig:inv_eh_tas3}
\end{figure}

\subsection{Transversal electron-hole fluctuations.}\label{sec:transversal_fluctuations}

\begin{figure}[!hptb]
    \centering
    \includegraphics[width=0.470\textwidth]{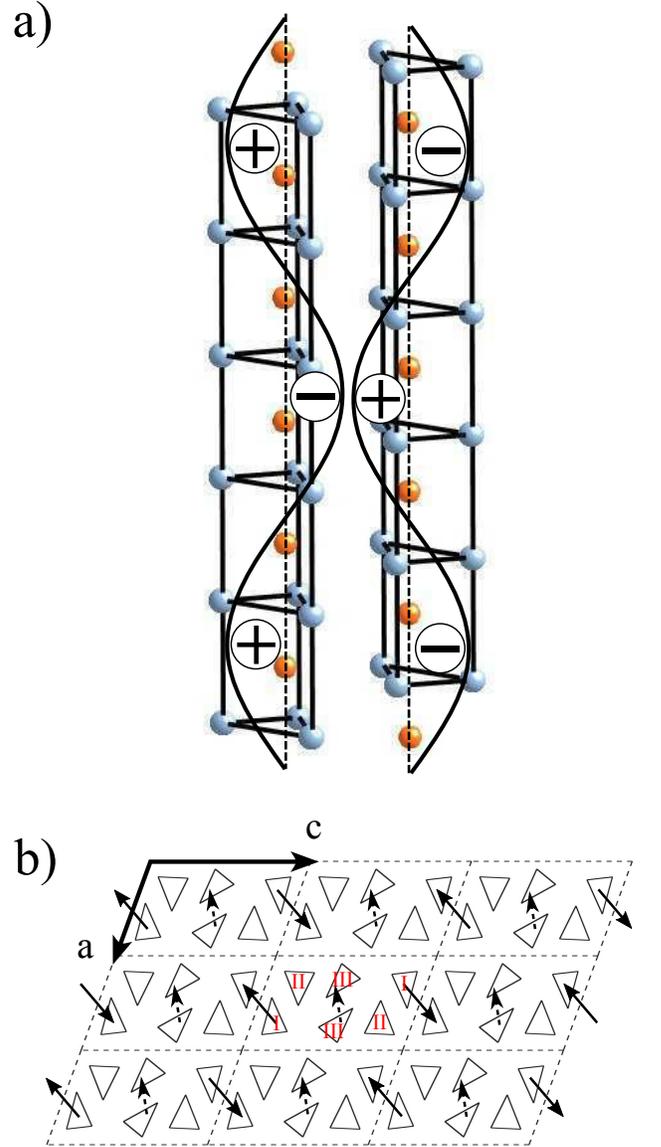}
    \caption{(a) Schematic representation of the 4$\times b$ out of phase modulation on type III \nbse~chains~\cite{Smaalen1992,Smaalen1993}. (b) Schematic representation of the two kinds of dipolar CDW and their orientation in the $a$ and $c$ directions. Solid (dashed) arrows are the dipole moments on the pairs of type I (type III) chains. As the relationship between 2$k_F^I$ and 2$k_F^{III}$ is incommensurate the phasing between the two CDW sublattices is arbitrary.}
    \label{fig:nbse3_dipolar_cdw}
\end{figure}

Let us first consider the electron-hole fluctuations in the $a$* direction. Fig.~\ref{fig:inv_eh_tas3} gives the thermal dependence of $1/\xi_{eh}^{a^*}$ on type III chains for both $m$-TaS$_3$ and NbSe$_3$. In both cases $1/\xi_{eh}^{a^*}$  decreases upon cooling. Also reported in this figure is $1/\xi_{eh}^{a^*}$ for type I chains of NbSe$_3$. For both compounds and for the whole temperature range, $\xi_{eh}^{a^*}$ is always larger than the lateral distance between pairs of chains of the same type, $\sim$ 3.75 \AA. This means that electron-hole pairs located on pairs made of the closest type III or type I chains are always coupled. Since the structural refinement of the two modulated structures of NbSe$_3$ shows that the CDW/BOW located on pairs of chains are out-of-phase~\cite{Smaalen1992}, it follows that pre-transitional CDW fluctuations involving coupled chains should be of dipolar nature, as schematically represented in Fig.~\ref{fig:nbse3_dipolar_cdw}a and previously considered in refs.~\cite{Canadell1990},~\cite{Rouziere1996} and~\cite{Pouget2016}. Thus, dipolar coupled electron-hole pairs are the basic units to consider when analyzing the inter-chain coupling mechanism achieving the 3D ordering for the Peierls transitions of these trichalcogenides. 

$\xi_{eh}^{a^*}$ for type III and type I chains of NbSe$_3$  reach the value of the unit cell parameter $a$ (distance between first neighbor pairs of identical chains) at about 340 K, well above the T$_{P1}$ Peierls transition. This means that dipolar electron-hole pairs are already well coupled beyond neighboring pairs along $a$ at T$_{P1}$. In contrast, $\xi_{eh}^{a^*}$ for type III chains in $m$-TaS$_3$ only reaches the unit cell parameter $a$ at about 190 K, which is in-between T$_{P1}$ and T$_{P2}$. Thus, in $m$-TaS$_3$ the dipolar electron-hole pairs neither are coupled along $a$ at T$_{P1}$= 240 K nor they are coupled at T$_{P2}$ since there is no visible maximum in the Lindhard function at (1/2, 2$k_F^I$, 1/2) (see Fig.~\ref{fig:tas3_lrf_maps}b).

The 1/$\xi_{eh}^{a^*}$ of NbSe$_3$ is compared in Fig.~\ref{fig:inv_eh_tas3} with the thermal dependence of the inverse CDW/BOW correlation length measured along the $a$* direction (1/$\xi_{BOW}^{a*}$) due to transverse pre-transitional fluctuations at T$_{P1}$~\cite{Moudden1990}. This quantity increases quickly upon heating above T$_{P1}$ and reaches the inverse electron-hole coherence length $1/\xi_{eh}^{a^*}$  around 200 K. Fig.~\ref{fig:inv_eh_tas3} also reports another measurement of the CDW/BOW correlation length (1/$\xi_{BOW}^{a}$) along the $a$ direction (proper direction of the tensor of correlation lengths)~\cite{Rouziere1996}, which reaches the inverse electron-hole coherence length $1/\xi_{eh}^{a^*}$  at about 170 K. The thermal dependence of the inverse CDW/BOW correlation length measured along $a$ (1/$\xi_{BOW}^{a}$) associated with the transverse pre-transitional fluctuations of the T$_{P2}$ transition which involve type I chains~\cite{Rouziere1996} is also shown in Fig.~\ref{fig:inv_eh_tas3}. This quantity increases rapidly upon heating above T$_{P2}$ and reaches the inverse electron-hole coherence length $1/\xi_{eh}^{a}$  around 80 K. The strong deviation between the thermal dependencies of 1/$\xi_{BOW}^{a}$ or 1/$\xi_{BOW}^{a^*}$ and 1/$\xi_{eh}^{a^*}$  is due to the critical effect of the inter-chain coupling due to either tunneling and/or Coulomb interactions (see Sect.~\ref{sec:CDW_BOW coupling}) in the vicinity of the Peierls transitions. 

For NbSe$_3$ $\xi_{eh}^{c^*} (\approx$ 11 \AA) is smaller than the distance between type III chains along $c$ ($c$ = 13.6 \AA) for all the temperature range. This means that dipolar electron-hole pairs located on neighboring type III chains are not coupled by tunneling along $c$ above T$_{P1}$. Similarly, the dipolar electron-hole pairs located on neighboring type I chains are never coupled by tunneling along $c$ above T$_{P2}$.  This conclusion remains true for $m$-TaS$_3$, where $\xi_{eh}^{c^*}$ cannot be measured above T$_{P1}$ and T$_{P2}$. 

\subsection{Inter-chain coupling mechanism between BOW and CDW.}\label{sec:CDW_BOW coupling}

Two main types of inter-chain coupling must be considered in 1D Peierls systems~\cite{Jerome1982,Pouget2016}:
\newline
\phantom{a}- Interchain tunneling causing the warping of the open FS, which leads to maxima of the Lindhard response for the best FS nesting transverse wave vector components, and
\newline
\phantom{a}- Coulomb coupling between quasi-1D CDW (here referred to as dipolar CDW).

\noindent Let us now discuss how both mechanisms operate coherently in the trichalcogenides.

NbSe$_3$ and $m$-TaS$_3$ trichalcogenides undergo two successive Peierls transition at T$_{P1}$ and T$_{P2}$ with the critical wave vectors $q_1$=  (0, 2$k_F^{III}$, 0) and $q_2$= (1/2, 2$k_F^I$, 1/2) respectively. In NbSe$_3$ each of these critical wave vectors corresponds to maxima of the electron-hole response function (Fig.~\ref{fig:nbse3_2d_trans_map}). In the case of $m$-TaS$_3$ only the (0, 2$k_F^{III}$, 0) wave vector corresponds to a maximum of the Lindhard response (Fig.~\ref{fig:tas3_lrf_maps}). Thus, the $transverse$ nesting process of the warped open FS calculated at 10 K can account for the 0$a$* or 1/2$a$* components of the  T$_{P1}$ and  T$_{P2}$ CDW modulations of NbSe$_3$ and the T$_{P1}$ of $m$-TaS$_3$ but not those of T$_{P2}$ of $m$-TaS$_3$. In addition, the efficiency of the transverse FS nesting process is less evident for the T$_{P2}$ transition of NbSe$_3$ because the FS sheets associated with type I chains are strongly hybridized. Clearly, a closer look at the interchain coupling mechanism is in order. The divergence of the electron-hole response function due to nesting is reduced by the thermal broadening of the FS. Such thermal effects are included in the calculation of the electron-hole response at T$_{P1}$ and T$_{P2}$ (Figs.~\ref{fig:tas3_lrf_maps} and \ref{fig:nbse3_2d_trans_map}). The occurrence of FS nesting breaking effects, as those clearly seen in NbSe$_3$, reduce the Peierls instability and even can suppress it if the gap remaining between electron and hole pockets after the nesting process closes the (mean-field) Peierls gap~\cite{Hasegawa1986}. This is for instance what occurs as the result of pressure application. Since pressure increases significantly the warping and hybridization of the different FSs, the nesting breaking effects become more important and lead to the vanishing of the lower/upper CDW of NbSe$_3$ at 0.75/4 GPa, respectively. As a consequence, a superconducting ground state is stabilized at high pressure (see Fig. 26 in ref.~\cite{Monceau2012}). Because of its best nested FSs, the Peierls transitions of $m$-TaS$_3$ are much less depressed under pressure. In contrast, strong magnetic fields should render the electronic motion more 1D. The associated decrease of nesting breaking effects should lead to an enhancement of the Peierls temperature under magnetic field. This is nicely illustrated by the 40\% increase of T$_{P2}$ in NbSe$_3$ under 30 T (see Fig. 164 in ref.~\cite{Monceau2012}). Nesting breaking terms are also responsible of the finite value of the inverse longitudinal electron-hole coherence length $1/\xi_{eh}^{b} (0)$ at 0 K (see Fig.~\ref{fig:tas3_thermal_dependence_inv_eh}), whose value amounts to the typical size along $b$* of the electron and hole pockets remaining after the nesting process.   

Quantitatively, the relevance of inter-chain tunneling effects can be appreciated by comparing the value of the electron-hole coherence length in a transverse direction with the inter-stack distances along this direction. In the case of $m$-TaS$_3$, it is found that $\xi_{eh}^{a^*}$ is not large enough compared to $a$ so that nesting effects cannot set the 0$a$* component of the modulation for T$_{P1}$. In addition, since $\xi_{eh}^{a^*}$ is not measurable for T$_{P2}$, nesting can not fix the 1/2$a$* component for this modulation. Finally one finds that for both NbSe$_3$ and $m$-TaS$_3$ $\xi_{eh}^{c^*}$ is not large enough so that nesting cannot achieve a relevant coupling between neighboring electron-hole pairs along $c$. So FS nesting effects cannot impose the 0$c$* and 1/2$c$* components of the modulations for T$_{P1}$ and T$_{P2}$ respectively, for both NbSe$_3$ and $m$-TaS$_3$. In conclusion, FS nesting effects are only relevant to fix the $a$* components in NbSe$_3$. Thus, for the transverse coupling along $c$* in NbSe$_3$ and along both $a$* and $c$* for $m$-TaS$_3$, one must consider another inter-chain coupling mechanism such as the Coulomb attraction between CDW located on neighboring stacks.~\cite{Saub1976}

In these trichalcogenides simple electrostatic considerations previously developped in Ref.~\cite{Rouziere1996} show that Coulomb coupling between dipolar CDWs located on pairs of identical chains can account for the (0$a$*, 0$c$*) transverse components between type III chains and the (1/2$a$*, 1/2$c$*) transverse components between type I chains. The result is schematically shown in Fig.~\ref{fig:nbse3_dipolar_cdw}b. Note that this electrostatic coupling leads to the same phasing between neighboring CDWs as do FS nesting mechanisms. Also, as the $\xi_{eh}^{c^*}$ associated with the inter-chain tunneling along $c$* is not relevant, the Coulomb interaction should be the dominant inter-chain coupling mechanism in the $c$* direction.

\subsection{Peierls transitions.}\label{sec:Peierls transitions}

Fig.~\ref{fig:tas3_thermal_dependence_inv_eh} shows that the thermal dependence of $\xi_{BOW}$ in the $b$ chain direction deviates from that of $\xi_{eh}^{b}$ below about 300 K in NbSe$_3$ and 400 K in $m$-TaS$_3$. Here the critical divergence of $\xi_{BOW}$ when approaching the Peierls transition is driven by the coupling of the quasi-1D electron-hole response with the phonon-field. A deviation between the thermal dependencies of $\xi_{BOW}$ in the transverse direction $a$ or $a$* and $\xi_{eh}^{a^*}$ occurs around 20-50 K above T$_{P1}$ and T$_{P2}$ in NbSe$_3$ when the inter-chain tunneling or Coulomb coupling becomes critical. Such features, expected for 3D coupled Peierls chain systems, are observed in many quasi-1D Peierls compounds as for instance the blue bronze~\cite{Guster2019}.
There is however a substantial difference between the Peierls transitions occurring in NbSe$_3$ and the blue bronze. According to the experimental measurement of $1/\xi_{BOW}$ for NbSe$_3$ ($\sim$ 0.1 \AA $^{-1}$  at 300 K ~\cite{Pouget1983Nb}) it can be found that the Peierls critical lattice fluctuations occupy $\sim$ 25\% of the Brillouin zone (BZ) volume, while in K$_{0.3}$MoO$_3$ this quantity amounts to $\sim$ 12\% of the BZ~\cite{Guster2019}. With a lattice softening occupying only 12\% of the BZ, the lattice entropy can be neglected when considering the Peierls mechanism for K$_{0.3}$MoO$_3$ which  justifies the weak coupling scenario. With a volume twice larger in NbSe$_3$ it is not quite clear that the lattice entropy can be neglected. The lattice entropy, first considered by McMillan for transition metal dichalcogenides~\cite{McMillan1977}, can affect significantly the weak-coupling mechanism of the Peierls transition making it more similar to those obtained in strong coupling theories. 

An important question to consider is that of the strength of the electron-phonon coupling at work in these transition metal trichalcogenides (this is also a recurrent question for the transition metal dichalcogenides~\cite{Rossnagel2011}). It has been proposed from the non detection of a pre-transitional Kohn anomaly in the phonon spectrum of NbSe$_3$ that the Peierls instability should be caused by a strong electron-phonon coupling~\cite{Monceau2012}. In such a case, the pretransitional BOW/CDW fluctuations should exhibit a quasi-elastic dynamics which corresponds to the formation of local clusters of quasi-static BOW/CDW. However note that in NbSe$_3$ such quasi-elastic (or order-disorder) scattering is observed only 10 K above T$_{P1}$\cite{Requardt2002} and not on the whole temperature range (between T$_{P1}$ and 300 K) where 1D fluctuations are detected~\cite{Pouget1983Nb}. This means that the Peierls transition of NbSe$_3$ cannot be described in the order-disorder limit. Quasi-elastic BOW/CDW clusters can be viewed as the formation of local chemical bonds. From the associated bonding energy gain one expects the occurrence of local modulations with large atomic displacements. In reciprocal space the modulated structure based on clusters of strongly modified chemical bonds should be described by an anharmonic modulation of large amplitude. Such features are observed in the BOW/CDW ground state of the large $m$ members of another family of CDW materials, the monophosphate tungsten bronzes (PO$_2$)$_4$(WO$_3$)$_{2m}$~\cite{Ottolenghi1996,Roussel2000}. This is apparently not the case for NbSe$_3$ because the amplitude of the Nb displacement and of the modulation of the Nb-Se distances in the T$_{P1}$ modulated structure, $\sim$ 0.05 \AA~\cite{Smaalen1992,Smaalen1993}, is comparable to that found for the Mo in the Peierls ground state of the blue bronze~\cite{Schutte1993} considered as a model Peierls system. This is sustained by the fact that local measurements in the Peierls ground states of NbSe$_3$, such as NMR~\cite{Ross1986} and STM~\cite{Brun2009} provide evidence for simple sinusoidal modulations.

\begin{figure}[!hptb]
    \centering
    \includegraphics[width=0.35\textwidth]{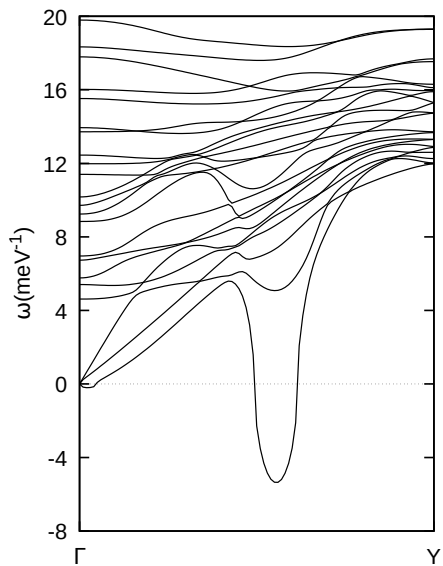}
    \caption{Phonon dispersion of the first 22 branches for $m$-TaS$_3$ in the $\Gamma$-Y segment of the BZ.}
    \label{fig:tas3_phonon dispersion}
\end{figure}

Another intriguing question concerns the role of phonons in the pre-transitional dynamics of the Peierls transition of NbSe$_3$ and $m$-TaS$_3$. The calculated phonon dispersion spectrum for $m$-TaS$_3$ at 10 K along $b$* is shown in Fig.~\ref{fig:tas3_phonon dispersion}. The calculation reveals the formation of a giant Kohn anomaly whose negative frequency around $2k_F$ implies a lattice instability. This broad phonon anomaly takes place in a longitudinal optical (LO) branch which, because of the screw axis symmetry, folds the $b$* dispersion of the longitudinal acoustic (LA) branch at the Brillouin zone boundary. We have checked that the Kohn anomaly involves mostly the displacement of Ta atoms located on chain III. The location of the Kohn anomaly in a LO branch implies out-of-phase longitudinal displacements of the Ta atoms between the two type III chains of the unit cell. This supports the formation of a dipolar CDW/BOW as schematically represented in Fig.~\ref{fig:nbse3_dipolar_cdw}a. Fig.~\ref{fig:tas3_phonon dispersion}. shows also that the LO branch hybridizes with the lower frequency acoustic branch for $q$ < $2k_F$ (transversal accoustic (TA) mode polarized along $c$*). If the LO mode hybridizes with a TA mode near $2k_F$ it can be expected that due to the $q$-dependent mixing of longitudinal and transverse atomic (basically Ta atoms) polarizations, the electron-phonon coupling should substantially vary in $q$ space in the vicinity of $2k_F$. More precisely if the electron-phonon coupling is less important for the TA mode than for the LO mode one expects an increase of the electron-phonon when $q$ increases on approaching $2k_F$. A similar feature was reported in the blue bronze for $q$>$2k_F$ suggesting a decrease of the electron-phonon when $q$ increases~\cite{Guster2019}. The decrease of the electron-phonon of the blue bronze with $q$ could explain the increase of the experimental $2k_F$ modulation wave vector upon cooling. $m$-TaS$_3$ shows an opposite variation of the electron-phonon coupling so that one expects a decrease of the experimental $2k_F$ modulation wave vector upon cooling. With certainly a similar phonon spectrum in NbSe$_3$, one expects a similar $q$ dependent electron-phonon coupling and thus, a thermal decrease of the $q_1$ modulation wave vector upon cooling, providing a suitable explanation for the experimental observation~\cite{Moudden1990}.

Finally, let us remark that in the case of a strong coupling scenario, the calculation of the Lindhard response should not bring reliable information concerning the Peierls mechanism compatible with the experimental data because in that case, the electron wave functions will be so strongly modified by the coupling with the phonon field that it would not be meaningful to use the unperturbed electronic wave function to calculate the electron-hole response function (Eq.~\ref{eq:chi}). The nice relationship between many of the results of the present work based on the unperturbed electron-hole response function  and the experimental results, such as the longitudinal and staggered maxima of the electron-hole response and BOW correlation lengths, suggests that one can exclude a strong electron-phonon coupling scenario to describe the mechanism of the Peierls transition in NbSe$_3$ and $m$-TaS$_3$. However our work shows that an intermediate or weak coupling scenario seems to be appropriate. This is also the case for the Lindhard function calculated for 2D oxydes and bronzes~\cite{Sandre2001,Guster2020a} where the CDW is triggered by nesting of differently oriented quasi-1D FS resulting from the hidden 1D nature of their electronic structure~\cite{Whangbo1991}. In other 1D conductors such as trichalcogenides (this paper) and the blue bronze~\cite{Guster2019}, both FS nesting and inter-chain coulomb coupling contribute to the stabilization the CDW ground state. Finally, note that the quantitative analysis of the Lindhard response of quasi-1D organic materials such as (TMTSF)$_2$PF$_6$~\cite{Guster2020} or $\alpha$-(BEDT-TTF)$_2$KHg(SCN)$_4$~\cite{Foury2010,Guster2020b} points out the importance of multi-nesting processes of their simply warped FS in the stabilization process of their spin density wave or CDW ground states. All these findings should be contrasted with those found in 2D metals such as transition metal dichalcogenides and tellurides~\cite{JMH06,JM08} where the Lindhard function calculation suggests that FS nesting does not trigger their CDW instabilities.    

\section{Concluding Remarks}\label{sec:conclusions}

The electron-hole Lindhard response function of the pseudo-1D trichalcogenides NbSe$_3$ and $m$-TaS$_3$ has been calculated and analyzed on the basis of the nesting features of their FS. Although both the FS and Lindhard function of NbSe$_3$ are considerably more complex as a result of the stronger inter-chain interactions, a common scheme can be put forward to understand the results. The intra-chain inter-band nesting processes dominate the strongest response for both chains I and III.  Two well-defined maxima of the Lindhard response for NbSe$_3$ are found with the  (0$a$*, 0$c$*) and (1/2$a$*, 1/2$c$*) transverse components whereas the second is not observed for $m$-TaS$_3$ at T$_{P2}$. Analysis of the different inter-chain coupling mechanisms leads to the conclusion that FS nesting effects are only relevant to set the $a$* components in NbSe$_3$. Thus, for the transverse coupling along $c$* in NbSe$_3$ and along both $a$* and $c$* for $m$-TaS$_3$, one must take into account an inter-chain Coulomb coupling mechanism. Note that Coulomb coupling between dipolar CDWs leads to the same transverse phasing between CDWs as do FS nesting processes Altogether, the present results of the Lindhard response calculation and the relevant experimental information at hand point out that even if a weak coupling scenario of the Peierls transition is not as perfectly suited as for the blue bronzes, a large body of experimental work can be well accounted for within this approach. Phonon calculations provide evidence for the formation of a giant $q_1$ Kohn anomaly at the upper CDW transition of $m$-TaS$_3$. Strong coupling scenarios as those apparently at work in 2D transition metal dichalcogenides do not seem relevant for these quasi-1D transition metal trichalcogenides.    

\section*{Acknowledgements}
This work was supported by Spanish MINECO (the Severo Ochoa Centers of Excellence Program under Grants No. SEV-2017-0706 and SEV-2015-0496), Spanish MICIU, AEI and EU FEDER (Grants No. PGC2018-096955-B-C43 and No. PGC2018-096955-B-C44), Generalitat de Catalunya (Grant No. 2017SGR1506 and the CERCA Programme), and the European Union  MaX Center of Excellence (EU-H2020 Grant No. 824143). Phonons computational resources have been provided by the supercomputing facilities of the Universit\'e catholique de Louvain (CISM/UCL) and the Consortium des \'Equipements de Calcul Intensif en F\'ed\'eration Wallonie Bruxelles (C\'ECI) funded by the Fond de la Recherche Scientifique de Belgique (F.R.S.-FNRS) under convention 2.5020.11 and by the Walloon Region. 

\
\appendix*
\section{Longitudinal phonon spectrum of $m$-TaS$_3$. }\label{sec:phonon spectrum}

 The full phonon dispersion for $m$-TaS$_3$ along the $b$* direction has been calculated. Below we specify several noticeable features  which are relevant for the discussion in Sect.~\ref{sec:Peierls transitions}:

1- From a total of 72 branches which disperse up to 60 meV (484.0 cm$^{-1}$), 22 phonon branches are present below 20 meV (161.3 cm$^{-1}$). These low frequency dispersions are those reported in Fig.~\ref{fig:tas3_phonon dispersion}.  

\
2-  The dispersion in Fig.~\ref{fig:tas3_phonon dispersion} has been calculated along the direction of the 2$_1$ screw axis symmetry of the structure. Thus, the structure projected in this direction has a $b$/2 periodicity which means that the dispersion has a 2$b$* periodicity. Also the energy of phonon branches is 2-fold degenerate at the Brillouin zone boundary ($\pm b$*/2). In particular, all the acoustic branches are continuously extended by an optical branch. This is in particular the case for the LO branch bearing the Kohn anomaly (see point 4) which continuously extends the LA branch.

\ 
3- The acoustic branches are quite anisotropic. The higher frequency acoustic mode is the LA branch polarized along $b$. The transverse acoustic mode of intermediate frequency is polarized along $a$. The lower frequency acoustic branch should be the TA branch polarized along $c$*. LO($b$) and TA($a$) branches of similar dispersion have been measured in NbSe$_3$~\cite{Requardt2002}. The sound velocity deduced for the slope of the calculated acoustic dispersions are 34 eV$\cdot$\AA$^{-1}$, 17.5 eV$\cdot$\AA$^{-1}$ and 11 eV$\cdot$\AA$^{-1}$ for the LA($b$), TA($a$) and TA($c$*) modes of $m$-TaS$_3$. The two highest velocities are consistently close to those measured in NbSe$_3$: 32 eV$\cdot$\AA$^{-1}$ LA($b$) and 16.5 eV$\cdot$\AA$^{-1}$ TA($a$), respectively~\cite{Requardt2002}. The anisotropy ratio of the sound velocity being 3 : 1.5 :1, one obtains a quite anisotropic ratio of elastic constants C$_{22}$ : C$_{66}$ : C$_{44}$ of 9 : 2.5 : 1. Note that this anisotropy does not follow the structural anisotropy quoted in the literature~\cite{Hodeau1978} because the smallest elastic constant corresponds to a shear deformation inside the strongly linked ($bc$) layers.

4- There is a giant Kohn anomaly in the LO branch dispersing from 5.6 meV (45.2 cm$^{-1}$)  at $\Gamma$ to 13.3 meV (107.3 cm$^{-1}$) at Y. The negative frequency for $q\sim$ 0.27$b$* (i.e. close to $q_1$) implies a lattice instability which is found to be located on chains III. 

5- As shown in Fig.~\ref{fig:tas3_phonon dispersion}, the gap formation at the crossing point clearly shows that the LO branch continuously extends the LA branch and that the LO branch hybridizes with the lower frequency TA ($c$*) branch (hybridization with the TA($a$) branch can be also guessed). This hybridization could favor the transverse deformation of the sulfur prisms surrounding Ta$_{III}$, such as observed in the $q_1$ CDW modulated state of NbSe$_3$~\cite{Smaalen1992,Smaalen1993}.

6- Another weaker phonon anomaly is observed at 5.3 meV (42.7 cm$^{-1}$) for $q\sim$ 0.26$b$*. This branch also hybridizes with the TA($c$*) and TA($a$) branches. Other phonon anomalies can be identified at higher energies.
  
7- The low frequency phonon branches exhibit frequent hybridizations for $q$ < $2k_F$. However there is nearly no inter-branch hybridization for $q$ > $2k_F$.


\bibliographystyle{apsrev4-1}
\bibliography{tmtds}

\begin{thebibliography}{66}%
\makeatletter
\providecommand \@ifxundefined [1]{%
 \@ifx{#1\undefined}
}%
\providecommand \@ifnum [1]{%
 \ifnum #1\expandafter \@firstoftwo
 \else \expandafter \@secondoftwo
 \fi
}%
\providecommand \@ifx [1]{%
 \ifx #1\expandafter \@firstoftwo
 \else \expandafter \@secondoftwo
 \fi
}%
\providecommand \natexlab [1]{#1}%
\providecommand \enquote  [1]{``#1''}%
\providecommand \bibnamefont  [1]{#1}%
\providecommand \bibfnamefont [1]{#1}%
\providecommand \citenamefont [1]{#1}%
\providecommand \href@noop [0]{\@secondoftwo}%
\providecommand \href [0]{\begingroup \@sanitize@url \@href}%
\providecommand \@href[1]{\@@startlink{#1}\@@href}%
\providecommand \@@href[1]{\endgroup#1\@@endlink}%
\providecommand \@sanitize@url [0]{\catcode `\\12\catcode `\$12\catcode
  `\&12\catcode `\#12\catcode `\^12\catcode `\_12\catcode `\%12\relax}%
\providecommand \@@startlink[1]{}%
\providecommand \@@endlink[0]{}%
\providecommand \url  [0]{\begingroup\@sanitize@url \@url }%
\providecommand \@url [1]{\endgroup\@href {#1}{\urlprefix }}%
\providecommand \urlprefix  [0]{URL }%
\providecommand \Eprint [0]{\href }%
\providecommand \doibase [0]{http://dx.doi.org/}%
\providecommand \selectlanguage [0]{\@gobble}%
\providecommand \bibinfo  [0]{\@secondoftwo}%
\providecommand \bibfield  [0]{\@secondoftwo}%
\providecommand \translation [1]{[#1]}%
\providecommand \BibitemOpen [0]{}%
\providecommand \bibitemStop [0]{}%
\providecommand \bibitemNoStop [0]{.\EOS\space}%
\providecommand \EOS [0]{\spacefactor3000\relax}%
\providecommand \BibitemShut  [1]{\csname bibitem#1\endcsname}%
\let\auto@bib@innerbib\@empty
\bibitem [{\citenamefont {Com\`{e}s}\ \emph {et~al.}(1973)\citenamefont
  {Com\`{e}s}, \citenamefont {Lambert}, \citenamefont {Launois},\ and\
  \citenamefont {Zeller}}]{Comes1973}%
  \BibitemOpen
  \bibfield  {author} {\bibinfo {author} {\bibfnamefont {R.}~\bibnamefont
  {Com\`{e}s}}, \bibinfo {author} {\bibfnamefont {M.}~\bibnamefont {Lambert}},
  \bibinfo {author} {\bibfnamefont {H.}~\bibnamefont {Launois}}, \ and\
  \bibinfo {author} {\bibfnamefont {H.~R.}\ \bibnamefont {Zeller}},\
  }\href@noop {} {\bibfield  {journal} {\bibinfo  {journal} {Phys. Rev. B}\
  }\textbf {\bibinfo {volume} {8}},\ \bibinfo {pages} {571} (\bibinfo {year}
  {1973})}\BibitemShut {NoStop}%
\bibitem [{\citenamefont {J\'erome}\ and\ \citenamefont
  {Schulz}(1982)}]{Jerome1982}%
  \BibitemOpen
  \bibfield  {author} {\bibinfo {author} {\bibfnamefont {D.}~\bibnamefont
  {J\'erome}}\ and\ \bibinfo {author} {\bibfnamefont {H.~J.}\ \bibnamefont
  {Schulz}},\ }\href@noop {} {\bibfield  {journal} {\bibinfo  {journal} {Adv.
  Phys.}\ }\textbf {\bibinfo {volume} {31}},\ \bibinfo {pages} {299} (\bibinfo
  {year} {1982})}\BibitemShut {NoStop}%
\bibitem [{\citenamefont {Wilson}\ \emph {et~al.}(1975)\citenamefont {Wilson},
  \citenamefont {Di~Salvo},\ and\ \citenamefont {Mahajan}}]{Wilson1975}%
  \BibitemOpen
  \bibfield  {author} {\bibinfo {author} {\bibfnamefont {J.}~\bibnamefont
  {Wilson}}, \bibinfo {author} {\bibfnamefont {F.}~\bibnamefont {Di~Salvo}}, \
  and\ \bibinfo {author} {\bibfnamefont {S.}~\bibnamefont {Mahajan}},\
  }\href@noop {} {\bibfield  {journal} {\bibinfo  {journal} {Adv. Phys.}\
  }\textbf {\bibinfo {volume} {24}},\ \bibinfo {pages} {117} (\bibinfo {year}
  {1975})}\BibitemShut {NoStop}%
\bibitem [{\citenamefont {Gor'kov}\ and\ \citenamefont
  {Gr\"uner}(1989)}]{Gorkov1989}%
  \BibitemOpen
  \bibinfo {editor} {\bibfnamefont {L.}~\bibnamefont {Gor'kov}}\ and\ \bibinfo
  {editor} {\bibfnamefont {G.}~\bibnamefont {Gr\"uner}},\ eds.,\ \href@noop {}
  {\emph {\bibinfo {title} {Charge Density Waves in Solids}}},\ Modern Problems
  in Condensed Matter Sciences, vol. 25,\ (\bibinfo  {publisher} {Elsevier,
  Netherlands},\ \bibinfo {year} {1989})\BibitemShut {NoStop}%
\bibitem [{\citenamefont {Monceau}(2012)}]{Monceau2012}%
  \BibitemOpen
  \bibfield  {author} {\bibinfo {author} {\bibfnamefont {P.}~\bibnamefont
  {Monceau}},\ }\href@noop {} {\bibfield  {journal} {\bibinfo  {journal} {Adv.
  Phys.}\ }\textbf {\bibinfo {volume} {61}},\ \bibinfo {pages} {325} (\bibinfo
  {year} {2012})}\BibitemShut {NoStop}%
\bibitem [{\citenamefont {Pouget}(2016)}]{Pouget2016}%
  \BibitemOpen
  \bibfield  {author} {\bibinfo {author} {\bibfnamefont {J.-P.}\ \bibnamefont
  {Pouget}},\ }\href@noop {} {\bibfield  {journal} {\bibinfo  {journal} {C. R.
  Phys.}\ }\textbf {\bibinfo {volume} {17}},\ \bibinfo {pages} {332} (\bibinfo
  {year} {2016})}\BibitemShut {NoStop}%
\bibitem [{\citenamefont {Peierls}(1955)}]{Peierls1955}%
  \BibitemOpen
  \bibfield  {author} {\bibinfo {author} {\bibfnamefont {R.}~\bibnamefont
  {Peierls}},\ }\href@noop {} {\emph {\bibinfo {title} {{Quantum Theory of
  Solids}}}}\ (\bibinfo  {publisher} {Oxford University Press, London},\
  \bibinfo {year} {1955})\BibitemShut {NoStop}%
\bibitem [{\citenamefont {Chan}\ and\ \citenamefont {Heine}(1973)}]{Chan1973}%
  \BibitemOpen
  \bibfield  {author} {\bibinfo {author} {\bibfnamefont {S.-K.}\ \bibnamefont
  {Chan}}\ and\ \bibinfo {author} {\bibfnamefont {V.}~\bibnamefont {Heine}},\
  }\href@noop {} {\bibfield  {journal} {\bibinfo  {journal} {J. Phys. F: Met.
  Phys.}\ }\textbf {\bibinfo {volume} {3}},\ \bibinfo {pages} {795} (\bibinfo
  {year} {1973})}\BibitemShut {NoStop}%
\bibitem [{\citenamefont {Fr\"{o}hlich}(1954)}]{Frohlich1954}%
  \BibitemOpen
  \bibfield  {author} {\bibinfo {author} {\bibfnamefont {H.}~\bibnamefont
  {Fr\"{o}hlich}},\ }\href@noop {} {\bibfield  {journal} {\bibinfo  {journal}
  {Proc. R. Soc. A}\ }\textbf {\bibinfo {volume} {223}} (\bibinfo {year}
  {1954})}\BibitemShut {NoStop}%
\bibitem [{\citenamefont {Monceau}\ \emph {et~al.}(1976)\citenamefont
  {Monceau}, \citenamefont {Ong}, \citenamefont {Portis}, \citenamefont
  {Meerschaut},\ and\ \citenamefont {Rouxel}}]{Monceau1976}%
  \BibitemOpen
  \bibfield  {author} {\bibinfo {author} {\bibfnamefont {P.}~\bibnamefont
  {Monceau}}, \bibinfo {author} {\bibfnamefont {N.~P.}\ \bibnamefont {Ong}},
  \bibinfo {author} {\bibfnamefont {A.~M.}\ \bibnamefont {Portis}}, \bibinfo
  {author} {\bibfnamefont {A.}~\bibnamefont {Meerschaut}}, \ and\ \bibinfo
  {author} {\bibfnamefont {J.}~\bibnamefont {Rouxel}},\ }\href@noop {}
  {\bibfield  {journal} {\bibinfo  {journal} {Phys. Rev. Lett.}\ }\textbf
  {\bibinfo {volume} {37}},\ \bibinfo {pages} {602} (\bibinfo {year}
  {1976})}\BibitemShut {NoStop}%
\bibitem [{\citenamefont {{J. Dumas and C. Schlenker and J. Marcus and R.
  Buder}}(1983)}]{Dumas1983}%
  \BibitemOpen
  \bibfield  {author} {\bibinfo {author} {\bibnamefont {{J. Dumas and C.
  Schlenker and J. Marcus and R. Buder}}},\ }\href@noop {} {\bibfield
  {journal} {\bibinfo  {journal} {Phys. Rev. Lett.}\ }\textbf {\bibinfo
  {volume} {50}},\ \bibinfo {pages} {757} (\bibinfo {year} {1983})}\BibitemShut
  {NoStop}%
\bibitem [{\citenamefont {Hodeau}\ \emph {et~al.}(1978)\citenamefont {Hodeau},
  \citenamefont {Marezio}, \citenamefont {Roucau}, \citenamefont {Ayroles},
  \citenamefont {Meerschaut}, \citenamefont {Rouxel},\ and\ \citenamefont
  {Monceau}}]{Hodeau1978}%
  \BibitemOpen
  \bibfield  {author} {\bibinfo {author} {\bibfnamefont {J.~L.}\ \bibnamefont
  {Hodeau}}, \bibinfo {author} {\bibfnamefont {M.}~\bibnamefont {Marezio}},
  \bibinfo {author} {\bibfnamefont {C.}~\bibnamefont {Roucau}}, \bibinfo
  {author} {\bibfnamefont {R.}~\bibnamefont {Ayroles}}, \bibinfo {author}
  {\bibfnamefont {A.}~\bibnamefont {Meerschaut}}, \bibinfo {author}
  {\bibfnamefont {J.}~\bibnamefont {Rouxel}}, \ and\ \bibinfo {author}
  {\bibfnamefont {P.}~\bibnamefont {Monceau}},\ }\href@noop {} {\bibfield
  {journal} {\bibinfo  {journal} {J. Phys. C.: Solid State Phys.}\ }\textbf
  {\bibinfo {volume} {11}},\ \bibinfo {pages} {4117} (\bibinfo {year}
  {1978})}\BibitemShut {NoStop}%
\bibitem [{\citenamefont {Meerschaut}\ \emph {et~al.}(1981)\citenamefont
  {Meerschaut}, \citenamefont {Guemas},\ and\ \citenamefont
  {Rouxel}}]{Meerschaut1981}%
  \BibitemOpen
  \bibfield  {author} {\bibinfo {author} {\bibfnamefont {A.}~\bibnamefont
  {Meerschaut}}, \bibinfo {author} {\bibfnamefont {L.}~\bibnamefont {Guemas}},
  \ and\ \bibinfo {author} {\bibfnamefont {J.}~\bibnamefont {Rouxel}},\
  }\href@noop {} {\bibfield  {journal} {\bibinfo  {journal} {J. Solid State
  Chem.}\ }\textbf {\bibinfo {volume} {36}},\ \bibinfo {pages} {118} (\bibinfo
  {year} {1981})}\BibitemShut {NoStop}%
\bibitem [{\citenamefont {Roucau}\ \emph {et~al.}(1980)\citenamefont {Roucau},
  \citenamefont {Ayroles}, \citenamefont {Monceau}, \citenamefont {Guemas},
  \citenamefont {Meerschaut},\ and\ \citenamefont {Rouxel}}]{Roucau1980}%
  \BibitemOpen
  \bibfield  {author} {\bibinfo {author} {\bibfnamefont {C.}~\bibnamefont
  {Roucau}}, \bibinfo {author} {\bibfnamefont {R.}~\bibnamefont {Ayroles}},
  \bibinfo {author} {\bibfnamefont {P.}~\bibnamefont {Monceau}}, \bibinfo
  {author} {\bibfnamefont {L.}~\bibnamefont {Guemas}}, \bibinfo {author}
  {\bibfnamefont {A.}~\bibnamefont {Meerschaut}}, \ and\ \bibinfo {author}
  {\bibnamefont {Rouxel}},\ }\href@noop {} {\bibfield  {journal} {\bibinfo
  {journal} {Phys. Stat. Sol. (a)}\ }\textbf {\bibinfo {volume} {62}},\
  \bibinfo {pages} {483} (\bibinfo {year} {1980})}\BibitemShut {NoStop}%
\bibitem [{\citenamefont {Fleming}\ \emph {et~al.}(1978)\citenamefont
  {Fleming}, \citenamefont {Moncton},\ and\ \citenamefont
  {McWhan}}]{Fleming1978}%
  \BibitemOpen
  \bibfield  {author} {\bibinfo {author} {\bibfnamefont {R.}~\bibnamefont
  {Fleming}}, \bibinfo {author} {\bibfnamefont {D.}~\bibnamefont {Moncton}}, \
  and\ \bibinfo {author} {\bibfnamefont {D.~B.}\ \bibnamefont {McWhan}},\
  }\href@noop {} {\bibfield  {journal} {\bibinfo  {journal} {Phys. Rev. B}\
  }\textbf {\bibinfo {volume} {18}},\ \bibinfo {pages} {5560} (\bibinfo {year}
  {1978})}\BibitemShut {NoStop}%
\bibitem [{\citenamefont {Devreux}(1982)}]{Devreux1982}%
  \BibitemOpen
  \bibfield  {author} {\bibinfo {author} {\bibfnamefont {F.}~\bibnamefont
  {Devreux}},\ }\href@noop {} {\bibfield  {journal} {\bibinfo  {journal} {J.
  Phys. (France)}\ }\textbf {\bibinfo {volume} {43}},\ \bibinfo {pages} {1489}
  (\bibinfo {year} {1982})}\BibitemShut {NoStop}%
\bibitem [{\citenamefont {Ross~Jr}\ \emph {et~al.}(1986)\citenamefont
  {Ross~Jr}, \citenamefont {Wang},\ and\ \citenamefont {Slichter}}]{Ross1986}%
  \BibitemOpen
  \bibfield  {author} {\bibinfo {author} {\bibfnamefont {J.~H.}\ \bibnamefont
  {Ross~Jr}}, \bibinfo {author} {\bibfnamefont {Z.}~\bibnamefont {Wang}}, \
  and\ \bibinfo {author} {\bibfnamefont {C.~P.}\ \bibnamefont {Slichter}},\
  }\href@noop {} {\bibfield  {journal} {\bibinfo  {journal} {Phys. Rev. Lett.}\
  }\textbf {\bibinfo {volume} {56}},\ \bibinfo {pages} {663} (\bibinfo {year}
  {1986})}\BibitemShut {NoStop}%
\bibitem [{\citenamefont {Brun}\ \emph {et~al.}(2009)\citenamefont {Brun},
  \citenamefont {Wang},\ and\ \citenamefont {Monceau}}]{Brun2009}%
  \BibitemOpen
  \bibfield  {author} {\bibinfo {author} {\bibfnamefont {C.}~\bibnamefont
  {Brun}}, \bibinfo {author} {\bibfnamefont {Z.~Z.}\ \bibnamefont {Wang}}, \
  and\ \bibinfo {author} {\bibfnamefont {P.}~\bibnamefont {Monceau}},\
  }\href@noop {} {\bibfield  {journal} {\bibinfo  {journal} {Phys. Rev. B}\
  }\textbf {\bibinfo {volume} {80}},\ \bibinfo {pages} {045423} (\bibinfo
  {year} {2009})}\BibitemShut {NoStop}%
\bibitem [{\citenamefont {van Smaalen}\ \emph {et~al.}(1992)\citenamefont {van
  Smaalen}, \citenamefont {de~Boer}, \citenamefont {Meetsma}, \citenamefont
  {Graafsma}, \citenamefont {Sheu}, \citenamefont {Daroskikh}, \citenamefont
  {Coppens},\ and\ \citenamefont {Levy}}]{Smaalen1992}%
  \BibitemOpen
  \bibfield  {author} {\bibinfo {author} {\bibfnamefont {S.}~\bibnamefont {van
  Smaalen}}, \bibinfo {author} {\bibfnamefont {J.~L.}\ \bibnamefont {de~Boer}},
  \bibinfo {author} {\bibfnamefont {A.}~\bibnamefont {Meetsma}}, \bibinfo
  {author} {\bibfnamefont {H.}~\bibnamefont {Graafsma}}, \bibinfo {author}
  {\bibfnamefont {H.-S.}\ \bibnamefont {Sheu}}, \bibinfo {author}
  {\bibfnamefont {A.}~\bibnamefont {Daroskikh}}, \bibinfo {author}
  {\bibfnamefont {P.}~\bibnamefont {Coppens}}, \ and\ \bibinfo {author}
  {\bibfnamefont {F.}~\bibnamefont {Levy}},\ }\href@noop {} {\bibfield
  {journal} {\bibinfo  {journal} {Phys. Rev. B}\ }\textbf {\bibinfo {volume}
  {45}},\ \bibinfo {pages} {3103} (\bibinfo {year} {1992})}\BibitemShut
  {NoStop}%
\bibitem [{\citenamefont {van Smaalen}\ \emph {et~al.}(1993)\citenamefont {van
  Smaalen}, \citenamefont {de~Boer},\ and\ \citenamefont
  {Coppens}}]{Smaalen1993}%
  \BibitemOpen
  \bibfield  {author} {\bibinfo {author} {\bibfnamefont {S.}~\bibnamefont {van
  Smaalen}}, \bibinfo {author} {\bibfnamefont {J.~L.}\ \bibnamefont {de~Boer}},
  \ and\ \bibinfo {author} {\bibfnamefont {P.}~\bibnamefont {Coppens}},\
  }\href@noop {} {\bibfield  {journal} {\bibinfo  {journal} {J. Phys. IV
  Colloques}\ }\textbf {\bibinfo {volume} {3}},\ \bibinfo {pages}
  {C2\textendash89} (\bibinfo {year} {1993})}\BibitemShut {NoStop}%
\bibitem [{\citenamefont {Johannes}\ \emph {et~al.}(2006)\citenamefont
  {Johannes}, \citenamefont {Mazin},\ and\ \citenamefont {Howells}}]{JMH06}%
  \BibitemOpen
  \bibfield  {author} {\bibinfo {author} {\bibfnamefont {M.~D.}\ \bibnamefont
  {Johannes}}, \bibinfo {author} {\bibfnamefont {I.~I.}\ \bibnamefont {Mazin}},
  \ and\ \bibinfo {author} {\bibfnamefont {C.~A.}\ \bibnamefont {Howells}},\
  }\href@noop {} {\bibfield  {journal} {\bibinfo  {journal} {Phys. Rev. B}\
  }\textbf {\bibinfo {volume} {73}},\ \bibinfo {pages} {205102} (\bibinfo
  {year} {2006})}\BibitemShut {NoStop}%
\bibitem [{\citenamefont {Guster}\ \emph
  {et~al.}(2019{\natexlab{a}})\citenamefont {Guster}, \citenamefont {Pruneda},
  \citenamefont {Ordej\'on}, \citenamefont {Canadell},\ and\ \citenamefont
  {Pouget}}]{Guster2019}%
  \BibitemOpen
  \bibfield  {author} {\bibinfo {author} {\bibfnamefont {B.}~\bibnamefont
  {Guster}}, \bibinfo {author} {\bibfnamefont {M.}~\bibnamefont {Pruneda}},
  \bibinfo {author} {\bibfnamefont {P.}~\bibnamefont {Ordej\'on}}, \bibinfo
  {author} {\bibfnamefont {E.}~\bibnamefont {Canadell}}, \ and\ \bibinfo
  {author} {\bibfnamefont {J.-P.}\ \bibnamefont {Pouget}},\ }\href@noop {}
  {\bibfield  {journal} {\bibinfo  {journal} {Phys. Rev. Mat.}\ }\textbf
  {\bibinfo {volume} {3}},\ \bibinfo {pages} {055001} (\bibinfo {year}
  {2019}{\natexlab{a}})}\BibitemShut {NoStop}%
\bibitem [{\citenamefont {Guster}\ \emph
  {et~al.}(2019{\natexlab{b}})\citenamefont {Guster}, \citenamefont
  {Rubio-Verd\'u}, \citenamefont {Robles}, \citenamefont {Zald\'ivar},
  \citenamefont {Dreher}, \citenamefont {Pruneda}, \citenamefont
  {Silva-Guill\'en}, \citenamefont {Choi}, \citenamefont {Pascual},
  \citenamefont {Ugeda}, \citenamefont {Ordej\'on},\ and\ \citenamefont
  {Canadell}}]{Guster2019NbSe2}%
  \BibitemOpen
  \bibfield  {author} {\bibinfo {author} {\bibfnamefont {B.}~\bibnamefont
  {Guster}}, \bibinfo {author} {\bibfnamefont {C.}~\bibnamefont
  {Rubio-Verd\'u}}, \bibinfo {author} {\bibfnamefont {R.}~\bibnamefont
  {Robles}}, \bibinfo {author} {\bibfnamefont {J.}~\bibnamefont {Zald\'ivar}},
  \bibinfo {author} {\bibfnamefont {P.}~\bibnamefont {Dreher}}, \bibinfo
  {author} {\bibfnamefont {M.}~\bibnamefont {Pruneda}}, \bibinfo {author}
  {\bibfnamefont {J.~A.}\ \bibnamefont {Silva-Guill\'en}}, \bibinfo {author}
  {\bibfnamefont {D.-J.}\ \bibnamefont {Choi}}, \bibinfo {author}
  {\bibfnamefont {J.~I.}\ \bibnamefont {Pascual}}, \bibinfo {author}
  {\bibfnamefont {M.~M.}\ \bibnamefont {Ugeda}}, \bibinfo {author}
  {\bibfnamefont {P.}~\bibnamefont {Ordej\'on}}, \ and\ \bibinfo {author}
  {\bibfnamefont {E.}~\bibnamefont {Canadell}},\ }\href@noop {} {\bibfield
  {journal} {\bibinfo  {journal} {Nano Letters}\ }\textbf {\bibinfo {volume}
  {19}},\ \bibinfo {pages} {3027} (\bibinfo {year}
  {2019}{\natexlab{b}})}\BibitemShut {NoStop}%
\bibitem [{\citenamefont {Bullett}(1979)}]{Bullett1979}%
  \BibitemOpen
  \bibfield  {author} {\bibinfo {author} {\bibfnamefont {D.~W.}\ \bibnamefont
  {Bullett}},\ }\href@noop {} {\bibfield  {journal} {\bibinfo  {journal} {J.
  Phys. C: Solid State Phys.}\ }\textbf {\bibinfo {volume} {12}},\ \bibinfo
  {pages} {277} (\bibinfo {year} {1979})}\BibitemShut {NoStop}%
\bibitem [{\citenamefont {Hoffmann}\ \emph {et~al.}(1980)\citenamefont
  {Hoffmann}, \citenamefont {Shaik}, \citenamefont {Scott}, \citenamefont
  {Whangbo},\ and\ \citenamefont {Foshee}}]{Hoffmann1980}%
  \BibitemOpen
  \bibfield  {author} {\bibinfo {author} {\bibfnamefont {R.}~\bibnamefont
  {Hoffmann}}, \bibinfo {author} {\bibfnamefont {S.}~\bibnamefont {Shaik}},
  \bibinfo {author} {\bibfnamefont {J.~D.}\ \bibnamefont {Scott}}, \bibinfo
  {author} {\bibfnamefont {M.-H.}\ \bibnamefont {Whangbo}}, \ and\ \bibinfo
  {author} {\bibfnamefont {M.~J.}\ \bibnamefont {Foshee}},\ }\href@noop {}
  {\bibfield  {journal} {\bibinfo  {journal} {J. Solid State Chem.}\ }\textbf
  {\bibinfo {volume} {34}},\ \bibinfo {pages} {263} (\bibinfo {year}
  {1980})}\BibitemShut {NoStop}%
\bibitem [{\citenamefont {Shima}(1982)}]{Shima1982}%
  \BibitemOpen
  \bibfield  {author} {\bibinfo {author} {\bibfnamefont {N.}~\bibnamefont
  {Shima}},\ }\href@noop {} {\bibfield  {journal} {\bibinfo  {journal} {J.
  Phys. Soc. Jpn.}\ }\textbf {\bibinfo {volume} {51}},\ \bibinfo {pages} {11}
  (\bibinfo {year} {1982})}\BibitemShut {NoStop}%
\bibitem [{\citenamefont {Shima}(1983)}]{Shima1983}%
  \BibitemOpen
  \bibfield  {author} {\bibinfo {author} {\bibfnamefont {N.}~\bibnamefont
  {Shima}},\ }\href@noop {} {\bibfield  {journal} {\bibinfo  {journal} {J.
  Phys. Soc. Jpn.}\ }\textbf {\bibinfo {volume} {52}},\ \bibinfo {pages} {578}
  (\bibinfo {year} {1983})}\BibitemShut {NoStop}%
\bibitem [{\citenamefont {Canadell}\ \emph {et~al.}(1990)\citenamefont
  {Canadell}, \citenamefont {Rachidi}, \citenamefont {Pouget}, \citenamefont
  {Gressier}, \citenamefont {Meerschaut}, \citenamefont {Rouxel}, \citenamefont
  {Jung}, \citenamefont {Evain},\ and\ \citenamefont {Whangbo}}]{Canadell1990}%
  \BibitemOpen
  \bibfield  {author} {\bibinfo {author} {\bibfnamefont {E.}~\bibnamefont
  {Canadell}}, \bibinfo {author} {\bibfnamefont {I.~E.-I.}\ \bibnamefont
  {Rachidi}}, \bibinfo {author} {\bibfnamefont {J.~P.}\ \bibnamefont {Pouget}},
  \bibinfo {author} {\bibfnamefont {P.}~\bibnamefont {Gressier}}, \bibinfo
  {author} {\bibfnamefont {A.}~\bibnamefont {Meerschaut}}, \bibinfo {author}
  {\bibfnamefont {J.}~\bibnamefont {Rouxel}}, \bibinfo {author} {\bibfnamefont
  {D.}~\bibnamefont {Jung}}, \bibinfo {author} {\bibfnamefont {M.}~\bibnamefont
  {Evain}}, \ and\ \bibinfo {author} {\bibfnamefont {M.-H.}\ \bibnamefont
  {Whangbo}},\ }\href@noop {} {\bibfield  {journal} {\bibinfo  {journal}
  {Inorg. Chem.}\ }\textbf {\bibinfo {volume} {29}},\ \bibinfo {pages} {1401}
  (\bibinfo {year} {1990})}\BibitemShut {NoStop}%
\bibitem [{\citenamefont {Sch\"afer}\ \emph {et~al.}(2001)\citenamefont
  {Sch\"afer}, \citenamefont {Rotenberg}, \citenamefont {Kevan}, \citenamefont
  {Blaha}, \citenamefont {Claessen},\ and\ \citenamefont
  {Thorne}}]{Schafer2001}%
  \BibitemOpen
  \bibfield  {author} {\bibinfo {author} {\bibfnamefont {J.}~\bibnamefont
  {Sch\"afer}}, \bibinfo {author} {\bibfnamefont {E.}~\bibnamefont
  {Rotenberg}}, \bibinfo {author} {\bibfnamefont {S.~D.}\ \bibnamefont
  {Kevan}}, \bibinfo {author} {\bibfnamefont {P.}~\bibnamefont {Blaha}},
  \bibinfo {author} {\bibfnamefont {R.}~\bibnamefont {Claessen}}, \ and\
  \bibinfo {author} {\bibfnamefont {R.~E.}\ \bibnamefont {Thorne}},\
  }\href@noop {} {\bibfield  {journal} {\bibinfo  {journal} {Phys. Rev. Lett.}\
  }\textbf {\bibinfo {volume} {87}},\ \bibinfo {pages} {196403} (\bibinfo
  {year} {2001})}\BibitemShut {NoStop}%
\bibitem [{\citenamefont {Nicholson}\ \emph {et~al.}(2017)\citenamefont
  {Nicholson}, \citenamefont {Berthod}, \citenamefont {Puppin}, \citenamefont
  {Berger}, \citenamefont {Wolf}, \citenamefont {Hoesch},\ and\ \citenamefont
  {Monney}}]{Nicholson2017}%
  \BibitemOpen
  \bibfield  {author} {\bibinfo {author} {\bibfnamefont {C.~W.}\ \bibnamefont
  {Nicholson}}, \bibinfo {author} {\bibfnamefont {C.}~\bibnamefont {Berthod}},
  \bibinfo {author} {\bibfnamefont {M.}~\bibnamefont {Puppin}}, \bibinfo
  {author} {\bibfnamefont {H.}~\bibnamefont {Berger}}, \bibinfo {author}
  {\bibfnamefont {M.}~\bibnamefont {Wolf}}, \bibinfo {author} {\bibfnamefont
  {M.}~\bibnamefont {Hoesch}}, \ and\ \bibinfo {author} {\bibfnamefont
  {C.}~\bibnamefont {Monney}},\ }\href@noop {} {\bibfield  {journal} {\bibinfo
  {journal} {Phys. Rev. Lett.}\ }\textbf {\bibinfo {volume} {118}},\ \bibinfo
  {pages} {206401} (\bibinfo {year} {2017})}\BibitemShut {NoStop}%
\bibitem [{\citenamefont {Valbuena}\ \emph {et~al.}(2019)\citenamefont
  {Valbuena}, \citenamefont {Chudzinski}, \citenamefont {Pons}, \citenamefont
  {Conejeros}, \citenamefont {Alemany}, \citenamefont {Canadell}, \citenamefont
  {Berger}, \citenamefont {Frantzeskakis}, \citenamefont {Avila}, \citenamefont
  {Asensio}, \citenamefont {Giamarchi},\ and\ \citenamefont
  {Grioni}}]{Valbuena2019}%
  \BibitemOpen
  \bibfield  {author} {\bibinfo {author} {\bibfnamefont {M.~A.}\ \bibnamefont
  {Valbuena}}, \bibinfo {author} {\bibfnamefont {P.}~\bibnamefont
  {Chudzinski}}, \bibinfo {author} {\bibfnamefont {S.}~\bibnamefont {Pons}},
  \bibinfo {author} {\bibfnamefont {S.}~\bibnamefont {Conejeros}}, \bibinfo
  {author} {\bibfnamefont {P.}~\bibnamefont {Alemany}}, \bibinfo {author}
  {\bibfnamefont {E.}~\bibnamefont {Canadell}}, \bibinfo {author}
  {\bibfnamefont {H.}~\bibnamefont {Berger}}, \bibinfo {author} {\bibfnamefont
  {E.}~\bibnamefont {Frantzeskakis}}, \bibinfo {author} {\bibfnamefont
  {J.}~\bibnamefont {Avila}}, \bibinfo {author} {\bibfnamefont {M.~C.}\
  \bibnamefont {Asensio}}, \bibinfo {author} {\bibfnamefont {T.}~\bibnamefont
  {Giamarchi}}, \ and\ \bibinfo {author} {\bibfnamefont {M.}~\bibnamefont
  {Grioni}},\ }\href@noop {} {\bibfield  {journal} {\bibinfo  {journal} {Phys.
  Rev. B}\ }\textbf {\bibinfo {volume} {99}},\ \bibinfo {pages} {075118}
  (\bibinfo {year} {2019})}\BibitemShut {NoStop}%
\bibitem [{\citenamefont {Sch\"afer}\ \emph {et~al.}(2003)\citenamefont
  {Sch\"afer}, \citenamefont {Sing}, \citenamefont {Claessen}, \citenamefont
  {Rotenberg}, \citenamefont {Zhou}, \citenamefont {Thorne},\ and\
  \citenamefont {Kevan}}]{Schafer2003}%
  \BibitemOpen
  \bibfield  {author} {\bibinfo {author} {\bibfnamefont {J.}~\bibnamefont
  {Sch\"afer}}, \bibinfo {author} {\bibfnamefont {M.}~\bibnamefont {Sing}},
  \bibinfo {author} {\bibfnamefont {R.}~\bibnamefont {Claessen}}, \bibinfo
  {author} {\bibfnamefont {E.}~\bibnamefont {Rotenberg}}, \bibinfo {author}
  {\bibfnamefont {X.~J.}\ \bibnamefont {Zhou}}, \bibinfo {author}
  {\bibfnamefont {R.~E.}\ \bibnamefont {Thorne}}, \ and\ \bibinfo {author}
  {\bibfnamefont {S.~D.}\ \bibnamefont {Kevan}},\ }\href@noop {} {\bibfield
  {journal} {\bibinfo  {journal} {Phys. Rev. Lett.}\ }\textbf {\bibinfo
  {volume} {91}},\ \bibinfo {pages} {066401} (\bibinfo {year}
  {2003})}\BibitemShut {NoStop}%
\bibitem [{\citenamefont {Johannes}\ and\ \citenamefont {Mazin}(2008)}]{JM08}%
  \BibitemOpen
  \bibfield  {author} {\bibinfo {author} {\bibfnamefont {M.~D.}\ \bibnamefont
  {Johannes}}\ and\ \bibinfo {author} {\bibfnamefont {I.~I.}\ \bibnamefont
  {Mazin}},\ }\href@noop {} {\bibfield  {journal} {\bibinfo  {journal} {Phys.
  Rev. B}\ }\textbf {\bibinfo {volume} {77}},\ \bibinfo {pages} {165135}
  (\bibinfo {year} {2008})}\BibitemShut {NoStop}%
\bibitem [{\citenamefont {Hohenberg}\ and\ \citenamefont
  {Kohn}(1964)}]{HohKoh1964}%
  \BibitemOpen
  \bibfield  {author} {\bibinfo {author} {\bibfnamefont {P.}~\bibnamefont
  {Hohenberg}}\ and\ \bibinfo {author} {\bibfnamefont {W.}~\bibnamefont
  {Kohn}},\ }\href@noop {} {\bibfield  {journal} {\bibinfo  {journal} {Phys.
  Rev.}\ }\textbf {\bibinfo {volume} {136}},\ \bibinfo {pages} {B864} (\bibinfo
  {year} {1964})}\BibitemShut {NoStop}%
\bibitem [{\citenamefont {Kohn}\ and\ \citenamefont {Sham}(1965)}]{KohSha1965}%
  \BibitemOpen
  \bibfield  {author} {\bibinfo {author} {\bibfnamefont {W.}~\bibnamefont
  {Kohn}}\ and\ \bibinfo {author} {\bibfnamefont {L.~J.}\ \bibnamefont
  {Sham}},\ }\href@noop {} {\bibfield  {journal} {\bibinfo  {journal} {Phys.
  Rev.}\ }\textbf {\bibinfo {volume} {140}},\ \bibinfo {pages} {A1133}
  (\bibinfo {year} {1965})}\BibitemShut {NoStop}%
\bibitem [{\citenamefont {Soler}\ \emph {et~al.}(2002)\citenamefont {Soler},
  \citenamefont {Artacho}, \citenamefont {Gale}, \citenamefont {Garc\'ia},
  \citenamefont {Junquera}, \citenamefont {Ordej\'on},\ and\ \citenamefont
  {S\'anchez-Portal}}]{SolArt2002}%
  \BibitemOpen
  \bibfield  {author} {\bibinfo {author} {\bibfnamefont {J.~M.}\ \bibnamefont
  {Soler}}, \bibinfo {author} {\bibfnamefont {E.}~\bibnamefont {Artacho}},
  \bibinfo {author} {\bibfnamefont {J.~D.}\ \bibnamefont {Gale}}, \bibinfo
  {author} {\bibfnamefont {A.}~\bibnamefont {Garc\'ia}}, \bibinfo {author}
  {\bibfnamefont {J.}~\bibnamefont {Junquera}}, \bibinfo {author}
  {\bibfnamefont {P.}~\bibnamefont {Ordej\'on}}, \ and\ \bibinfo {author}
  {\bibfnamefont {D.}~\bibnamefont {S\'anchez-Portal}},\ }\href@noop {}
  {\bibfield  {journal} {\bibinfo  {journal} {J. Phys.: Condens. Matter}\
  }\textbf {\bibinfo {volume} {14}},\ \bibinfo {pages} {2745} (\bibinfo {year}
  {2002})}\BibitemShut {NoStop}%
\bibitem [{\citenamefont {Artacho}\ \emph {et~al.}(2008)\citenamefont
  {Artacho}, \citenamefont {Anglada}, \citenamefont {Di\'eguez}, \citenamefont
  {Gale}, \citenamefont {Garc\'ia}, \citenamefont {Junquera}, \citenamefont
  {Martin}, \citenamefont {Ordej\'on}, \citenamefont {Pruneda}, \citenamefont
  {S\'anchez-Portal},\ and\ \citenamefont {Soler}}]{ArtAng2008}%
  \BibitemOpen
  \bibfield  {author} {\bibinfo {author} {\bibfnamefont {E.}~\bibnamefont
  {Artacho}}, \bibinfo {author} {\bibfnamefont {E.}~\bibnamefont {Anglada}},
  \bibinfo {author} {\bibfnamefont {O.}~\bibnamefont {Di\'eguez}}, \bibinfo
  {author} {\bibfnamefont {J.~D.}\ \bibnamefont {Gale}}, \bibinfo {author}
  {\bibfnamefont {A.}~\bibnamefont {Garc\'ia}}, \bibinfo {author}
  {\bibfnamefont {J.}~\bibnamefont {Junquera}}, \bibinfo {author}
  {\bibfnamefont {R.~M.}\ \bibnamefont {Martin}}, \bibinfo {author}
  {\bibfnamefont {P.}~\bibnamefont {Ordej\'on}}, \bibinfo {author}
  {\bibfnamefont {J.~M.}\ \bibnamefont {Pruneda}}, \bibinfo {author}
  {\bibfnamefont {D.}~\bibnamefont {S\'anchez-Portal}}, \ and\ \bibinfo
  {author} {\bibfnamefont {J.~M.}\ \bibnamefont {Soler}},\ }\href@noop {}
  {\bibfield  {journal} {\bibinfo  {journal} {J. Phys.: Condens. Matter}\
  }\textbf {\bibinfo {volume} {20}},\ \bibinfo {pages} {064208} (\bibinfo
  {year} {2008})}\BibitemShut {NoStop}%
\bibitem [{\citenamefont {Perdew}\ \emph {et~al.}(1996)\citenamefont {Perdew},
  \citenamefont {Burke},\ and\ \citenamefont {Ernzerhof}}]{PBE96}%
  \BibitemOpen
  \bibfield  {author} {\bibinfo {author} {\bibfnamefont {J.~P.}\ \bibnamefont
  {Perdew}}, \bibinfo {author} {\bibfnamefont {K.}~\bibnamefont {Burke}}, \
  and\ \bibinfo {author} {\bibfnamefont {M.}~\bibnamefont {Ernzerhof}},\
  }\href@noop {} {\bibfield  {journal} {\bibinfo  {journal} {Phys. Rev. Lett.}\
  }\textbf {\bibinfo {volume} {77}},\ \bibinfo {pages} {3865} (\bibinfo {year}
  {1996})}\BibitemShut {NoStop}%
\bibitem [{\citenamefont {Troullier}\ and\ \citenamefont
  {Martins}(1991)}]{tro91}%
  \BibitemOpen
  \bibfield  {author} {\bibinfo {author} {\bibfnamefont {N.}~\bibnamefont
  {Troullier}}\ and\ \bibinfo {author} {\bibfnamefont {J.~L.}\ \bibnamefont
  {Martins}},\ }\href@noop {} {\bibfield  {journal} {\bibinfo  {journal} {Phys.
  Rev. B}\ }\textbf {\bibinfo {volume} {43}},\ \bibinfo {pages} {1993}
  (\bibinfo {year} {1991})}\BibitemShut {NoStop}%
\bibitem [{\citenamefont {Kleinman}\ and\ \citenamefont
  {Bylander}(1982)}]{klby82}%
  \BibitemOpen
  \bibfield  {author} {\bibinfo {author} {\bibfnamefont {L.}~\bibnamefont
  {Kleinman}}\ and\ \bibinfo {author} {\bibfnamefont {D.~M.}\ \bibnamefont
  {Bylander}},\ }\href@noop {} {\bibfield  {journal} {\bibinfo  {journal}
  {Phys. Rev. Lett.}\ }\textbf {\bibinfo {volume} {48}},\ \bibinfo {pages}
  {1425} (\bibinfo {year} {1982})}\BibitemShut {NoStop}%
\bibitem [{\citenamefont {Louie}\ \emph {et~al.}(1982)\citenamefont {Louie},
  \citenamefont {Froyen},\ and\ \citenamefont {Cohen}}]{LFC82}%
  \BibitemOpen
  \bibfield  {author} {\bibinfo {author} {\bibfnamefont {S.~G.}\ \bibnamefont
  {Louie}}, \bibinfo {author} {\bibfnamefont {S.}~\bibnamefont {Froyen}}, \
  and\ \bibinfo {author} {\bibfnamefont {M.~L.}\ \bibnamefont {Cohen}},\
  }\href@noop {} {\bibfield  {journal} {\bibinfo  {journal} {Phys. Rev. B}\
  }\textbf {\bibinfo {volume} {26}},\ \bibinfo {pages} {1738} (\bibinfo {year}
  {1982})}\BibitemShut {NoStop}%
\bibitem [{\citenamefont {Artacho}\ \emph {et~al.}(1999)\citenamefont
  {Artacho}, \citenamefont {S\'{a}nchez-Portal}, \citenamefont {Ordej\'{o}n},
  \citenamefont {Garc\'{i}a},\ and\ \citenamefont {Soler}}]{arsan99}%
  \BibitemOpen
  \bibfield  {author} {\bibinfo {author} {\bibfnamefont {E.}~\bibnamefont
  {Artacho}}, \bibinfo {author} {\bibfnamefont {D.}~\bibnamefont
  {S\'{a}nchez-Portal}}, \bibinfo {author} {\bibfnamefont {P.}~\bibnamefont
  {Ordej\'{o}n}}, \bibinfo {author} {\bibfnamefont {A.}~\bibnamefont
  {Garc\'{i}a}}, \ and\ \bibinfo {author} {\bibfnamefont {J.~M.}\ \bibnamefont
  {Soler}},\ }\href@noop {} {\bibfield  {journal} {\bibinfo  {journal} {Phys.
  Stat. Solidi (b)}\ }\textbf {\bibinfo {volume} {215}},\ \bibinfo {pages}
  {809} (\bibinfo {year} {1999})}\BibitemShut {NoStop}%
\bibitem [{\citenamefont {Monkhorst}\ and\ \citenamefont
  {Pack}(1976)}]{MonPac76}%
  \BibitemOpen
  \bibfield  {author} {\bibinfo {author} {\bibfnamefont {H.~J.}\ \bibnamefont
  {Monkhorst}}\ and\ \bibinfo {author} {\bibfnamefont {J.~D.}\ \bibnamefont
  {Pack}},\ }\href@noop {} {\bibfield  {journal} {\bibinfo  {journal} {Phys.
  Rev. B}\ }\textbf {\bibinfo {volume} {13}},\ \bibinfo {pages} {5188}
  (\bibinfo {year} {1976})}\BibitemShut {NoStop}%
\bibitem [{\citenamefont {Ziman}(1972)}]{Ziman1972}%
  \BibitemOpen
  \bibfield  {author} {\bibinfo {author} {\bibfnamefont {J.~M.}\ \bibnamefont
  {Ziman}},\ }\href@noop {} {\emph {\bibinfo {title} {{Principles f the Theory
  of Solids}}}}\ (\bibinfo  {publisher} {Cambridge University Press,
  Cambridge},\ \bibinfo {year} {1972})\BibitemShut {NoStop}%
\bibitem [{not()}]{note1}%
  \BibitemOpen
  \bibinfo {note} {{Although the associated fifth band occurs in most DFT
  studies it can slightly shift above the Fermi level depending on some
  computational details; see ref~\cite{Valbuena2019}.}}\BibitemShut {Stop}%
\bibitem [{\citenamefont {Moudden}\ \emph {et~al.}(1990)\citenamefont
  {Moudden}, \citenamefont {Axe}, \citenamefont {Monceau},\ and\ \citenamefont
  {Levy}}]{Moudden1990}%
  \BibitemOpen
  \bibfield  {author} {\bibinfo {author} {\bibfnamefont {A.~H.}\ \bibnamefont
  {Moudden}}, \bibinfo {author} {\bibfnamefont {J.~D.}\ \bibnamefont {Axe}},
  \bibinfo {author} {\bibfnamefont {P.}~\bibnamefont {Monceau}}, \ and\
  \bibinfo {author} {\bibfnamefont {F.}~\bibnamefont {Levy}},\ }\href@noop {}
  {\bibfield  {journal} {\bibinfo  {journal} {Phys. Rev. Lett.}\ }\textbf
  {\bibinfo {volume} {65}},\ \bibinfo {pages} {223} (\bibinfo {year}
  {1990})}\BibitemShut {NoStop}%
\bibitem [{\citenamefont {Rouzi\`ere}\ \emph {et~al.}(1996)\citenamefont
  {Rouzi\`ere}, \citenamefont {Ravy}, \citenamefont {Pouget},\ and\
  \citenamefont {Thorne}}]{Rouziere1996}%
  \BibitemOpen
  \bibfield  {author} {\bibinfo {author} {\bibfnamefont {S.}~\bibnamefont
  {Rouzi\`ere}}, \bibinfo {author} {\bibfnamefont {S.}~\bibnamefont {Ravy}},
  \bibinfo {author} {\bibfnamefont {J.~P.}\ \bibnamefont {Pouget}}, \ and\
  \bibinfo {author} {\bibfnamefont {R.~E.}\ \bibnamefont {Thorne}},\
  }\href@noop {} {\bibfield  {journal} {\bibinfo  {journal} {Solid State
  Comm.}\ }\textbf {\bibinfo {volume} {97}},\ \bibinfo {pages} {1073 }
  (\bibinfo {year} {1996})}\BibitemShut {NoStop}%
\bibitem [{\citenamefont {Divilov}\ \emph {et~al.}(2020)\citenamefont
  {Divilov}, \citenamefont {Mayo}, \citenamefont {Soler},\ and\ \citenamefont
  {Yndurain}}]{Divilov2020}%
  \BibitemOpen
  \bibfield  {author} {\bibinfo {author} {\bibfnamefont {S.}~\bibnamefont
  {Divilov}}, \bibinfo {author} {\bibfnamefont {S.~G.}\ \bibnamefont {Mayo}},
  \bibinfo {author} {\bibfnamefont {J.~M.}\ \bibnamefont {Soler}}, \ and\
  \bibinfo {author} {\bibfnamefont {F.}~\bibnamefont {Yndurain}},\ }\href@noop
  {} {\  (\bibinfo {year} {2020})},\ \Eprint {http://arxiv.org/abs/2007.04737}
  {arXiv:2007.04737 [cond-mat.mtrl-sci]} \BibitemShut {NoStop}%
\bibitem [{\citenamefont {Heil}\ \emph {et~al.}(2014)\citenamefont {Heil},
  \citenamefont {Sormann}, \citenamefont {Boeri}, \citenamefont {Aichhorn},\
  and\ \citenamefont {von~der Linden}}]{Heil2014}%
  \BibitemOpen
  \bibfield  {author} {\bibinfo {author} {\bibfnamefont {C.}~\bibnamefont
  {Heil}}, \bibinfo {author} {\bibfnamefont {H.}~\bibnamefont {Sormann}},
  \bibinfo {author} {\bibfnamefont {L.}~\bibnamefont {Boeri}}, \bibinfo
  {author} {\bibfnamefont {M.}~\bibnamefont {Aichhorn}}, \ and\ \bibinfo
  {author} {\bibfnamefont {W.}~\bibnamefont {von~der Linden}},\ }\href@noop {}
  {\bibfield  {journal} {\bibinfo  {journal} {Phys. Rev. B}\ }\textbf {\bibinfo
  {volume} {90}},\ \bibinfo {pages} {115143} (\bibinfo {year}
  {2014})}\BibitemShut {NoStop}%
\bibitem [{\citenamefont {Pouget}\ and\ \citenamefont
  {Com\`es}(1989)}]{Pouget1989}%
  \BibitemOpen
  \bibfield  {author} {\bibinfo {author} {\bibfnamefont {J.~P.}\ \bibnamefont
  {Pouget}}\ and\ \bibinfo {author} {\bibnamefont {Com\`es}},\ }in\ \href@noop
  {} {\emph {\bibinfo {booktitle} {Charge Density Waves in Solids}}},\ \bibinfo
  {editor} {edited by\ \bibinfo {editor} {\bibfnamefont {L.}~\bibnamefont
  {Gor'kov}}\ and\ \bibinfo {editor} {\bibfnamefont {G.}~\bibnamefont
  {Gr\"uner}}}\ (\bibinfo  {publisher} {Modern Problems in Condensed Matter
  Sciences, vol. 25, Elsevier},\ \bibinfo {year} {1989})\ Chap.~\bibinfo
  {chapter} {3}, pp.\ \bibinfo {pages} {85--136}\BibitemShut {NoStop}%
\bibitem [{Mor()}]{Moret}%
  \BibitemOpen
  \bibinfo {note} {{R. Moret, unpublished results.}}\BibitemShut {Stop}%
\bibitem [{\citenamefont {Pouget}\ \emph {et~al.}(1983)\citenamefont {Pouget},
  \citenamefont {Moret}, \citenamefont {Meerschaut}, \citenamefont {Guemas},\
  and\ \citenamefont {Rouxel}}]{Pouget1983Nb}%
  \BibitemOpen
  \bibfield  {author} {\bibinfo {author} {\bibfnamefont {J.~P.}\ \bibnamefont
  {Pouget}}, \bibinfo {author} {\bibfnamefont {R.}~\bibnamefont {Moret}},
  \bibinfo {author} {\bibfnamefont {A.}~\bibnamefont {Meerschaut}}, \bibinfo
  {author} {\bibfnamefont {L.}~\bibnamefont {Guemas}}, \ and\ \bibinfo {author}
  {\bibfnamefont {J.}~\bibnamefont {Rouxel}},\ }\href@noop {} {\bibfield
  {journal} {\bibinfo  {journal} {J. Phys. Colloques}\ }\textbf {\bibinfo
  {volume} {44}},\ \bibinfo {pages} {C3\textendash1729} (\bibinfo {year}
  {1983})}\BibitemShut {NoStop}%
\bibitem [{\citenamefont {Guster}\ \emph {et~al.}(2020)\citenamefont {Guster},
  \citenamefont {Pruneda}, \citenamefont {Ordej\'on}, \citenamefont
  {Canadell},\ and\ \citenamefont {Pouget}}]{Guster2020}%
  \BibitemOpen
  \bibfield  {author} {\bibinfo {author} {\bibfnamefont {B.}~\bibnamefont
  {Guster}}, \bibinfo {author} {\bibfnamefont {M.}~\bibnamefont {Pruneda}},
  \bibinfo {author} {\bibfnamefont {P.}~\bibnamefont {Ordej\'on}}, \bibinfo
  {author} {\bibfnamefont {E.}~\bibnamefont {Canadell}}, \ and\ \bibinfo
  {author} {\bibfnamefont {J.-P.}\ \bibnamefont {Pouget}},\ }\href@noop {}
  {\bibfield  {journal} {\bibinfo  {journal} {Journal of Physics: Condensed
  Matter}\ }\textbf {\bibinfo {volume} {32}},\ \bibinfo {pages} {345701}
  (\bibinfo {year} {2020})}\BibitemShut {NoStop}%
\bibitem [{\citenamefont {Hasegawa}\ and\ \citenamefont
  {Fukuyama}(1986)}]{Hasegawa1986}%
  \BibitemOpen
  \bibfield  {author} {\bibinfo {author} {\bibfnamefont {Y.}~\bibnamefont
  {Hasegawa}}\ and\ \bibinfo {author} {\bibfnamefont {H.}~\bibnamefont
  {Fukuyama}},\ }\href@noop {} {\bibfield  {journal} {\bibinfo  {journal} {J.
  Phys. Soc. Jpn.}\ }\textbf {\bibinfo {volume} {55}},\ \bibinfo {pages} {3978}
  (\bibinfo {year} {1986})}\BibitemShut {NoStop}%
\bibitem [{\citenamefont {\v{S}aub}\ \emph {et~al.}(1976)\citenamefont
  {\v{S}aub}, \citenamefont {Bari\v{s}i\'c},\ and\ \citenamefont
  {Friedel}}]{Saub1976}%
  \BibitemOpen
  \bibfield  {author} {\bibinfo {author} {\bibfnamefont {K.}~\bibnamefont
  {\v{S}aub}}, \bibinfo {author} {\bibfnamefont {S.}~\bibnamefont
  {Bari\v{s}i\'c}}, \ and\ \bibinfo {author} {\bibfnamefont {J.}~\bibnamefont
  {Friedel}},\ }\href@noop {} {\bibfield  {journal} {\bibinfo  {journal} {Phys.
  Lett.}\ }\textbf {\bibinfo {volume} {56A}},\ \bibinfo {pages} {302} (\bibinfo
  {year} {1976})}\BibitemShut {NoStop}%
\bibitem [{\citenamefont {McMillan}(1977)}]{McMillan1977}%
  \BibitemOpen
  \bibfield  {author} {\bibinfo {author} {\bibfnamefont {W.~L.}\ \bibnamefont
  {McMillan}},\ }\href@noop {} {\bibfield  {journal} {\bibinfo  {journal}
  {Phys. Rev. B}\ }\textbf {\bibinfo {volume} {16}},\ \bibinfo {pages} {643}
  (\bibinfo {year} {1977})}\BibitemShut {NoStop}%
\bibitem [{\citenamefont {Rossnagel}(2011)}]{Rossnagel2011}%
  \BibitemOpen
  \bibfield  {author} {\bibinfo {author} {\bibfnamefont {K.}~\bibnamefont
  {Rossnagel}},\ }\href@noop {} {\bibfield  {journal} {\bibinfo  {journal} {J.
  Phys.: Condens. Matter}\ }\textbf {\bibinfo {volume} {23}},\ \bibinfo {pages}
  {213001} (\bibinfo {year} {2011})}\BibitemShut {NoStop}%
\bibitem [{\citenamefont {Requardt}\ \emph {et~al.}(2002)\citenamefont
  {Requardt}, \citenamefont {Lorenzo}, \citenamefont {Currat}, \citenamefont
  {Monceau},\ and\ \citenamefont {Krisch}}]{Requardt2002}%
  \BibitemOpen
  \bibfield  {author} {\bibinfo {author} {\bibfnamefont {H.}~\bibnamefont
  {Requardt}}, \bibinfo {author} {\bibfnamefont {J.~E.}\ \bibnamefont
  {Lorenzo}}, \bibinfo {author} {\bibfnamefont {R.}~\bibnamefont {Currat}},
  \bibinfo {author} {\bibfnamefont {P.}~\bibnamefont {Monceau}}, \ and\
  \bibinfo {author} {\bibfnamefont {M.}~\bibnamefont {Krisch}},\ }\href@noop {}
  {\bibfield  {journal} {\bibinfo  {journal} {Phys. Rev. B}\ }\textbf {\bibinfo
  {volume} {66}},\ \bibinfo {pages} {214303} (\bibinfo {year}
  {2002})}\BibitemShut {NoStop}%
\bibitem [{\citenamefont {Ottolenghi}\ and\ \citenamefont
  {Pouget}(1996)}]{Ottolenghi1996}%
  \BibitemOpen
  \bibfield  {author} {\bibinfo {author} {\bibfnamefont {A.}~\bibnamefont
  {Ottolenghi}}\ and\ \bibinfo {author} {\bibfnamefont {J.-P.}\ \bibnamefont
  {Pouget}},\ }\href@noop {} {\bibfield  {journal} {\bibinfo  {journal} {J.
  Phys. I}\ }\textbf {\bibinfo {volume} {6}},\ \bibinfo {pages} {1059}
  (\bibinfo {year} {1996})}\BibitemShut {NoStop}%
\bibitem [{\citenamefont {Roussel}\ \emph {et~al.}(2000)\citenamefont
  {Roussel}, \citenamefont {Labb\'e}, \citenamefont {Leligny}, \citenamefont
  {Groult}, \citenamefont {Foury-Leylekian},\ and\ \citenamefont
  {Pouget}}]{Roussel2000}%
  \BibitemOpen
  \bibfield  {author} {\bibinfo {author} {\bibfnamefont {P.}~\bibnamefont
  {Roussel}}, \bibinfo {author} {\bibfnamefont {P.}~\bibnamefont {Labb\'e}},
  \bibinfo {author} {\bibfnamefont {H.}~\bibnamefont {Leligny}}, \bibinfo
  {author} {\bibfnamefont {D.}~\bibnamefont {Groult}}, \bibinfo {author}
  {\bibfnamefont {P.}~\bibnamefont {Foury-Leylekian}}, \ and\ \bibinfo {author}
  {\bibfnamefont {J.~P.}\ \bibnamefont {Pouget}},\ }\href@noop {} {\bibfield
  {journal} {\bibinfo  {journal} {Phys. Rev. B}\ }\textbf {\bibinfo {volume}
  {62}},\ \bibinfo {pages} {176} (\bibinfo {year} {2000})}\BibitemShut
  {NoStop}%
\bibitem [{\citenamefont {Schutte}\ and\ \citenamefont
  {Boer}(1993)}]{Schutte1993}%
  \BibitemOpen
  \bibfield  {author} {\bibinfo {author} {\bibfnamefont {W.~J.}\ \bibnamefont
  {Schutte}}\ and\ \bibinfo {author} {\bibfnamefont {J.~L.~D.}\ \bibnamefont
  {Boer}},\ }\href@noop {} {\bibfield  {journal} {\bibinfo  {journal} {Acta
  Crystallogr. B}\ }\textbf {\bibinfo {volume} {49}},\ \bibinfo {pages} {579}
  (\bibinfo {year} {1993})}\BibitemShut {NoStop}%
\bibitem [{\citenamefont {Sandre}\ \emph {et~al.}(2001)\citenamefont {Sandre},
  \citenamefont {Foury-Leylekian}, \citenamefont {Ravy},\ and\ \citenamefont
  {Pouget}}]{Sandre2001}%
  \BibitemOpen
  \bibfield  {author} {\bibinfo {author} {\bibfnamefont {E.}~\bibnamefont
  {Sandre}}, \bibinfo {author} {\bibfnamefont {P.}~\bibnamefont
  {Foury-Leylekian}}, \bibinfo {author} {\bibfnamefont {S.}~\bibnamefont
  {Ravy}}, \ and\ \bibinfo {author} {\bibfnamefont {J.-P.}\ \bibnamefont
  {Pouget}},\ }\href@noop {} {\bibfield  {journal} {\bibinfo  {journal} {Phys.
  Rev. Lett.}\ }\textbf {\bibinfo {volume} {86}},\ \bibinfo {pages} {5100}
  (\bibinfo {year} {2001})}\BibitemShut {NoStop}%
\bibitem [{Gus({\natexlab{a}})}]{Guster2020a}%
  \BibitemOpen
  \bibinfo {note} {{B. Guster, M. Pruneda, P. Ordej\'on, E. Canadell, and J.-P.
  Pouget, work in progress.}}\BibitemShut {Stop}%
\bibitem [{\citenamefont {Whangbo}\ \emph {et~al.}(1991)\citenamefont
  {Whangbo}, \citenamefont {Canadell}, \citenamefont {Foury},\ and\
  \citenamefont {Pouget}}]{Whangbo1991}%
  \BibitemOpen
  \bibfield  {author} {\bibinfo {author} {\bibfnamefont {M.-H.}\ \bibnamefont
  {Whangbo}}, \bibinfo {author} {\bibfnamefont {E.}~\bibnamefont {Canadell}},
  \bibinfo {author} {\bibfnamefont {P.}~\bibnamefont {Foury}}, \ and\ \bibinfo
  {author} {\bibfnamefont {J.-P.}\ \bibnamefont {Pouget}},\ }\href@noop {}
  {\bibfield  {journal} {\bibinfo  {journal} {Science}\ }\textbf {\bibinfo
  {volume} {252}},\ \bibinfo {pages} {96} (\bibinfo {year} {1991})}\BibitemShut
  {NoStop}%
\bibitem [{\citenamefont {Foury-Leylekian}\ \emph {et~al.}(2010)\citenamefont
  {Foury-Leylekian}, \citenamefont {Pouget}, \citenamefont {Lee}, \citenamefont
  {Nieminen}, \citenamefont {Ordej\'on},\ and\ \citenamefont
  {Canadell}}]{Foury2010}%
  \BibitemOpen
  \bibfield  {author} {\bibinfo {author} {\bibfnamefont {P.}~\bibnamefont
  {Foury-Leylekian}}, \bibinfo {author} {\bibfnamefont {J.-P.}\ \bibnamefont
  {Pouget}}, \bibinfo {author} {\bibfnamefont {Y.-J.}\ \bibnamefont {Lee}},
  \bibinfo {author} {\bibfnamefont {R.}~\bibnamefont {Nieminen}}, \bibinfo
  {author} {\bibfnamefont {P.}~\bibnamefont {Ordej\'on}}, \ and\ \bibinfo
  {author} {\bibfnamefont {E.}~\bibnamefont {Canadell}},\ }\href@noop {}
  {\bibfield  {journal} {\bibinfo  {journal} {Phys. Rev. B}\ }\textbf {\bibinfo
  {volume} {82}},\ \bibinfo {pages} {134116} (\bibinfo {year}
  {2010})}\BibitemShut {NoStop}%
\bibitem [{Gus({\natexlab{b}})}]{Guster2020b}%
  \BibitemOpen
  \bibinfo {note} {{B. Guster, M. Pruneda, P. Ordej\'on, E. Canadell, and J.-P.
  Pouget, work in progress.}}\BibitemShut {Stop}%
\end{thebibliography}%

\onecolumngrid
\pagebreak
\begin{center}
\textbf{\large Supplementary Information for "Electron-hole response function of transition metal trichalcogenides NbSe$_3$ and monoclinic-TaS$_3$"}
\end{center}

\setcounter{equation}{0}
\setcounter{figure}{0}
\setcounter{table}{0}
\setcounter{page}{1}
\makeatletter
\renewcommand{\theequation}{S\arabic{equation}}
\renewcommand{\thefigure}{S\arabic{figure}}
\renewcommand{\bibnumfmt}[1]{[S#1]}
\renewcommand{\citenumfont}[1]{S#1}

\vspace{50pt}

\begin{center}
    \textbf{\Large Contents}
\end{center}

\noindent
\begin{itemize}
    \item[S1.]{{Thermal dependence of the transverse scans of the Lindhard response of $m$-\tas~along the two diagonal ($a^{*}\pm c^{*}$) directions. Note that these scans link the two transverse (0 $a^{*}$,0  $c^{*}$) and (1/2 $a^{*}$,1/2 $c^{*}$) low temperature maxima of the Lindhard response.}}
    \item[S2.]{{Thermal dependence of the transverse scans of the Lindhard response of \nbse~along the two diagonal ($a^{*}\pm c^{*}$) directions. Note that these scans link the two transverse (0 $a^{*}$,0  $c^{*}$) and (1/2 $a^{*}$,1/2 $c^{*}$) low temperature maxima of the Lindhard response.}} 
    \item[S3.]{{Separate contribution of (a) the two outer bands (chains I based) and (b) the two inner bands (chains III based) to the longitudinal scans of the (0, \textit{q}, 0) and (1/2, \textit{q}, 1/2) Lindhard responses of $m$-\tas~at 10 K.}} 
\end{itemize}

\begin{figure}[!hptb]
    \centering
    \includegraphics[width=0.45\textwidth]{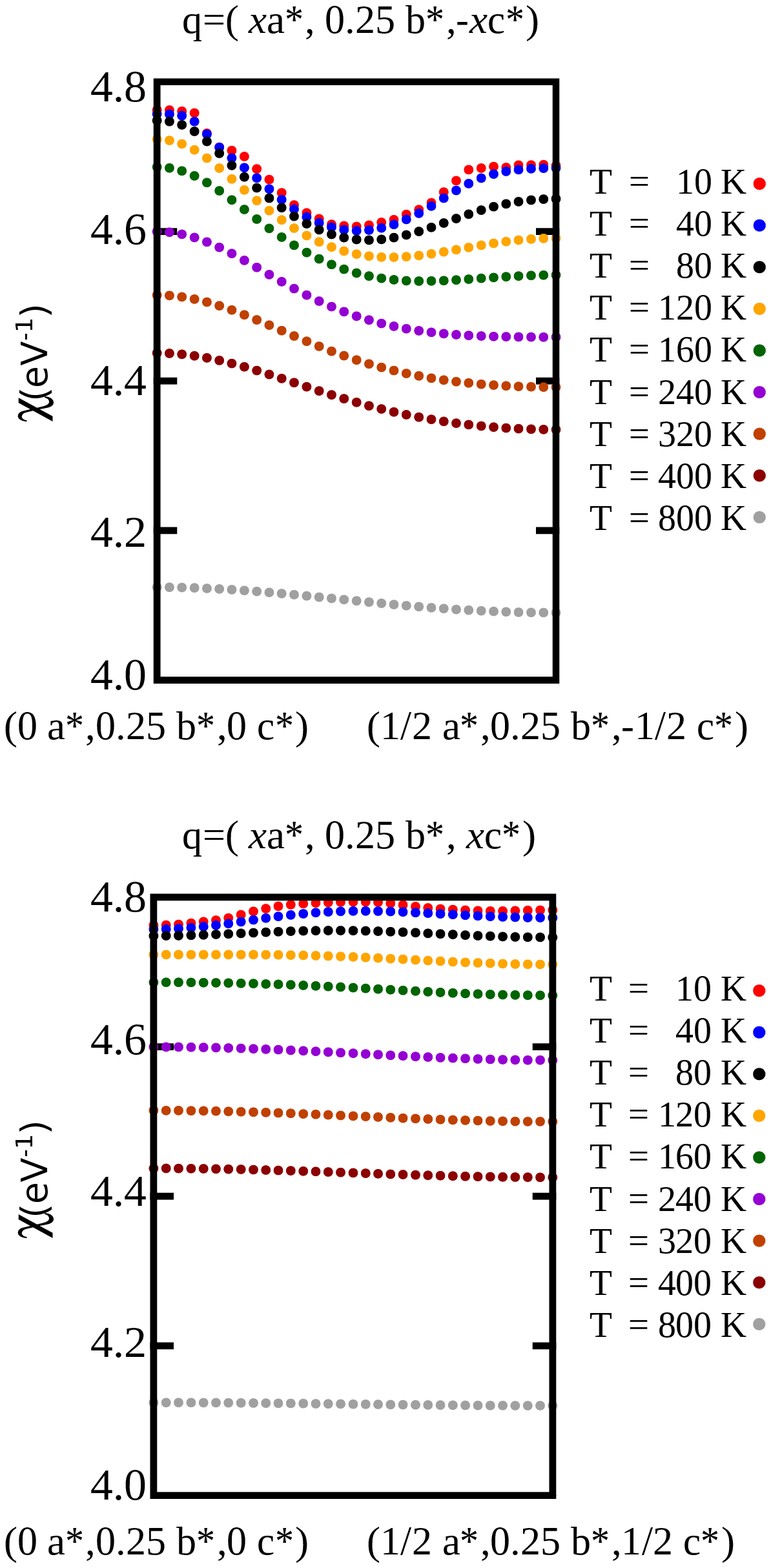}
    \caption{Thermal dependence of the transverse scans of the Lindhard response of $m$-\tas~along the two diagonal ($a^{*}\pm c^{*}$) directions. Note that these scans link the two transverse (0 $a^{*}$,0  $c^{*}$) and (1/2 $a^{*}$,1/2 $c^{*}$) low temperature maxima of the Lindhard response.}
    \label{fig:tas3_lrf_trans_scan_ac}
\end{figure}

\begin{figure}[!hptb]
    \centering
    \includegraphics[width=0.45\textwidth]{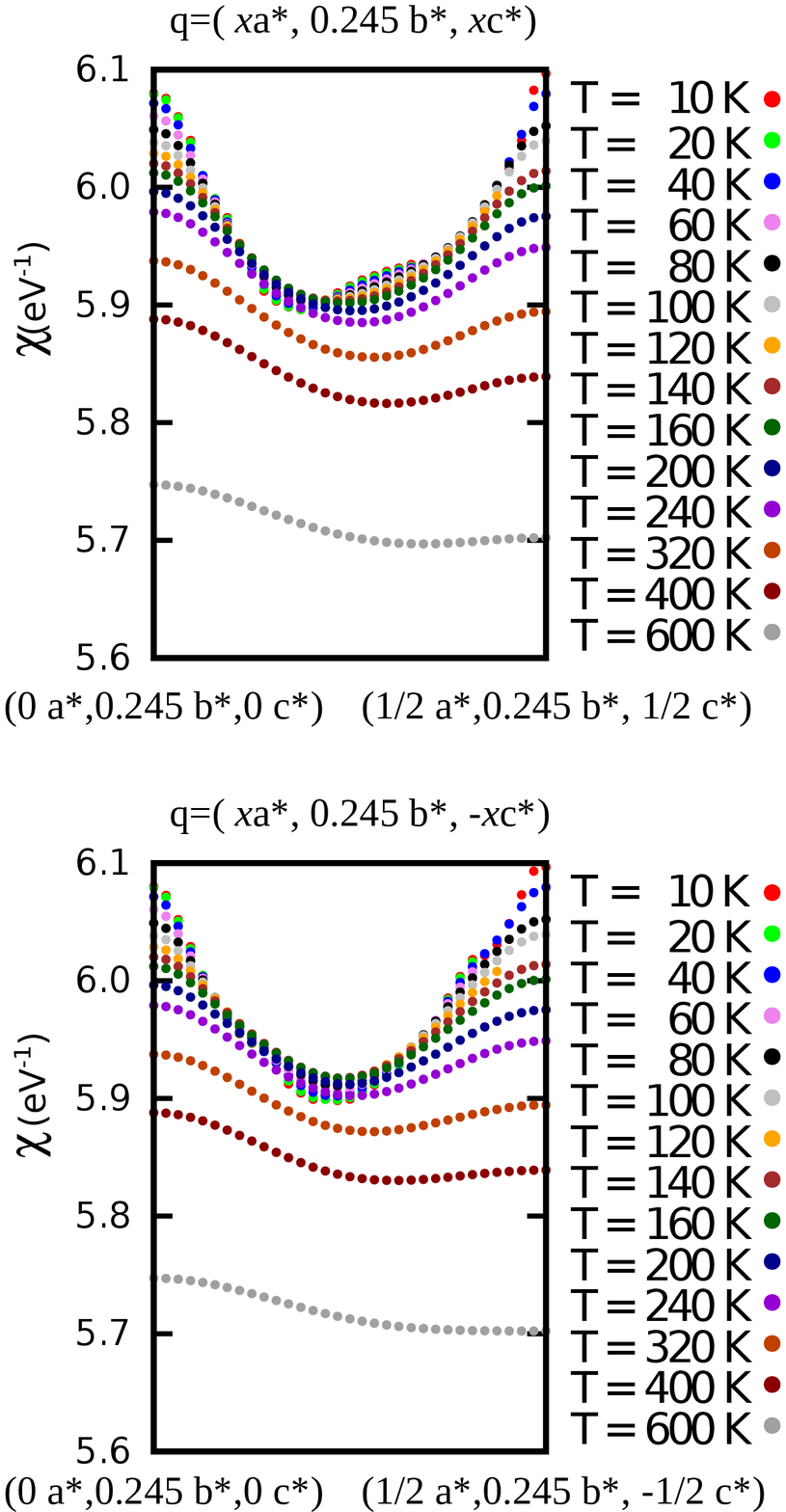}
    \caption{Thermal dependence of the transverse scans of the Lindhard response of \nbse~along the two diagonal ($a^{*}\pm c^{*}$) directions. Note that these scans link the two transverse (0 $a^{*}$,0  $c^{*}$) and (1/2 $a^{*}$,1/2 $c^{*}$) low temperature maxima of the Lindhard response.}
    \label{fig:nbse3_lrf_trans_scan_ac}
\end{figure}

\begin{figure}[!hptb]
    \centering
    \includegraphics[width=0.45\textwidth]{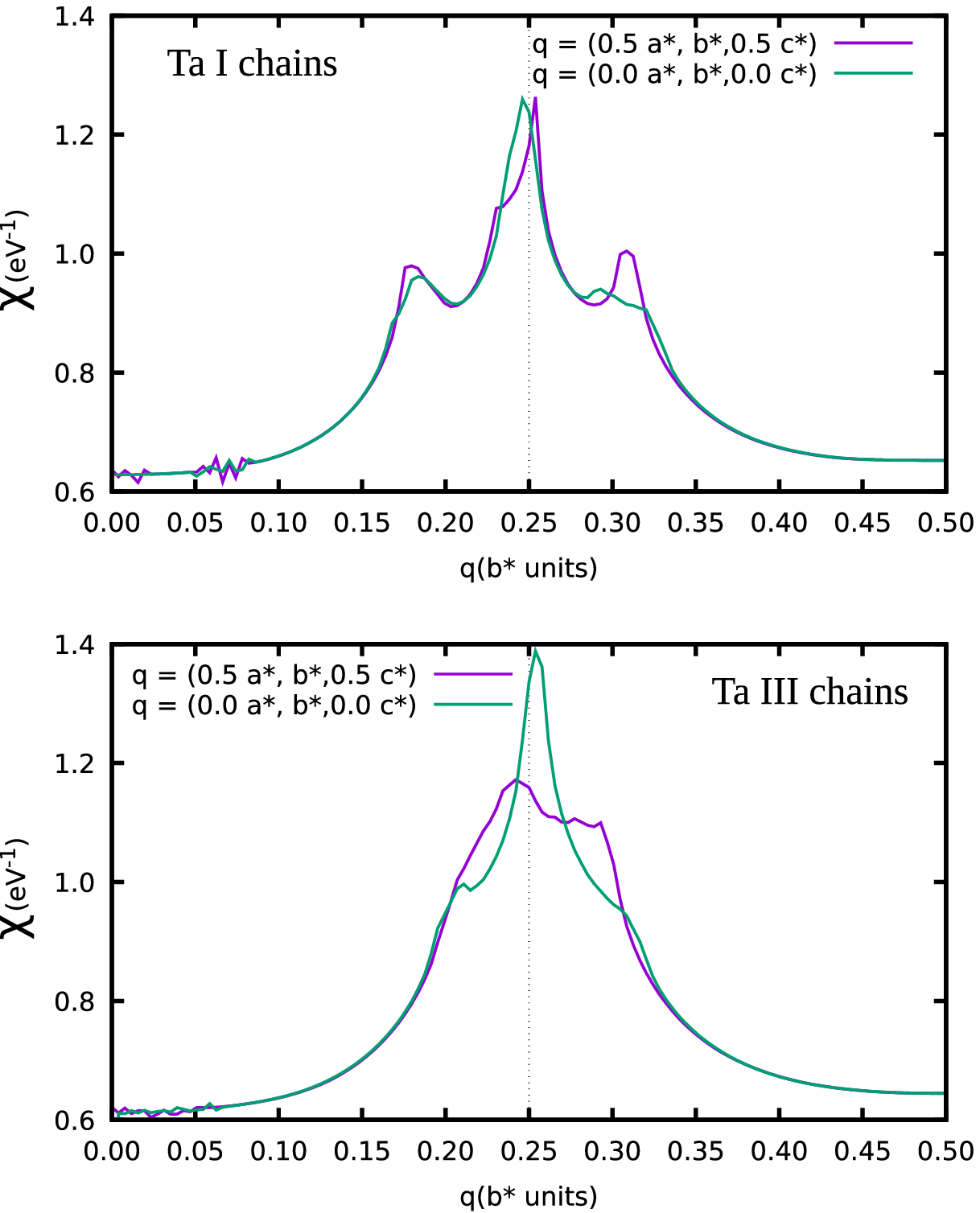}
    \caption{Separate contribution of (a) the two outer bands (chains I based) and (b) the two inner bands (chains III based) to the longitudinal scans of the (0, \textit{q}, 0) and (1/2, \textit{q}, 1/2) Lindhard responses of $m$-\tas~at 10 K.}
    \label{fig:lrf_band_to_band}
\end{figure}

\end{document}